\documentclass[11pt, oneside]{article}
\usepackage{geometry}                		
\RequirePackage[OT1]{fontenc}
\RequirePackage{amsthm,amsmath}
\usepackage[authoryear]{natbib}
\RequirePackage[colorlinks,linkcolor={blue},citecolor={blue},urlcolor={blue}]{hyperref}

\usepackage[english]{babel}
\usepackage[utf8]{inputenc}
\usepackage{graphicx}
\usepackage{epsfig}

\usepackage{url}            
\usepackage{booktabs}       
\usepackage{amsfonts}       
\usepackage{nicefrac}       
\usepackage{microtype}      
\usepackage{multirow}
\usepackage{amssymb}          
\usepackage{subcaption}
\usepackage{psfrag}
\usepackage{siunitx}
\usepackage{caption}
\usepackage{float}
\usepackage{mathtools}
\usepackage{enumitem}
\usepackage{algorithmic}
\usepackage{algorithm}
\usepackage{color}

\newcommand{\bs}[1]{{\boldsymbol{#1}}}
\newcommand{\norm}[1]{\left\lVert#1\right\rVert}
\newcommand{\mean}[1]{\left\langle#1\right\rangle}
\newcommand{\abs}[1]{\left\lvert#1\right\rvert}
\newcommand{\tL}{\text{L}}
\newcommand{\AL}{\text{AL}}
\newcommand{\enet}{\text{E}}
\newcommand{\aenet}{\text{AE}}
\newcommand{\saenet}{\text{SAE}}

\newcommand{\prob}{\mathbb{P}}
\newcommand{\E}{\mathbb{E}}
\newcommand{\R}{\mathbb{R}}
\newcommand{\xmat}{\bs{X}}
\newcommand{\y}{\bs{y}}
\newcommand{\aux}{\bs{U}}
\newcommand{\e}{\bs{e}}

\theoremstyle{plain}
\newtheorem{theorem}{Theorem}[section]
\newtheorem{lemma}{Lemma}[section]
\newtheorem{ass}{Assumption}[section]

\newtheorem{prop}{Proposition}[section]
\newtheorem{cor}{Corollary}[section]
\newtheorem{rem}{\textit{Remark}}[section]

\def\T{{ \mathrm{\scriptscriptstyle T} }}

\DeclareMathOperator*{\argmin}{arg\,min\,\,}

\title{Structure Adaptive Elastic-Net}
\author{Sandipan Pramanik\thanks{Corresponding author. Email: \href{mailto:sandy.pramanik@gmail.com}{sandy.pramanik@gmail.com}} and Xianyang Zhang\thanks{Email: \href{mailto:zhangxiany@stat.tamu.edu}{zhangxiany@stat.tamu.edu}}\\
Department of Statistics, Texas A\&M University}
\date{\ }							

\begin{document}

\maketitle

\begin{abstract}
Penalized linear regression is of fundamental importance in high-dimensional statistics and has been routinely used to regress a response on a high-dimensional set of predictors. In many scientific applications, there exists external information that encodes the predictive power and sparsity structure of the predictors. In this article, we propose the Structure Adaptive Elastic-Net (SA-Enet), which provides a new framework for incorporating potentially useful side information into a penalized regression. The basic idea is to translate the external information into different penalization strengths for the regression coefficients. We particularly focus on group and covariate-dependent structures and study the risk properties of the resulting estimator. To this, we generalize the state evolution framework recently introduced for the analysis of the approximate message-passing algorithm to the SA-Enet framework. We show that the finite sample risk of the SA-Enet estimator is consistent with the theoretical risk predicted by the state evolution equation. Our theory suggests that the SA-Enet with an informative group or covariate structure can outperform the Lasso, Adaptive Lasso, Sparse Group Lasso, Feature-weighted Elastic-Net, and Graper. 
This evidence is further confirmed in our numerical studies. We also demonstrate the usefulness and the superiority of our method for leukemia data from molecular biology and precision medicine.
\end{abstract}



\tableofcontents


\section{Introduction}\label{sec: Introduction}

High-dimensional data occur very frequently and are especially common in genomics studies, where one of the important scientific interests is to find genomic features that yield good predictions for the response. In this paper, we focus on the high-dimensional linear regression problem where univariate responses are observed together with a high-dimensional set of predictors. To cope with the high-dimensionality of predictors, a common approach is to restrict the complexity of the model by penalizing the regression coefficients; see, for example, \cite{FL01,tib96, Z10, ZH05}. These approaches improve prediction performance and often yield a sparse estimate that facilitates feature selection.

{\color{black}Conventional penalization methods are often agnostic to auxiliary structural information of features. The features are either assumed to be of similar importance and are penalized equally, or they are assumed to have varying importance and are penalized with strengths all different from each other. Thus all the features are treated according to a common principle.} 
Real data, however, often consists of a collection of heterogeneous features, for which such an approach does not account. In particular, traditional methods ignore external information and structural differences that may be present among the features. In genomics studies, there are rich covariates that are potentially informative of the importance of a predictor in explaining the response. In transcriptomics studies using RNA-Seq, the sum of read counts per gene across all samples is a statistical covariate informative of the predictive power since the low-count genes are subject to more sampling variability. In genomics, there are genes that belong to one or more genetic pathways and we may expect genes belonging to the same pathway to be correlated with each other. The minor allele frequency and the prevalence of the bacterial species can be considered external covariates for genome-wide association studies (GWAS) and microbiome-wide association studies (MWAS), respectively. The average methylation level of a CpG site in epigenome-wide association studies (EWAS) can be an informative external covariate due to the fact that differential methylation frequently occurs in the highly or lowly methylated region depending on the biological context. Other examples include group structures, structural differences, spatiotemporal information, differences in the scales in which the predictors are measured, different assay types in ``multi-omics'' data, and so on. 

In the context of multiple hypothesis testing, it is possible to make use of such side information to increase the statistical power of the tests \citep{dobriban15,ferkingstad08,ignatiadis16,lei18,li19,zhang19,cao2022optimal}. The inclusion of such information makes the testing procedure significantly more powerful while exactly or approximately maintaining the error rate at a target level. So it is natural to ask the question of how can one incorporate such external information flexibly and robustly in the high-dimensional regression framework. 

To address this, we introduce the Structure Adaptive Elastic-Net (SA-Enet) to incorporate the external structure of the predictors in high-dimensional linear regression. The basic idea behind the SA-Enet is to translate the external information into different penalization
strengths for the regression coefficients. More precisely, at each iteration of the proposed algorithm, the penalization strength is jointly determined by the external information and the current estimates of the regression coefficients. When no external information is provided, our method reduces to the (iterative) Adaptive Elastic-Net (A-Enet) \citep{zou2009}. This is similar to structure-adaptive multiple testing where we relax the $p$ value thresholds for hypotheses that are more likely to be non-null while tightening the thresholds for the other hypotheses so that the overall error measure can be controlled.

The group Lasso and the fused Lasso are two conventional approaches for incorporating group and order information \citep{simon2013,tib05,gl}. A critical difference between the SA-Enet and these variants of the Lasso is that the SA-Enet imposes a ``soft'' constraint on the regression coefficients through varying penalization strengths as compared to the ``hard'' constraints imposed by the group Lasso and the fused Lasso.  For example, under a group structure, the SA-Enet does not force all the regression coefficients within the same group to be simultaneously zero, which is in sharp contrast to the group Lasso. So the SA-Enet is expected to be more robust to misspecified or less informative external information. It is a desirable feature from a practical viewpoint as the informativeness of the external covariates is often unknown to researchers.

\cite{tay2020} discussed the potential benefit of harnessing the ``feature of the features'', which is referred to here as the external or auxiliary structural information on the predictors. They propose the Feature-weighted Elastic-Net (``Fwelnet''). It assigns differential penalty weights similar to that we propose here, but importantly SA-Enet uses the adaptive Elastic-Net penalty as proposed in \cite{zou2009} and is different from the Elastic-Net penalty used in Fwelnet. When $p$ diverges with the sample size $n$, as we assume here, the adaptive penalty used in SA-Enet is known to achieve the desired oracle property according to \cite{zou2009}. In presence of group structural information on the covariates, \cite{velten19} took the Bayesian paradigm and assumed a reparameterized \textit{spike-and-slab} prior on the regression coefficient. For scalability, they propose Graper which adopts a variational inference framework under the ``mean field approximation''. As we demonstrate through simulation and real data application, SA-Enet in general performs as well as others in the worst case and often leads to substantial improvement in performance, particularly feature inclusion probability. But as the correlation among the predictors increases, Graper starts to break down possibly due to the apriori mean-field assumption.

Another novel contribution of the paper is that we introduce the approximate message-passing (AMP) algorithm and the corresponding state evolution theory to the SA-Enet framework. The AMP algorithm was inspired by belief propagation in graphical models and has made a significant impact on compressed sensing; see, for example, \cite{bayati11,bayati12,dmm09,dmm10a,dmm10b}. 
Here we develop the AMP algorithm to study the asymptotic behavior of SA-Enet. Suppose, we observe responses from $n$ samples and corresponding to each of them covariate information of $p_n$ features is also available. In the AMP framework with the large system limit as $n/p_n\rightarrow \delta\in (0,\infty)$, 
we study the asymptotic risk of the estimator using the state evolution equations associated with the AMP algorithm. These results shed new light on the applicability of the AMP and the state evolution theory in the structure-adaptive framework. Our numerical study confirms the practical relevance of the theory in predicting the finite sample risk of the SA-Enet. 
The usefulness and the superiority of our method are demonstrated through both simulations and a real data application.


The rest of the article is organized as follows. In Section~\ref{sec: Methodology}, we define the SA-Enet estimator, provide some motivation behind it, and discuss ways of incorporating different structural information. Section~\ref{sec: Approximate message passing algorithm and state evolution} provides the AMP algorithms for the SA-Enet under the group and the covariate-dependent structures which is a novel contribution to the best of our knowledge. Finally, in Sections \ref{sec: Simulation Study}--\ref{sec: Drug response prediction in leukemia samples} we validate our claims through a wide variety of simulation studies and a motivating application to the chronic lymphocytic leukemia (CLL) data in molecular biology and precision medicine. Section~\ref{sec: Conclusion} concludes with a discussion.

\section{Methodology}\label{sec: Methodology}

\subsection{Setup}\label{subsec: Setup}
Suppose we observe $n$ samples, denoted by $(\y, \xmat)$, satisfying a linear model
\begin{equation}\label{model}
\y = \xmat \bs{\beta} + \bs{\varepsilon},
\end{equation}
where $ \y_{n \times 1} = {( y_1, y_2, \cdots, y_n )}^\T $ is a response vector, $ \xmat_{n \times p_n} $ is a design matrix and $ \bs{\varepsilon}_{n \times 1}={(\varepsilon_1, \varepsilon_2, \cdots, \varepsilon_n )}^\T $ is a vector of random errors. We further assume that (\ref{model}) holds exactly for some true parameter value $\bs{\beta}_0$ of $\bs{\beta}$. Throughout the article, we focus on the high dimensional regime where $p_n$ grows with $n$ and $ \bs{\beta}_0 $ is assumed to be sparse.

\subsection{Lasso, Elastic-Net, and their adaptive variants}\label{subsec: Lasso, Elastic-net, and their adaptive variants}
Under the above setup, one of the most popular methods for simultaneous variable selection and estimation is the Lasso \citep{tib96}. Specifically, the Lasso estimator is defined as
\begin{equation*}\label{eqlasso}
\hat{\bs{\beta}}^{\tL} = \argmin_{\bs{\beta} \in \mathbb{R}^{p_n}} {(2n)}^{-1} \norm{ \y - \xmat \bs{\beta}}^2_2 + \lambda \norm{\bs{\beta}}_1 ,
\end{equation*}
where $\norm{\bs{a}}_q =(\sum_j \abs{a_j}^q )^{1/q}$ denotes the $ \ell_q $ norm of any real vector $ \bs{a} $. 

Despite its popularity Lasso has two key drawbacks. \cite{FL01} showed that the Lasso estimator incurs a bias in estimating the nonzero coefficients which cannot be ignored. \cite{zou06} showed that due to this incurrence of bias, the Lasso does not have the oracle property as defined in \cite{FL01}, and is also inconsistent for model selection unless the design matrix satisfies a strong condition. To address this, \cite{zou06} proposed the adaptive Lasso (A-Lasso) estimator defined as
\begin{equation*}\label{eqalasso}
\hat{\bs{\beta}}^{\AL} = \argmin_{\bs{\beta} \in \mathbb{R}^{p_n}} {(2n)}^{-1} \norm{ \y - \xmat \bs{\beta}}^2_2 + \lambda \sum_{j=1}^{p_n} \hat{w}_j \abs{\beta_j} .
\end{equation*}
Here $ \hat{\bs{w}} = {(\hat{w}_1 , \hat{w}_2 , \ldots, \hat{w}_{p_n} )}^\T $ is a data-dependent vector of non-negative weights. The basic difference between the Lasso and A-Lasso is that the individual non-negative weights $ \hat{w}_j $'s are assigned to each $ \beta_j $'s in addition to the common $ \lambda $, which allows differential shrinkage of the components of $\bs{\beta}$. This enables the resulting estimator to achieve a consistent variable selection and to correct for the bias incurred by the Lasso estimator. When $p_n$ does not grow with $n$, it has been proved that the A-Lasso is an oracle estimator (in the sense of \cite{FL01} and \cite{FP04}) and it enjoys a near-minimax optimality \cite{zou06}. For $ \gamma>0 $, a recommended choice is to set $\hat{w}_j =|\hat{\beta}_j |^{-\gamma}$, where $\hat{\bs{\beta}} $ is a ``well-behaved'' preliminary estimator of $ \bs{\beta}_0 $ \citep{zou06}. 

A well-known issue of the $\ell_1$ penalization is that its performance degrades when the number of predictors or the collinearity among them increases. \cite{ZH05} showed that the Lasso paths become unstable under multicollinearity. To address this, they proposed the Elastic-Net (Enet) estimator defined as
\begin{equation}\label{eqenet}
\hat{\bs{\beta}}^{\enet} = \argmin_{\bs{\beta} \in \mathbb{R}^{p_n}} {(2n)}^{-1} \norm{ \y - \xmat \bs{\beta}}^2_2 + \lambda_2 \norm{\bs{\beta}}^2_2 + \lambda_1 \norm{\bs{\beta}}_1 .
\end{equation}
The $\ell_1$ term $\lambda_1 \norm{\bs{\beta}}_1$ encourages automatic variable selection and the $\ell_2$ term $\lambda_2 \norm{\bs{\beta}}_2$ stabilizes the solution path which improves the prediction accuracy.

In the spectrum of desirability and improvement, A-Lasso and Enet lie at two opposite extremes. On the one hand, the A-Lasso achieves the oracle property because of the adaptive penalties and the Enet can better deal with collinearity. On the other hand, the A-Lasso acquires the instability of the Lasso in high-dimensional data and the Enet lacks the oracle property. To reduce this gap in the spectrum, \cite{zou2009} proposed adaptive Elastic-Net (A-Enet) which penalizes the squared error loss using a combination of $\ell_2$ and adaptive $\ell_1$ penalties. The estimator is defined as
\begin{equation*}\label{eqaenet}
\hat{\bs{\beta}}^{\aenet} = \argmin_{\bs{\beta} \in \mathbb{R}^{p_n}} {(2n)}^{-1} \norm{ \y - \xmat \bs{\beta}}^2_2 + \lambda_2 \norm{\bs{\beta}}^2_2 + \lambda_1 \sum_{j=1}^{p_n} \hat{w}_j \abs{\beta_j} ,
\end{equation*}
where $\hat{\bs{w}} = {(\hat{w}_1 , \hat{w}_2 , \ldots, \hat{w}_{p_n} )}^\T$, like in the A-Lasso, is a data-dependent vector of non-negative weights. For $ \gamma>0 $, \cite{zou2009} recommended the choice $\hat{w}_j =|\hat{\beta}^{\enet}_j |^{-\gamma}$, where $\hat{\bs{\beta}}^{\enet} $ is the Enet estimator.  When $\lambda_2$ equals 0 or the design matrix is orthogonal, the A-Enet reduces to the A-Lasso. This coherence is desirable as in that case the A-Lasso is known to achieve minimax optimal risk bound. In other cases, the $\ell_2$ term stabilizes the A-Lasso path in the possible presence of collinearity and the $\ell_1$ term provides sparsity through adaptively weighted shrinkage. To our interest, as $p_n$ diverges with $n$, the A-Enet achieves the oracle property under some regularity conditions \citep{zou2009}.

\subsection{Structure Adaptive Elastic-Net}\label{subsec: Structure Adaptive Elastic-net}
In many real applications, it is possible to have some external information on the importance of each $ \beta_j $'s in predicting $\y$. Here our goal is to make use of such additional knowledge in guiding us to choose $ \hat{\bs{w}} $ in a data-dependent fashion. In general, let us refer to such external information as the \textit{structural information} and denote them by $\aux$. Some common examples of such a structure within the components of $\bs{\beta}$ include group information, a monotonic ordering of their magnitudes, graph-based information, extrinsic covariate information, and so on. Here we propose the Structure Adaptive Elastic-Net (SA-Enet) as a method for utilizing the auxiliary information $\aux$ combined with $(\y, \xmat)$ in choosing $\hat{\bs{w}}$. In what follows, we first provide the motivation and then introduce the algorithm for obtaining the proposed estimator. We also discuss some examples of structural information and derive the data-adaptive weights $\hat{\bs{w}}$ in each of those cases.

\subsubsection{Notation}\label{subsubsec: Notation}
For $K \in \mathbb{N}$, let $ \bs{a} = { \big( a_1, a_2, \cdots, a_K \big) }^\T \in \mathbb{R}^K $ and $ S $ be any subset of $ \{ 1,2, \cdots, K \} $. Then, (a) $ \abs{\bs{a}} := { \big( \abs{a_1}, \abs{a_2}, \cdots, \abs{a_K} \big) }^\T $. (b) $ \bs{a}_{S} := { \big( a_{1,S}, a_{2,S}, \cdots, a_{K,S} \big) }^\T $ where $ a_{j,S} := a_j \, \mathbb{I} \left\{j \in S\right\} $ for all $ j = 1,2, \cdots, K $. (c) $ \abs{\bs{a}_S} := { \big( \abs{a_{1,S}}, \abs{a_{2,S}}, \cdots, \abs{a_{K,S}} \big) }^\T = \abs{\bs{a}}_S $. (d) $ \mean{ \bs{a}_S } := |S|^{-1} \sum_{j \in S} a_j $, where $|S|$ denotes the cardinality of the set $S$. (e) Consider a scalar constant $ b $, $\bs{c} = { \big( c_1, c_2, \cdots, c_K \big) }^\T \in \R^K$ and $g: \R^2 \mapsto \R$. Then $g(\bs{a},\bs{c}) = ( g(a_1 , c_1) , \cdots, g(a_K , c_K ))^\T$ and $g(\bs{a},b) = ( g(a_1 ,b) , \cdots, g(a_K ,b))^\T$. Write $a\wedge b=\min(a,b)$
for $a,b\in\mathbb{R}$.

\subsubsection{Motivation}\label{subsubsec: Motivation}
Suppose the random errors in (\ref{model}) are independent and identically distributed Gaussian random variables with mean 0 and variance $\sigma^2$ (note that the Gaussian assumption is only used to motivate our procedure). 
Then the negative log-likelihood of $ \bs{\beta} $ is proportional to
\begin{equation}
{(2 \sigma^2)}^{-1} \norm{ \y - \xmat \bs{\beta}}^2_2 .
\end{equation}
To encourage sparsity we apriori assume that given $\bs{w}$, $\beta_1, \ldots, \beta_{p_n}$ are independent of each other and the negative log-likelihood of $\beta_j$ is proportional to $\beta_j^2 + w_j \abs{\beta_j}$, where $w_j$'s are non-negative. Thus the negative log-likelihood of $\bs{\beta}$ given $\bs{w}$ is given by
\begin{equation*}
\norm{\bs{\beta}}^2_2 + \sum_{j=1}^{p_n} \left[ w_j \abs{\beta_j} - \log C\left(w_j\right) \right] ,
\end{equation*}
where $C\left(w_j\right)$ is the proportionality constant in the conditional prior on $\beta_j$. We complete the hierarchy by specifying a prior on $\bs{w}$. To this, we constrain $\bs{w}$ to lie in $\mathcal{M} \subseteq [0,C_{U}]^{p_n}$ for some $0<C_U<\infty$. $\mathcal{M}$ encodes the structural information and its relevance is further discussed in Section~\ref{subsubsec: Structural information}. 
In practice, we set $C_U$ to be a sufficiently large positive number, for example, $10^{30}$. Under this constraint, we assume that the joint prior density of $\bs{w}$ is proportional to
\begin{equation*}
\prod_{j=1}^{p_n} h(w_j; \gamma) \quad \text{for} \,\, \bs{w} \in \mathcal{M} ,
\end{equation*}
where $\gamma$ is the hyperparameter and $h(\cdot; \gamma)$ is such that the joint prior specified above is a probability density over $\mathcal{M}$. Throughout the article, we assume that $h$ has the following form:
\begin{equation*}
h(w_j ; \gamma) =
\begin{cases}
C_1 \, C\left(w_j\right)^{-1} \exp \left[ w_j^{1- \gamma^{-1}} /\left(1- \gamma^{-1}\right) \right], & \text{if $0<\gamma<1$} , \\
C_2 \, w_j C\left(w_j\right)^{-1}, & \text{if $\gamma=1$} , \\
\end{cases}
\end{equation*}
where $C_1$ and $C_2$ are positive constants such that the joint density on $\bs{w}$ over $\mathcal{M}$ integrates to 1. When no structural information is available, this choice of $h$ leads to adaptive weight updates in the SA-Enet which are the same as that have been proposed for the A-Enet. 
Combining the model and priors, the negative logarithm of the joint posterior of $\left(\bs{\beta},\bs{w}\right)$ given the data $\left(\y,\xmat,\aux\right)$ becomes proportional to
\begin{equation}\label{jointpost}
{(2n)}^{-1} \norm{\y - \xmat \bs{\beta}}^2_2 + \lambda_{2n} \norm{\bs{\beta}}^2_2 + \lambda_{1n} \sum_{j=1}^{p_n} \left[ w_j \abs{\beta_j} - \log g(w_j ;\gamma) \right] \, \mathbb{I}\left\{\bs{w}\in\mathcal{M}\right\} ,
\end{equation}
where $\lambda_1 = \lambda_2 = \sigma^2/n$, and $ g(w_j ;\gamma) = C\left(w_j\right) h(w_j ; \gamma) $. If $\sigma$ is known, we interpret $ \bs{w} $ in (\ref{jointpost}) as a vector of hyper-parameters and aim to estimate it (together with $ \bs{\beta} $) by maximizing the joint posterior density. But even in this case, $\lambda_1 = \lambda_2 = \sigma^2 /n$ may not be a good choice from a theoretical point of view. For a general purpose, we replace the terms with some positive tuning parameters $\lambda_1$ and $\lambda_2$, respectively. By doing this, we treat the minimization of (\ref{jointpost}) as a frequentist approach similar to the Lasso or the A-Lasso and estimate $\left(\lambda_1,\lambda_2,\gamma\right)$ using cross-validation. This provides a direct way of incorporating external information and makes our setup widely applicable.

\subsubsection{Estimator and algorithm}\label{subsubsec: Estimator and algorithm}
Given $\lambda_1, \lambda_2 >0 $ and $\gamma \in (0,1]$, we define the SA-Enet estimator $\hat{\bs{\beta}}^{\saenet}$ as
\begin{equation}\label{eqsaenet}
\Big( \hat{\bs{\beta}}^{\saenet} , \hat{\bs{w}} \Big) = \argmin_{ \bs{\beta} \in \mathbb{R}^{p_n}, \bs{w} \in \mathcal{M} } Q^{\saenet} (\bs{\beta},\bs{w}),
\end{equation}
where $Q^{\saenet} (\bs{\beta},\bs{w})$ equals to
\begin{equation}\label{saenet objective function}
    {(2n)}^{-1} \norm{\y - \xmat \bs{\beta}}^2_2 + \lambda_2 \norm{\bs{\beta}}^2_2 + \lambda_1 \sum_{j=1}^{p_n} \left[ w_j \abs{\beta_j} - \log g(w_j ;\gamma) \right],
\end{equation}
and the definition of $g$ is the same as in (\ref{jointpost}). Note that (\ref{saenet objective function}) is not jointly convex in $\left( \bs{\beta}, \bs{w}\right)$. {\color{black}We propose Algorithm \ref{algorithm saenet} as an iterative approach for the optimization. Starting with initialization of either $\bs{\beta}$ or $\bs{w}$, the algorithm iteratively updates $\bs{w}$ and $\bs{\beta}$ by minimizing (\ref{saenet objective function}), accordingly.
\begin{algorithm}[ht]
	\caption{\textbf{: Iterative algorithm for the SA-Enet}}\label{algorithm saenet}
	\vspace{1mm}
	\begin{enumerate}
	    \item Fix the maximum number iterations $T \, (\geq 1)$.
	    
		\item {\bf Initial Step:} Initialize the weights to 1. Then $\bs{\beta}$ is updated by solving the Enet problem
		\begin{equation}\label{enet step}
            \hat{\bs{\beta}}^{\saenet}_0 = \argmin_{ \bs{\beta} \in \mathbb{R}^{p_n} } {(2n)}^{-1} \norm{ \bs{y} - \mathbf{X} \bs{\beta}}^2_2 + \lambda_{10} \norm{\bs{\beta}}_1 + \lambda_{20} \norm{\bs{\beta}}^2_2 .
		\end{equation}
		$\lambda_{10}$ and $\lambda_{20}$ are positive and prefixed.
	    
		\item {\bf Iteration 1 to $\bs{T}$:} At iteration $k = 1, \ldots, T$, $\bs{\beta}$ is updated by repeating the following two steps:
		
		\begin{itemize}
		
		\hypertarget{optimization step}{}
		\item {\bf Update $\bs{w}$:} Given $\hat{\bs{\beta}}^{\saenet}_{k-1}$ from the previous iteration, update the weights by solving the constrained optimization problem
		\begin{equation}\label{optimization step}
		\hat{\bs{w}}_k = \argmin_{ \bs{w} \in \mathcal{M} } \sum_{j=1}^{p_n} \left[ w_j \abs{\hat{\beta}^{\saenet}_{k-1,j}} - \log g(w_j ; \gamma_k ) \right].
		\end{equation}
		
		\hypertarget{aenet step}{}
		\item {\bf Update $\bs{\beta}$:} Given $\hat{\bs{w}}_k$, update $\bs{\beta}$ by solving the A-Enet problem 
		\begin{equation}\label{aenet step}
		\hat{\bs{\beta}}^{\saenet}_k = \argmin_{ \bs{\beta} \in \mathbb{R}^{p_n} } {(2n)}^{-1} \norm{ \bs{y} - \mathbf{X} \bs{\beta}}^2_2 + \lambda_{1k} \sum_{j=1}^{p_n} \hat{w}_{kj} \abs{\beta_j} + \lambda_{2k} \norm{\bs{\beta}}^2_2 .
		\end{equation}
		
		\end{itemize}
		$\lambda_{1k}$, $\lambda_{2k}$, and $\gamma_k$ are positive and prefixed.
	\end{enumerate}
\end{algorithm}
In particular, Algorithm~\ref{algorithm saenet} initializes all the weights to 1. For a prefixed number of iterations $T (\geq 1)$, the algorithm can then be narrated as follows.
\begin{enumerate}
    \item[(0)] Given the initial weights, we update $\bs{\beta}$ according to (\ref{enet step}) and get the initial estimate $\hat{\bs{\beta}}^{\saenet}_0$. Since the weights are initialized at 1, this is the Enet estimate.
    
    \item[(1)] Given $\hat{\bs{\beta}}^{\saenet}_0$ we first update the adaptive weights according to (\ref{optimization step}), and obtain $\hat{\bs{w}}_1$. Then given $\hat{\bs{w}}_1$, $\bs{\beta}$ is updated according to (\ref{aenet step}) and we get $\hat{\bs{\beta}}^{\saenet}_1$. This is the A-Enet estimate with the adaptive weights $\hat{\bs{w}}_1$.
    
    \item[(2)] Given $\hat{\bs{\beta}}^{\saenet}_1$ we first update the adaptive weights according to (\ref{optimization step}), and obtain $\hat{\bs{w}}_2$. Then given $\hat{\bs{w}}_2$, $\bs{\beta}$ is updated according to (\ref{aenet step}) and we get $\hat{\bs{\beta}}^{\saenet}_2$. This is the A-Enet estimate with the adaptive weights $\hat{\bs{w}}_2$.
\end{enumerate}
We repeat this until iteration $T$ to get the SA-Enet estimate $\hat{\bs{\beta}}^{\saenet}_T$. From here on, the estimator is referred to as SA-Enet($T$).} We find that the improvement in performance for larger iterations is negligible. So, for default implementations, we recommend $T=5$. Numerical results supporting this are deferred to Sections \ref{sec app: Robustness across iterations} and \ref{subsec app: Robustness across iterations} in the appendix.


\subsubsection{Structural information}\label{subsubsec: Structural information}

Here we present examples of structural information that are motivated by specific applications. We show that when no structural information is provided, the weight updates from (\ref{optimization step}) simplify to the adaptive weights as recommended by \cite{zou2009} for the A-Enet. For ease of notation, we suppress the dependence of $p$ on $n$ for the rest of the discussion.

\paragraph{Group structure.} In microarray experiments, different genes may be clustered into several groups along biological pathways or based on phenotype information and gene ontology, and so on. This implies that the set of predictors can be partitioned into $ D $ mutually exclusive blocks $ { \{S_d\} }_{d=1}^D $ with $\abs{S_d} = p_d$ and the signals belonging to the same group are likely to appear together. So it seems natural to consider the following set of $\bs{w}$:
\begin{equation}\label{group info}
\begin{split}
    \mathcal{M}_{\text{G}} = \Big\{ & \bs{w}\subseteq [0,C_{U}]^p \Big| w_i = w_j \text{ if } i,j \in S_d \text{ for } i,j \in \{ 1,2, \cdots, p \} \\ 
    & \text{and} \,\,  d \in \{ 1,2, \cdots, D \} \Big\} .\\
\end{split}
\end{equation}
Under this assumption the objective function in (\ref{optimization step}) is convex and the minimizer can be analytically obtained. Fix $j \in \{ 1,2, \cdots, p \}$, $d \in \{ 1,2, \cdots, D \}$, and $\gamma \in (0,1]$. For $j \in S_d$, the minimizer for a given $\bs{\beta}$ is
\begin{equation}\label{weight update group}
\begin{split}
 \hat{w}_{j} (\bs{\beta}) & = \hat{w}_{j} (\bs{\beta}_{S_d}) =  \begin{cases}
    C_U, & \text{if $\beta_j = 0$  $\forall j \in S_d$}, \\
    \mean{ \abs{ \bs{\beta}}_{S_d} }^{-\gamma}\wedge C_U, & \text{otherwise} . \\
    \end{cases}   
\end{split}
\end{equation}

\paragraph{Covariate-dependent structure.} 
In genomics studies, there are rich covariates that are potentially informative on the importance of a predictor in explaining the response. Examples include, but are not restricted to, the sum of read counts per gene across all samples in transcriptomics studies using RNA-Seq, the minor allele frequency in genome-wide association studies (GWAS), the prevalence of the bacterial species in microbiome-wide association studies (MWAS), and the average methylation level of a CpG site in epigenome-wide association studies (EWAS). 

Mathematically, let $ \bs{u}_j $ denote the external covariate associated with the $j^{th}$ feature lying in some generic space $\mathcal{U} \subseteq \R^q$. The external covariate can bear information on the predictor variable $x_j$ being a signal or not, or has to do with the strength of the regression coefficient $\beta_j$. But importantly, the true nature of this relationship is not known and has to be learned from the data. To incorporate the covariate information, we define the set of $\bs{w}$ as follows:
\begin{equation}\label{covariate info}
\mathcal{M}_{\text{Cov}} = \Big\{ \bs{w} \subseteq [0,C_{U}]^p \Big| w_j = f( \bs{u}_j ; \bs{\rho} ) \text{ for } \bs{\rho} \in \mathcal{B} \text{ and } j \in \{ 1,2, \cdots, p \} \Big\},
\end{equation}
where $ f : \mathcal{U} \mapsto [0, \infty) $ is a smooth non-negative valued function parameterized by $\bs{\rho}$, and $\mathcal{B}$ is a compact subset of $\mathbb{R}^{q+1}$. In particular, letting $\bs{\rho} = \left( \rho_0 , \bs{\rho}_1 \right)^\T$, we assume the parameterization
\begin{equation}\label{weight parameterization covariate}
    \log w_j = \rho_0 + \bs{u}_j^\T \bs{\rho}_1 .
\end{equation}
In this case, the minimizer in (\ref{optimization step}) given $\bs{\beta}$ is 
\begin{equation}\label{weight update covariate}
    \hat{w}_j (\bs{\beta}) = \exp{\Big( \hat{\rho}_0 (\bs{\beta}) + \bs{u}_j^{\T} \, \hat{\bs{\rho}}_1 (\bs{\beta}) \Big)}, \quad \mbox{for } j = 1, \ldots,p,
\end{equation}
where
\begin{equation}\label{rhoopt covariate}
    \hat{\bs{\rho}} (\bs{\beta}) = \argmin_{\bs{\rho} \in \mathcal{B}} \,\, \sum_{j=1}^{ p } \Big[ f( \bs{u}_j ; \bs{\rho} )\abs{\beta_j} - \log g \Big( f( \bs{u}_j ; \bs{\rho} ) ; \gamma \Big) \Big],
\end{equation}
and $\max_{j}f( \bs{u}_j ; \bs{\rho} )\leq C_U$. Note that, for a given $\bs{\beta}$, when $\mathcal{B}$ is a convex set, the objective function in (\ref{rhoopt covariate}) is convex in $\bs{\rho}$. 

{\color{black}
\paragraph{Ordered structure.}  In genomic studies, researchers can use prior information (for example, $P$-values from previous/related studies) to generate a ranked list of the genomic features even before performing the experiment. A natural way to incorporate such structure into our framework is by considering
\begin{equation}
    \mathcal{M}_{\text{Order}} = \Big\{\bs{w}\subseteq [0,C_{U}]^p \,\Big|\, 0\leq w_1\leq w_2\leq \cdots \leq w_p\leq C_U \Big\} .
\end{equation}
A larger $w_i$ corresponds to a potentially less significant variable, and vice versa.

\paragraph{Graph structure.}  Suppose an underlying graph governs the similarity among regression coefficients. Let $\mathcal{G} = (\mathcal{V},\mathcal{E})$ denote the undirected graph where $\mathcal{V} = \{1, \dots,p\}$ is the set of nodes and $\mathcal{E}$ is the set of edges. To translate the graph structure into constraints on the adaptive weights, we can assume
\begin{equation}
    \mathcal{M}_{\text{Graph}} = \Big\{\bs{w}\subseteq [0,C_{U}]^p \,\Big|\, w_i=\exp(v_i) \text{ and } \sum_{(i,j)\in \mathcal{E}} a_{ij}|v_i-v_j|\leq \kappa \Big\}.
\end{equation}
Here $a_{ij}>0$ are prespecified and reflect the apriori importance of an edge between nodes $i$ and $j$ compared to all the edges, and $\kappa>0$ is a tuning parameter. Without loss of generality, we can constraint $a_{ij}$'s to satisfy $\sum_{(i,j)\in \mathcal{E}} a_{ij} = 1$.}

\paragraph{No structural information.} Suppose we want to use adaptive weights but we do not have any prior structural information on $ \bs{\beta} $ that we can take advantage of. Then the set of $\bs{w}$ that we are interested in is $ \mathcal{M}_{\text{NS}} = [0,C_U]^p$. The objective function (\ref{optimization step}) in this case is convex. For $j = 1, \ldots, p$ and $\gamma \in (0,1]$, the minimizer for given $\bs{\beta}$ can be obtained analytically and it is given by
\begin{equation}\label{weight update no info}
\hat{w}_j (\bs{\beta}) = \hat{w}_j (\beta_j) =
\begin{cases}
C_U, & \text{if $\beta_j = 0$}, \\
{\abs{\beta_j}}^{-\gamma}\wedge C_U, & \text{if $\beta_j \neq 0$}. \\
\end{cases}
\end{equation}
So in the absence of structural information, SA-Enet reduces to A-Enet as proposed in \cite{zou2009}.
{\color{black}The framework also includes Elastic-Net as a special case where $ \mathcal{M}_\enet = \{\bs{1}\}$.
}

\begin{rem}\label{rem2}
{\color{black}For brevity, the rest of the article only focuses on group and covariate-dependent structures. For these two types of structural information, we discuss the theoretical properties of the SA-Enet estimator and compare their performances through numerical studies.} For these structures the objective function in (\ref{optimization step}) is convex. This ensures that both (\ref{optimization step}) and (\ref{aenet step}) in Algorithm~\ref{algorithm saenet} are convex optimizations. 
\end{rem}


\begin{rem}
The group structure can be viewed as a special case of the covariate-dependent structure where the covariate $\bs{u}_j$ denotes the index of the group to which $\beta_j$ belongs.
\end{rem}

\section{Approximate message passing algorithm and state evolution}\label{sec: Approximate message passing algorithm and state evolution}

In this section, we theoretically analyze the risk of the SA-Enet estimator. To this, we note that the initial estimate is the Enet estimate, while at every subsequent step, we calculate the A-Enet estimates. Thus following Algorithm~\ref{algorithm saenet}, under group and covariate-dependent structural information, it boils down to analyzing the risks of the Enet estimator and the A-Enet estimator where for the latter the data-adaptive weights are obtained using the Enet or A-Enet estimates.

Before getting into the AMP framework, we briefly review some existing theories and highlight their differences from the AMP approach. For a fixed $p$, \cite{geer11} and \cite{zou06} provided theoretical guarantees for the A-Lasso. \cite{huang08} extended this and analyzed the asymptotic properties of the estimator for a sparse high-dimensional linear regression model with a fixed design matrix. Given a suitable initial estimator, they proved that, under some conditions, the A-Lasso correctly selects the true nonzero coefficients with probability converging to one. The authors further show that the asymptotic distribution of the estimator is the same as that they would have if the zero coefficients were known in advance. Although this is an oracle property in the sense of \cite{FL01} and \cite{FP04}, they require a fairly strong condition on the design matrix \citep{zhou09}. Along these lines, \cite{zhou09} has defined a two-step A-Lasso procedure for linear regression and has described general model selection properties of the second stage weighted procedure for variable selection. Finally, to our interest, similar strategies have also been applied to analyze the A-Enet estimator \citep{zou2009}. {\color{black} A common practice in the literature for theoretically analyzing a regularized estimator in a high-dimensional setting is obtaining an oracle inequality that provides a high-probability upper bound to the $\ell_q$ error of the estimator. Along this line of argument, demonstrating the superiority of the SA-Enet estimator would require us to show an improved upper bound of its $\ell_q$ error. However, it is often unclear how tight these upper bounds are in real-life applications. We take a different route and utilize the AMP machinery in theoretically analyzing the estimator. In this framework, we propose the AMP algorithm that constructs a theoretical estimate, known as the AMP estimate, for the same problem. Under some conditions, the asymptotic behavior of the AMP estimate can be obtained by a one-dimensional recursion, known as the \textit{state evolution}. This lets us obtain the exact asymptotic risk of the AMP estimates for the group and covariate-dependent structure which is not obtainable in the traditional theoretical framework. Under the same conditions as required by the state evolution, this is also the risk of the SA-Enet estimator, because for prefixed AMP parameters $(\alpha_{k1}, \alpha_{k2})$ the AMP estimate at any AMP iteration equals to the SA-Enet($k$) estimate corresponding to some $(\lambda_{k1}, \lambda_{k2})$ where their relationship is given by the correspondence equations. Thus, we take advantage of the AMP framework only as an intermediate technical tool in deriving the asymptotic risk of the SA-Enet estimator. Although the predicted risks are asymptotic, the numerical results presented in Section~\ref{subsec: State evolution prediction as a finite sample approximation} indicate that the prediction closely matches their finite sample performances when $p$ is as small as 500. \cite{bayati12} observed similar results for the Lasso.}

Next, we provide a brief background on the AMP algorithm. The framework is inspired by belief propagation in graphical models and it has made a significant impact on compressed sensing, referring to a collection of signal processing techniques that focus on reconstructing high-dimensional signals in ``undersampled'' settings \citep{baraniuk08}. In a nutshell, compressed sensing aims at finding solutions to under-determined linear systems. 
In a high-dimensional linear regression, since the sample size is smaller than the number of parameters, the methods developed in the earlier stage require nonlinear and relatively expensive reconstruction schemes. One popular class of these schemes is based on linear programming (LP) methods. In spite of the theory being elegant and promising, solving the LPs in applications are more expensive than the standard linear reconstruction schemes. To reduce the computational cost and shed new light on the theoretical performance of the LP-based schemes, \cite{dmm09} first proposed the AMP algorithm as a special type of iterative thresholding algorithm, and showed that its performance is equivalent to the corresponding convex optimization procedure. Under the assumption that the design matrix $\xmat$ consists of independent and identically distributed Gaussian entries (``iid-design'' from here on), the reconstruction quality of the AMP algorithm has been proven to be identical to the LP-based methods while offering a significant decrease in computational cost \citep{bayati11,bayati12,dmm09,dmm10a,dmm10b}. To our interest, \cite{bayati12} proposed an AMP algorithm for analyzing the Lasso estimator. Under the assumption of an iid-design, it records two important findings. In the large system limit, that is as $n/p \rightarrow \delta \in (0, \infty)$, (i) the solution from the AMP algorithm (referred to as the AMP estimates) coincides with the Lasso estimator as the number of iterations grows to infinity, (ii) the normalized risk of the Lasso estimator converges to a quantity determined by the fixed point of an equation, defined as the state evolution. Following similar steps therein, we take advantage of the general recursion algorithm proposed in \cite{bayati11} and analyze the risk of the SA-Enet estimator. We make two contributions on this front. (1) We describe the AMP algorithm for the Enet. This corresponds to a specific choice of thresholding function in the general recursion algorithm from \cite{bayati11}. (2) We propose the AMP algorithm for the A-Enet and derive its state evolution. This lets us analyze the AMP estimates by taking the large system limit at any given iteration of the AMP algorithm. Finally, letting the number of AMP iterations go to infinity provides us with the asymptotic standardized risk of the SA-Enet estimator. The AMP algorithms and theoretical results associating the AMP algorithm for the proposed A-Enet are presented in the following subsections.

\subsection{AMP algorithm under group structure}\label{sec: AMP algorithm under group structure}

In this section, we propose the AMP algorithm for the SA-Enet under group structure. We assume that the true data generating parameter $\bs{\beta}_0$ has the underlying group structure as described in Section~\ref{subsubsec: Structural information}. Following the notations therein, we present the AMP algorithm of the SA-Enet in Algorithm~\ref{amp algorithm group}. The successive recursion that we propose here is an extension of the algorithm proposed in \cite{bayati12}. The function $\eta$ in Algorithm~\ref{amp algorithm group} is the proximal operator of the Elastic-Net penalty. Specifically, for $\bs{x}, \bs{b} \in \R^p$, the proximal operator $\eta$ is defined as
\begin{align}
    & \eta(x_i ; \theta_1, \theta_2) = \hat{b}_i, \quad \text{where}, \\
    & \hat{\bs{b}} = \argmin_{\bs{b}} \frac{1}{2} \norm{ \bs{x} - \bs{b}}^2_2 + \theta_1 \norm{\bs{b}}_1 + \theta_2 \norm{\bs{b}}^2_2 .
\end{align}
Following \cite{ZH05}, this corresponds to the naive Enet solution in the case of an orthogonal design. Thus for positive $\theta_1$ and $\theta_2$, $\eta : \R \mapsto \R$ is given by
\begin{equation}
     \eta(x ; \theta_1, \theta_2)= \frac{\left(\abs{x} - \theta_1\right)_+ \text{sgn}(x)}{1 + 2\theta_2} .
\end{equation}
Here $a_+ = \max\left(a, 0\right)$, and $\text{sgn}(a) = a/\abs{a}$ with $\text{sgn}(0) = 0$. Thus Algorithm~\ref{amp algorithm group} applies a scaled-soft thresholding rule with possibly different thresholds for different groups. This enables the SA-Enet to be more adaptive. By $\eta^\prime$, we denote the derivative of $\eta$ with respect to its first argument $x$. Whenever $\eta$ and $\eta^\prime$ are expressed with vectors $x$, $\theta_1$, and $\theta_2$ (all of the same length), this denotes a vector where the functions are applied element-wise to each vector.

\begin{algorithm}[h!]
	\caption{\textbf{: AMP algorithm for SA-Enet($\bs{T}$) under group structure}}\label{amp algorithm group}
	\vspace{1mm}
	\begin{enumerate}
 
        \item[(1)] Fix the maximum number of iterations $T$.
        
        \item[(2)] {\bf AMP for the SA-Enet(0).}  Initialize $\bs{\beta}_0^0 = 0$ and $\e_0^{-1}=0$. For $t\geq0$ the algorithm constructs the following recursion until convergence:
		\begin{equation}\label{amp recursion enet group} 
		    \begin{split}
		        \e_0^t & = \, \bs{y} - \xmat \bs{\beta}_0^t + \frac{\e_0^{t-1}}{\delta} \mean{ \eta^\prime \left( \, \xmat^\T \e_0^{t-1} + \bs{\beta}_0^{t-1} \, ; \theta_{10}^{t-1},\theta_{20}^{t-1} \right) },\\
		        \bs{\beta}_0^{t+1} & = \, \eta \left( \, \xmat^\T \e_0^t + \bs{\beta}_0^t \, ; \theta_{10}^{t},\theta_{20}^{t} \right).
		    \end{split}
		\end{equation}
		
		\item[(3)]  At iteration $k = 1, \ldots, T$, consider the following algorithm.\\

            \vspace{1mm}
            {\bf AMP for the SA-Enet($k$).}
  Define $\bs{\omega}_k = {( \omega_{k1},\dots,\omega_{kD})}^\T$ such that $\forall \, d=1,\dots,D$,
    \begin{equation}
        \omega_{0d} = 1, \quad \text{and} \quad \omega_{kd} = {\left( \E \abs{\eta \left( \, B_{0d} + \tau_{k-1}^* Z \, ; \theta_{1,k-1}^{*} \, \omega_{k-1,d}, \theta_{2,k-1}^{*} \right) } \right)}^{-\gamma} .
    \end{equation}
    For group $d$, define the sequence of thresholds $\left\{ \theta_{1k}^t \, \omega_{kd},\theta_{2k}^t \right\}_{t \geq 0}$, and initialize $\bs{\beta}_k^0 = 0$ and $\e_k^{-1}=0$. Then, for $t\geq0$ the algorithm constructs the following recursion until convergence:
		\begin{equation}\label{amp recursion saenet group}
		    \begin{split}
		        \e_k^t  & = \, \y - \xmat \bs{\beta}_k^t + \frac{\e_k^{t-1}}{\delta} \sum_{d=1}^D \frac{p_d}{p} \, \mean{ \eta^\prime \left( \, \left( \xmat^\T \e_k^{t-1} + \bs{\beta}_k^{t-1} \right)_{S_d} \, ; \theta_{1k}^{t-1} \, \omega_{kd},\theta_{2k}^{t-1} \right) },\\
		        \left(\bs{\beta}_k^{t+1}\right)_{S_d}  & = \, \eta \left( \left( \xmat^\T \e_k^{t} + \bs{\beta}_k^t \right)_{S_d} \, ; \theta_{1k}^{t} \, \omega_{kd},\theta_{2k}^t \right), \quad \forall \, d=1,\dots,D.
		    \end{split}
		\end{equation}
	\end{enumerate}
\end{algorithm}

Algorithm~\ref{amp algorithm group} comprises two key steps: (1) obtaining AMP estimates for the Enet or the A-Enet, and (2) determining data-adaptive weights based on the AMP estimates for either the Enet or the A-Enet. For an illustration, let $T=1$. The three steps in Algorithm~\ref{amp algorithm group} are: $(i)$ obtain the AMP estimates for the Enet; $(ii)$ determine data-adaptive weights based on the AMP Enet estimates, and $(iii)$ obtain the AMP estimates for the A-Enet using the data-adaptive weights. Below we take a closer look at these steps and provide a theoretical analysis of the SA-Enet(1) estimator. The framework can be recursively applied to develop the AMP Algorithm for the SA-Enet($T$). We conclude by summarizing the theoretical components for the SA-Enet($k$) estimator for $k=2,\dots,T$.

\paragraph{Interpreting the AMP algorithm for the SA-Enet(0).} For an arbitrary sequence of thresholds $\{ \theta_{10}^t,\theta_{20}^t \}_{t \geq 0}$, the recursions in (\ref{amp recursion enet group}) identifies the AMP estimates for the SA-Enet(0) (or the Enet). For a converging sequence of instances (according to Definition 1 in \cite{bayati12}), consider the sequence of vectors $\left\{ \bs{\beta}_0 (p), \bs{\varepsilon} (p) \right\}_{p \geq 0}$. Let us assume that their empirical distributions converge to the probability measures $\prob_{B_0}$ and $\prob_W$, respectively. Then, under the iid-design, the asymptotic behavior of the recursion (\ref{amp recursion enet group}) can be tracked by a one-dimensional recursion defined by the sequence $\{\tau_0^t\}_{t\geq0}$ as
\begin{equation}\label{state evolution enet}
\begin{split}
    {(\tau_0^0)}^2 & = \sigma^2 + \frac{1}{\delta} \E \left(B_0^2\right) \quad \mbox{and},\\
    {(\tau_0^{t+1})}^2 & = \sigma^2 + \frac{1}{\delta} \E \left[ \eta \big( \, B_0 + \tau_0^t Z \, ; \theta_{10}^t, \theta_{20}^t \big) - B_0 \right]^2 \quad \mbox{for } \, t\geq0,
\end{split}
\end{equation}
where $\sigma^2 = \E_{\prob_W} \left(W^2\right)$ and $Z \sim N(0,1)$ is independent of $B_0$. The fixed point equation (\ref{state evolution enet}) is defined as the \textit{state evolution} for the SA-Enet(0) and it characterizes the AMP algorithm. At each AMP iteration $t$, the recursions in (\ref{amp recursion enet group}) constructs a vector of ``effective observations'' $ \xmat^\T \e_0^t + \bs{\beta}_0^t $. When aggregated over components, the observations are distributed asymptotically as $B_0 + \tau_0^t Z$. Thus the effective observations can be thought of as a noisy version of the true signal $\bs{\beta}_0$, where each entry is corrupted by Gaussian noise with mean $0$ and standard deviation $\tau_0^t$. This is where $\eta$ plays a crucial role in the AMP algorithm and works as a denoiser on the vector. It treats effective observations with absolute values below $\theta_{10}^t$ as pure noises and shrinks them to 0. The theoretical guarantees follow from the general theorem in \cite{bayati11} as $\eta$ is Lipschitz (Please refer to Section III-B therein).

\paragraph{Correspondence between the AMP algorithm and the SA-Enet(0).} In order to provide an explicit connection between the SA-Enet(0) and its AMP algorithm in (\ref{amp recursion enet group}), we need a specific choice for the thresholds $\left\{ \theta_{10}^t, \theta_{20}^t \right\}_{t\geq0}$. The interpretation of the AMP algorithm presented above provides an intuition for this. At each AMP iteration $t$, since $ \xmat^\T \e_0^t + \bs{\beta}_0^t $ and $B_0 + \tau_0^t Z$ have the same distribution, $\big( \tau_0^t \big)^2$ can be interpreted as the mean square error (MSE) of the effective observations. Since $\theta_{10}^t$ provides a distinction between a noise and a signal, it intuitively makes sense to choose $\theta_{10}^t$ to be proportional to $\tau_0^t$. In the case of Lasso (that is, when $\theta_{20}^t = 0$), this choice is known to be minimax optimal for a suitable proportionality constant \citep{bayati12,donoho&johnstone1994,donoho&johnstone1998,dmm09}. So we set $\theta_{10}^t = \alpha_{10} \tau_0^t$ and $\theta_{20}^t = \alpha_{20} \tau_0^t$ where $\alpha_{10}$ and $\alpha_{20}$ are positive. Then, the AMP estimate $\bs{\beta}_0^t$ at any AMP iteration $t$ equals to the SA-Enet(0) estimate corresponding to $\lambda_{10} = \theta_{10}^t \left(1- \phi_0^t \right)$ and $\lambda_{20} = \theta_{20}^t \left(1- \phi_0^t \right)$, where
\begin{equation}
    \phi_0^t = \frac{1}{\delta} \mean{ \eta^\prime \left( \bs{\beta}_0^t + \xmat^\T \e_0^t ; \theta_{10}^t, \theta_{20}^t \right) } .
\end{equation}
In the large system limit, as the AMP iteration $t\uparrow\infty$, the correspondence is given by the functions $\lambda_{10} = \theta_{10}^* \left(1- \phi_0^* \right)$ and $\lambda_{20} = \theta_{20}^* \left(1- \phi_0^* \right)$, where for positive quantities $\alpha_{10}$ and $\alpha_{20}$
\begin{equation}\label{correspondence enet tau, theta_1, theta_2}
    \tau_0^* \equiv \tau_0^* (\alpha_{10}, \alpha_{20}) = \lim_{t\uparrow\infty} \tau_0^t, \quad \theta_{10}^* = \alpha_{10} \tau_0^*, \quad \theta_{20}^* = \alpha_{20} \tau_0^*, \quad \text{and}
\end{equation}
\begin{equation}\label{correspondence enet phi}
    \phi_0^* = \frac{1}{\delta} \, \E \left[\eta^\prime \left( B_0 + \tau_0^* Z ; \theta_{10}^*, \theta_{20}^* \right) \right] .
\end{equation}
This follows from Lemma~1(b) in \cite{bayati11}.

\paragraph{Adaptive weights based on the AMP SA-Enet(0) estimates.} Let $\bs{\beta}_{0}^*$ denote the limiting AMP SA-Enet(0) estimates. It is the limiting value of $\bs{\beta}_{0}^t$ in (\ref{amp recursion enet group}) as $t\uparrow\infty$. Following (\ref{weight update group}), in SA-Enet(1) it makes sense to choose the shrinkage threshold in group $d$ proportional to $\mean{\abs{\bs{\beta}_{0, S_d}^*}}^{-\gamma}$, the group average of the limiting AMP estimates. Following Lemma~1(b) in \cite{bayati11} under iid design, in the large system limit the group average converges to $\E \abs{\eta \left( \, B_{0d} + \tau_0^* Z \, ; \theta_{10}^{*}, \theta_{20}^{*} \right) }$. So we define the vector of adaptive weights $\bs{\omega}_1 = {( \omega_{11}, \cdots, \omega_{1D} )}^\T$ where $\omega_{1d} = {\left( \E \abs{\eta \left( \, B_{0d} + \tau_0^* Z \, ; \theta_{10}^{*}, \theta_{20}^{*} \right) } \right)}^{-\gamma}$ for any prefixed $\gamma \in (0,1]$ (it is worth noting that this choice of weight is not feasible and in practice, we estimate the weights using (\ref{weight update group})). Using $\bs{\omega}_1$, for group $d$ we propose the sequence of adaptive thresholds $\left\{ \theta_{11}^t \, \omega_{1d},\theta_{21}^t \right\}_{t \geq 0}$. The term $\theta_{11}^t \, \omega_{1d}$ plays the same role as $\theta_{10}^t$ in the SA-Enet(0), except now $\omega_{1d}$ allows the threshold of the denoiser $\eta$ to vary across groups encouraging adaptive shrinkage. This makes the AMP algorithm adaptive to the group structure.

\paragraph{Interpreting the AMP algorithm for the SA-Enet(1).} For the sequence of thresholds $\left\{ \theta_{11}^t \, \omega_{1d},\theta_{21}^t \right\}_{t \geq 0}$ in group $d$, the recursions in (\ref{amp recursion saenet group}) for $k=1$ identifies the AMP estimates for the SA-Enet(1). Assume the converging sequence of instances $\left\{ \bs{\beta}_0 (p), \bs{\varepsilon} (p) \right\}_{p \geq 0}$ as in the AMP algorithm for the SA-Enet(0). To formalize the group structure with $D$ groups, we further assume that $\prob_{B_0} = \sum_{d=1}^D c_d \, \prob_{B_{0d}}$ where $c_d$'s are non-negative and $\sum_{d=1}^D c_d = 1$. Then under the iid-design and in the large system limit as $p_d\uparrow\infty$, the asymptotic behavior of (\ref{amp recursion saenet group}) 
can be characterized by the state evolution defined by the sequence $\{\tau^t_1\}_{t\geq 0}$ as
\begin{equation}\label{state evolution saenet group}
\begin{split}
    {(\tau_1^0)}^2 & = \sigma^2 + \frac{1}{\delta} \E \left(B_0^2\right), \quad \mbox{and}\\
    {(\tau_1^{t+1})}^2 &= \sigma^2 + \frac{1}{\delta} \sum_{d=1}^D c_d \, \E \left[ \eta \left( \, B_{0d} + \tau_1^t Z \, ; \theta_{11}^t \, \omega_{1d}, \theta_{21}^t \right) - B_{0d} \right]^2 \quad \mbox{for } \, t\geq0,
\end{split}
\end{equation}
where $\sigma^2 = \E_{\prob_W} \left(W^2\right)$ and $Z \sim N(0,1)$ is independent of $B_{0d}$ for all $d$. At each AMP iteration $t$, the algorithm constructs the same vector of effective observations $ \xmat^\T \e_1^t + \bs{\beta}_1^t $, except that now under the assumption of a group structure the observations in group $d$ (when aggregated over components in that group) are distributed asymptotically as $B_{0d} + \tau_1^t Z$ with $B_{0d}$ and $Z \sim N(0,1)$ being independent of each other. The theoretical guarantees follow from Theorem~\ref{ampthm saenet group}.

\paragraph{Correspondence between the AMP and the SA-Enet(1).} To explicitly connect SA-Enet(1) to its AMP algorithm (\ref{amp recursion saenet group}), we similarly set the thresholds as $\theta_{11}^t = \alpha_{11} \tau_1^t$ and $\theta_{21}^t = \alpha_{21} \tau_1^t$. At any AMP iteration $t$, the AMP estimate $\bs{\beta}_1^t$ equal to the SA-Enet(1) corresponding to $\lambda_{11} = \theta_{11}^t \left(1- \phi_1^t \right)$ and $\lambda_{21} = \theta_{21}^t \left(1- \phi_1^t \right)$, where
\begin{equation}
    \phi_1^t = \frac{1}{\delta} \sum_{d=1}^D c_d \, \mean{\eta^\prime \left( \left( \xmat^\T \e_1^t + \bs{\beta}_1^t \right)_{S_d} \, ; \theta_{11}^t \, \omega_{1d},\theta_{21}^t \right)} .
\end{equation}
In the large system limit as the AMP iteration $t\uparrow\infty$ and $p_d\uparrow\infty$, the correspondence between the SA-Enet(1) and its AMP algorithm is given by the functions $\lambda_{11} = \theta_{11}^* \left(1- \phi_1^* \right)$ and $\lambda_{21} = \theta_{21}^* \left(1- \phi_1^* \right)$, where for positive quantities $\alpha_{11}$ and $\alpha_{21}$,
\begin{equation}\label{correspondence saenet group tau, theta_1, theta_2}
    \tau_1^* \equiv \tau_1^* (\alpha_{11}, \alpha_{21}) = \lim_{t\uparrow\infty} \tau_1^t, \quad \theta_{11}^* = \alpha_{11} \tau_1^*, \quad \theta_{21}^* = \alpha_{21} \tau_1^*, \quad \text{and}
\end{equation}
\begin{equation}\label{correspondence saenet group phi}
    \phi_1^* = \frac{1}{\delta} \sum_{d=1}^D c_d \, \E \left[ \eta^\prime \left( B_{0d} + \tau_1^* Z ; \theta_{11}^* \, \omega_{1d}, \theta_{21}^* \right) \right] .
\end{equation}

\begin{theorem}\label{ampthm saenet group}
Suppose model (\ref{model}) holds true for the observed data $(\y, \xmat)$. Let the true signal $\bs{\beta}_0 \in \R^p$ can be partitioned into $D$ mutually exclusive groups of sizes $p_1,\dots,p_D$. Consider the recursion (\ref{amp recursion saenet group}) for $k=1$ and let $\psi_d : \R^2 \mapsto \R$ is pseudo-Lipschitz of order $\nu$ for all $d=1,\dots,D$. Also, assume the following conditions hold:
\begin{itemize}

\item[(A1)] \textbf{``iid design''.} $\left\{ \xmat (p) \right\}_{p \geq 0}$ is a sequence of design matrices $\xmat \in \R^{n \times p} $ indexed by $p$ with iid entries $\xmat_{ij} \sim \bs{N} (0, 1/n)$.

\item[(A2)]\hypertarget{ampgrp-a2}{} \textbf{ Large system limit.} $\{p_d\}_{d=1}^D$ and $n \equiv n(p)$ increase to $\infty$ such that $n/p \rightarrow \delta \in (0,\infty)$, and $p_d/p \rightarrow c_d \in (0,1)$ for all $d$.

\item[(A3)] \textbf{Weak convergence of signals.} In group $d$, the empirical distribution of the sequence of signals $\{\bs{\beta}_{0, S_d} (p_d)\}_{p_d \geq 0}$ converge weakly to a probability measure $B_{0d} \sim \prob_{B_{0d}}$ with
bounded ${(2\nu-2)}^{th}$ moment.

\item[(A4)] \textbf{Weak convergence of noise.} The noise $\bs{\varepsilon}$ has iid entries and its empirical distribution weakly converges to a probability measure $\prob_W$ with bounded ${(2\nu-2)}^{th}$ moment.
\end{itemize}
Then, for all $d=1,\dots,D$ and $t \geq 0$,
\begin{equation}\label{amprisk indiv group}
    \lim_{p_d \rightarrow \infty} \, \frac{1}{p_d} \sum_{j \in S_d} \psi_d \left( \beta_{1j}^{t+1} , \beta_{0j} \right) \overset{a.s.}{=} \E \left[ \psi_d \left( \eta \left( B_{0d} + \tau_1^t Z \, ; \theta_{11}^t \, \omega_{1d}, \theta_{21}^t \right), B_{0d} \right) \right],
\end{equation}
where $Z \sim N (0,1)$ is independent of $B_{0d}$, and $\left\{ \tau_1^t \right\}_{t \geq 0}$ is defined by the state evolution in (\ref{state evolution saenet group}).
\end{theorem}

\begin{rem}
    Following \cite{bayati11}, we prove the result for a general recursion and identify the AMP recursions (\ref{amp recursion saenet group}) for $k=1$ as a special case. We present this in Section \ref{sec app: Proof of Theorem 3.1} in the appendix.
\end{rem}

{\color{black}
\begin{rem}
{\rm
Assume the conditions of Theorem~\ref{ampthm saenet group}. Emphasizing the dependence on $p$, let $\left\{\bs{\beta}_1^{t+1} (p)\right\}_{t\geq0}$ denote the sequence of AMP estimates corresponding to the AMP parameters $(\alpha_{11},\alpha_{21})$ and $\hat{\bs{\beta}}^{\saenet}_1 (p)$ be the SA-Enet estimator corresponding to $(\lambda_{11},\lambda_{21})$. Also, $(\alpha_{11},\alpha_{21})$ and $(\lambda_{11},\lambda_{21})$ satisfy the correspondence (\ref{correspondence saenet group tau, theta_1, theta_2})--(\ref{correspondence saenet group phi}). As $t\uparrow\infty$, the AMP estimate satisfies the KKT conditions for the SA-Enet. Like in the proof of Theorem~\ref{ampthm saenet group}, the techniques of Theorem 1.8 in \cite{bayati12} can be similarly adapted to our setting with some modifications and we get
\begin{equation}\label{amp saenet characterization}
    \lim_{t\uparrow\infty} \, \lim_{p\uparrow\infty} \, \frac{1}{p} \norm{\bs{\beta}_1^t(p) - \hat{\bs{\beta}}^{\saenet}_1 (p)}_2^2 = 0 \quad \text{almost surely.}
\end{equation}
As \cite{bayati12} points out in their theorem, this result requires taking the limit of $p\uparrow\infty$ first before taking the limit of $t\uparrow\infty$ and thus presents a high-dimensional limit behavior of the estimator for a large-but-finite number of AMP iterations. The interpretation of the result is the same as presented by them. It implies that for any finite tolerance $\zeta>0$, there exists a finite AMP iteration $t_* (\zeta)$ such that for any $t \geq t_* (\zeta)$ the difference between the MSEs of the AMP estimate and the SA-Enet estimator is at most $\zeta$ with high probability as $p\uparrow\infty$. Numerical results presented in Section~\ref{subsec: State evolution prediction as a finite sample approximation} confirm this finding for $p$ as low as 500.
}
\end{rem}
}



\begin{rem}
Suppose $\psi : \R^2 \mapsto \R$ is pseudo-Lipschitz of order $\nu$. For a prefixed $(\alpha_{11} , \alpha_{21})$, under the conditions of Theorem~\ref{ampthm saenet group} and following (\ref{amprisk indiv group}), the asymptotic risk of the AMP estimate $\bs{\beta}^{t+1}_1$ at any AMP iteration $t\geq0$ is given by
\begin{equation}\label{amprisk saenet group}
    \lim_{p \rightarrow \infty} \, \frac{1}{p} \sum_{j=1}^p \psi \left( \beta_{1j}^{t+1} , \beta_{0j} \right) \overset{a.s.}{=} \sum_{d=1}^D c_d \, \E \left[ \psi \left( \eta \left( B_{0d} + \tau_1^t Z \, ; \theta_{11}^t \, \omega_{1d}, \theta_{21}^t \right), B_{0d} \right) \right] .
\end{equation}
Assuming the squared error loss $\psi (a,b) = (a-b)^2$ and following (\ref{state evolution saenet group}) and (\ref{amprisk saenet group}), the risk at any iteration $t$ simplifies to $\delta \left( ({\tau^{t+1}_1})^2 - \sigma^2 \right)$ and equals to $\delta \left( ({\tau^*_1})^2 - \sigma^2 \right)$ as $t\uparrow\infty$.
To our interest, due to the correspondence (\ref{correspondence saenet group tau, theta_1, theta_2})--(\ref{correspondence saenet group phi}), the AMP estimate is the SA-Enet(1) estimate corresponding to $\lambda_{11} = \theta_{11}^* \left(1- \phi_1^* \right)$ and $\lambda_{21} = \theta_{21}^* \left(1- \phi_1^* \right)$. So, under a group structure, we expect $\delta \left( ({\tau^*_1})^2 - \sigma^2 \right)$ to accurately  approximate the squared error risk of the SA-Enet(1).
\end{rem}

{\color{black}
\begin{rem}
The AMP arguments for the SA-Enet(1) can be recursively applied to develop the AMP framework in theoretically analyzing the risk of the SA-Enet($T$). For $k = 2, \dots, T$, the key components in the analyses are summarized below.
\begin{itemize}
    \item \textbf{Asymptotic Adaptive Weights.} Based on the limiting AMP SA-Enet($k-1$) estimates, the vector of adaptive weights is defined as $\bs{\omega}_k = {( \omega_{k1}, \cdots, \omega_{kD} )}^\T$ with
    \begin{equation}
        \omega_{kd} = {\left( \E \abs{\eta \left( \, B_{0d} + \tau_{k-1}^* Z \, ; \theta_{1,k-1}^{*} \omega_{k-1,d}, \theta_{2,k-1}^{*} \right) } \right)}^{-\gamma} ,
    \end{equation}
    where $\theta_{1,k-1}^* = \alpha_{1,k-1} \tau_{k-1}^*$ and $\theta_{2,k-1}^* = \alpha_{2,k-1} \tau_{k-1}^*$. We note that, $\omega_{kd}$ is the asymptote of the finite-sample adaptive weights $\hat{w}_{kd}$ from (\ref{optimization step}) in Algorithm~\ref{algorithm saenet} as $p \uparrow \infty$. Here $\bs{\omega}_k$ is only used to define the AMP algorithm for theoretically analyzing the SA-Enet estimator. For observed data, Algorithm~\ref{algorithm saenet} is used to obtain the SA-Enet estimates.
    
    \item \textbf{AMP Recursions.} The AMP estimates are defined through recursions (\ref{amp recursion saenet group}), where for group $d$ the sequence of thresholds is $\left\{\theta_{1k}^t \, \omega_{kd}, \theta_{2k}^t\right\}_{t\geq0}$.
    
    \item \textbf{State Evolution.} The state evolution characterizing the behavior of AMP estimate is given by
    \begin{equation}\label{state evolution saenet_k group}
    \begin{split}
    {(\tau_k^0)}^2 & = \sigma^2 + \frac{1}{\delta} \E \left(B_0^2\right), \quad \mbox{and}\\
    {(\tau_k^{t+1})}^2 &= \sigma^2 + \frac{1}{\delta} \sum_{d=1}^D c_d \, \E \left[ \eta \left( \, B_{0d} + \tau_k^t Z \, ; \theta_{1k}^t \, \omega_{kd}, \theta_{2k}^t \right) - B_{0d} \right]^2, \mbox{ for } t\geq0 .
    \end{split}
    \end{equation}
    
    \item \textbf{Correspondence between the AMP and the SA-Enet($k$).} Define, $\theta_{1k}^t = \alpha_{1k} \tau_k^t$ and $\theta_{2k}^t = \alpha_{2k} \tau_k^t$. Then for prefixed positive constants $\left(\alpha_{1k} , \alpha_{2k} \right)$, the AMP estimate $\bs{\beta}_k^t$ at any AMP iteration $t$ equals to the $\mbox{SA-Enet}(k)$ estimate corresponding to $\lambda_{1k} = \theta_{1k}^t \left(1- \phi_k^t \right)$ and $\lambda_{2k} = \theta_{2k}^t \left(1- \phi_k^t \right)$, where
    \begin{equation}
        \phi_k^t = \frac{1}{\delta} \sum_{d=1}^D c_d \, \mean{\eta^\prime \left( \left( \xmat^\T \e_k^t + \bs{\beta}_k^t \right)_{S_d} \, ; \theta_{1k}^t \, \omega_{kd},\theta_{2k}^t \right)} .
    \end{equation}
In the large system limit, the limiting AMP estimate $\bs{\beta}_k^*$ matches with the $\mbox{SA-Enet}(k)$ estimate corresponding to $\lambda_{1k} = \theta_{1k}^* \left(1- \phi_k^* \right)$ and $\lambda_{2k} = \theta_{2k}^* \left(1- \phi_k^* \right)$ where
\begin{equation}\label{correspondence saenet_k group tau, theta_1, theta_2}
    \tau_k^* \equiv \tau_k^* (\alpha_{1k}, \alpha_{2k}) = \lim_{t\uparrow\infty} \tau_k^t, \quad \theta_{1k}^* = \alpha_{1k} \tau_k^*, \quad \theta_{2k}^* = \alpha_{2k} \tau_k^*, \quad \text{and}
\end{equation}
\begin{equation}\label{correspondence saenet_k group phi}
    \phi_k^* = \frac{1}{\delta} \sum_{d=1}^D c_d \, \E \left[ \eta^\prime \left( B_{0d} + \tau_k^* Z ; \theta_{1k}^* \, \omega_{kd}, \theta_{2k}^* \right) \right] .
\end{equation}
    
    \item \textbf{Asymptotic Risk of the SA-Enet($k$) Estimator.} Suppose, the assumptions in Theorem~\ref{ampthm saenet group} holds true. So for all $d=1,\dots,D$ and $t \geq 0$, applying (\ref{amprisk indiv group}) for a pseudo-Lipschitz function $\psi_d : \R^2 \mapsto \R$ of order $\nu$, we get
\begin{equation}\label{amprisk saenet_k indiv group}
    \lim_{p_d \rightarrow \infty} \, \frac{1}{p_d} \sum_{j \in S_d} \psi_d \left( \beta_{kj}^{t+1} , \beta_{0j} \right) \overset{a.s.}{=} \E \left[ \psi_d \left( \eta \left( B_{0d} + \tau_k^t Z \, ; \theta_{1k}^t \, \omega_{kd}, \theta_{2k}^t \right), B_{0d} \right) \right],
\end{equation}
where $Z \sim N (0,1)$ is independent of $B_{0d}$, and $\left\{ \tau_k^t \right\}_{t \geq 0}$ is defined by the state evolution in (\ref{state evolution saenet_k group}). Using this for a pseudo-Lipschitz function $\psi : \R^2 \mapsto \R$ of order $\nu$, the asymptotic risk of the AMP estimate $\bs{\beta}^{t+1}_k$ at any AMP iteration $t\geq0$ is given by
\begin{equation}\label{amprisk saenet_k group}
    \lim_{p \rightarrow \infty} \, \frac{1}{p} \sum_{j=1}^p \psi \left( \beta_{kj}^{t+1} , \beta_{0j} \right) \overset{a.s.}{=} \sum_{d=1}^D c_d \, \E \left[ \psi \left( \eta \left( B_{0d} + \tau_k^t Z \, ; \theta_{1k}^t \, \omega_{kd}, \theta_{2k}^t \right), B_{0d} \right) \right] .
\end{equation}
Assuming the squared error loss, the asymptotic squared error risk simplifies to $\delta \left( ({\tau^{t+1}_k})^2 - \sigma^2 \right)$, which in the limit equals to $\delta \left( ({\tau^*_k})^2 - \sigma^2 \right)$ as $t \uparrow \infty$. Because of the correspondence, under the group structure, we expect this to provide an accurate approximation of the squared error risk of the SA-Enet($k$) estimator. This is empirically confirmed in Section~\ref{subsec: State evolution prediction as a finite sample approximation} through simulation studies.
\end{itemize}
\end{rem}
}

\subsection{AMP algorithm under covariate-dependent structure}\label{sec: AMP algorithm under covariate-dependent structure}

In this section, we extend the arguments in Section~\ref{sec: AMP algorithm under group structure} and propose the AMP algorithm for the SA-Enet when auxiliary covariate information is available for the features.
Note that, the SA-Enet(0) does not depend on any structural information. So its AMP framework is the same as in the group structure. Fixing $T=1$, we first discuss the adaptive weight updates based on AMP SA-Enet(0) estimates and the AMP algorithm for the SA-Enet(1) under the covariate-dependent structure. Then the results can be recursively applied to develop the AMP framework for the SA-Enet($T$).
\begin{algorithm}[h!]
	\caption{\textbf{: AMP algorithm for SA-Enet($\bs{T}$) under covariate-dependent structure}}\label{amp algorithm cov}
	\vspace{1mm}
	\begin{enumerate}
 
        \item[(1)] Fix the maximum number of iterations $T$.
        
        \item[(2)] {\bf AMP for the SA-Enet(0).}  Initialize $\bs{\beta}_0^0 = 0$ and $\e_0^{-1}=0$. For $t\geq0$ the algorithm constructs the following recursion until convergence:
		\begin{equation}\label{amp recursion enet cov} 
		    \begin{split}
		        \e_0^t & = \, \bs{y} - \xmat \bs{\beta}_0^t + \frac{\e_0^{t-1}}{\delta} \mean{ \eta^\prime \left( \, \xmat^\T \e_0^{t-1} + \bs{\beta}_0^{t-1} \, ; \theta_{10}^{t-1},\theta_{20}^{t-1} \right) },\\
		        \bs{\beta}_0^{t+1} & = \, \eta \left( \, \xmat^\T \e_0^t + \bs{\beta}_0^t \, ; \theta_{10}^{t},\theta_{20}^{t} \right).
		    \end{split}
		\end{equation}
		
		\item[(3)]  At iteration $k = 1, \ldots, T$, consider the following algorithm.\\

            \vspace{1mm}
            {\bf AMP for the SA-Enet($k$).}
  Define $\bs{\omega}_k = {( \omega_{k1},\dots,\omega_{kp})}^\T$ such that $\omega_{kj} = f( \bs{u}_j ; {\bs{\rho}}^*_k )$ with $\omega_{0j} = 1$ $\forall \, j=1,\dots,p$, and $\Omega_k = f( U ; {\bs{\rho}}^*_k )$ with $\Omega_1=1$. Here
    \begin{equation*}
	\bs{\rho}^*_k = \argmin_{\bs{\rho}\in\mathcal{B}} L_k^*(\bs{\rho}; \gamma), \quad \text{and}
    \end{equation*}
    \begin{equation*}
	L^*_k ( \bs{\rho} ; \gamma) = \mathbb{E} \Big[ f( U ; \bs{\rho} ) \, \abs{\eta \left( B_0 + \tau_{k-1}^* Z \, ; \theta_{1,k-1}^* \Omega_{k-1}, \theta_{2,k-1}^* \right)} \Big]  - \mathbb{E} \Big[ \log g \Big( f( U ; \bs{\rho} ) ; \gamma \Big) \Big] .
    \end{equation*}
    For feature $j$, define the sequence of thresholds $\left\{ \theta_{1k}^t \, \omega_{kj},\theta_{2k}^t \right\}_{t \geq 0}$. Also, initialize $\bs{\beta}_k^0 = 0$ and $\e_k^{-1}=0$. Then, for $t\geq0$ the algorithm constructs the following recursion until convergence:
		\begin{equation}\label{amp recursion saenet cov}
		    \begin{split}
		        \e_k^t  & = \, \y - \xmat \bs{\beta}_k^t + \frac{\e_k^{t-1}}{\delta} \, \mean{ \eta^\prime \left( \, \xmat^\T \e_k^{t-1} + \bs{\beta}_k^{t-1} \, ; \theta_{1k}^{t-1} \, \bs{\omega}_k,\theta_{2k}^{t-1} \right) },\\
		        \bs{\beta}_k^{t+1}  & = \, \eta \left( \xmat^\T \e_k^{t} + \bs{\beta}_k^t \, ; \theta_{1k}^{t} \, \bs{\omega}_k,\theta_{2k}^t \right) .
		    \end{split}
		\end{equation}
	\end{enumerate}
\end{algorithm}

\paragraph{Adaptive weights based on the AMP SA-Enet(0) estimates.} Let $\bs{\beta}_{0}^*$ denote the limiting AMP SA-Enet(0) estimates. It is the limiting value of $\bs{\beta}_0^t$ in (\ref{amp recursion enet cov}) (the same as (\ref{amp recursion enet group})) as $t\uparrow\infty$. Following (\ref{weight update covariate}), in SA-Enet(1) it makes sense to choose the shrinkage threshold for the $j^{th}$ feature proportional to $w_j \left(\bs{\beta}_{0}^* \right)$. Let us assume that the empirical joint distribution of $\left(\bs{u}_j , \beta_{0j} \right)$ weakly converges to $(U , B_0 ) \sim \mathbb{P}_{U , B_0}$, and define
\begin{equation}\label{opt rho amp}
    \hat{\bs{\rho}}_1 = \argmin_{\bs{\rho}\in\mathcal{B}} L_{1p} (\bs{\rho}; \gamma), \quad \text{and} 
    \quad \bs{\rho}^*_1 =\argmin_{\bs{\rho}\in\mathcal{B}} L^*_1 (\bs{\rho}; \gamma) ,
\end{equation}
where
\begin{equation}\label{L_p function}
    L_{1p} ( \bs{\rho} ; \gamma) = \frac{1}{p} \sum_{j=1}^{ p } \Big[ f( \bs{u}_j ; \bs{\rho} )\abs{\beta^*_{0j}} - \log g \Big( f( \bs{u}_j ; \bs{\rho} ) ; \gamma \Big) \Big], \,\, \text{and}
\end{equation}
\begin{equation}\label{L function}
    L^*_1 ( \bs{\rho} ; \gamma) = \mathbb{E} \Big[ f( U ; \bs{\rho} ) \, \abs{\eta \left( B_0 + \tau_0^* Z \, ; \theta_{10}^{*}, \theta_{20}^{*} \right)} \Big]  - \mathbb{E} \Big[ \log g \Big( f( U ; \bs{\rho} ) ; \gamma \Big) \Big] .
\end{equation}
Under iid-design, following similar steps in the proof of Lemma~1(b) in \cite{bayati11} together with some assumptions on $f$ and $U$, it can be shown that $\hat{\bs{\rho}}_1 \overset{P}{\rightarrow} {\bs{\rho}}^*_1$ (A sketch of the proof is deferred to Section \ref{sec app: Adaptive weights in AMP algorithm under covariate-dependent structure} in the appendix). Then for a prefixed $\gamma$, we define the vector of adaptive weights $\bs{\omega}_1 = {( \omega_{11}, \cdots, \omega_{1p} )}^\T$ with $\omega_{1j} = f( \bs{u}_j ; {\bs{\rho}}^*_1 )$. Using $\bs{\omega}_1$, for feature $j$ we define the sequence of thresholds $\left\{ \theta_{11}^t \, \omega_{1j}, \theta_{21}^t \right\}_{t \geq 0}$. $\theta_{11}^t \, \omega_{1j}$ plays the same role as $\theta_{10}^t$ in the SA-Enet(0), except that $\omega_{1j}$ allows the threshold of the denoiser $\eta$ to vary across the features encouraging adaptive shrinkage. This makes the AMP algorithm adaptive to the auxiliary covariate information.

\paragraph{Interpreting the AMP algorithm for the SA-Enet(1).} For the sequence of thresholds $\left\{ \theta_{11}^t \, \omega_{1j},\theta_{21}^t \right\}_{t \geq 0}$ for the feature $j$, the recursions in (\ref{amp recursion saenet cov}) identifies the AMP estimates for the SA-Enet(1). Similarly assuming a weakly converging sequence of instances $\left\{ \bs{\beta}_0 (p), \bs{\varepsilon} (p) \right\}_{p \geq 0}$ where their empirical distributions converge to the probability measures $\prob_{B_0}$ and $\prob_W$, the asymptotic behavior of (\ref{amp recursion saenet cov}) under the iid-design
can be tracked by the state evolution defined as
\begin{equation}\label{state evolution saenet cov}
\begin{split}
    {(\tau_1^0)}^2 & = \sigma^2 + \frac{1}{\delta} \E \left(B_0^2\right) \quad \mbox{and},\\
    {(\tau_1^{t+1})}^2 &= \sigma^2 + \frac{1}{\delta} \E \Big[ \eta \left( \, B_0 + \tau_1^t Z \, ; \theta_{11}^t \, \Omega_1, \theta_{21}^t \right) - B_0 \Big ]^2 \quad \mbox{for } \, t\geq0,
\end{split}
\end{equation}
where $\sigma^2 = \E_{\prob_W} \left(W^2\right)$, $\Omega_1 = f( U ; {\bs{\rho}}^*_1 )$, and $Z \sim N(0,1)$ is independent of $B_0$. This characterizes the AMP algorithm in (\ref{amp recursion saenet cov}). With mild assumptions on $f$ and $U$ mentioned above, the theoretical guarantees follow by essentially following the same steps in the proof of the general theorem in \cite{bayati11}. At each AMP iteration $t$, the algorithm constructs the same vector of effective observations $ \xmat^\T \e_1^t + \bs{\beta}_1^t $. Under the assumption of a covariate-dependent structure, the observations (when aggregated over components) are asymptotically distributed as $B_{0} + \tau_1^t Z$. 
The $j^{th}$ effective observation can be thought of as a noisy version of the true signal $\beta_{0j}$ where each entry is corrupted by Gaussian noise with mean $0$ and standard deviation $\tau_1^t$. The function $\eta$ in (\ref{amp recursion saenet cov}) works as a denoiser and shrinks the element to 0 if the absolute value falls within $\theta_{11}^t \, \omega_{1j}$. This makes the AMP algorithm adaptive to the external covariate structural information.

\paragraph{Correspondence between the AMP and the SA-Enet(1).} For the correspondence, we similarly set the thresholds as $\theta_{11}^t = \alpha_{11} \tau_1^t$ and $\theta_{21}^t = \alpha_{21} \tau_1^t$. Then at any AMP iteration $t$, the fixed point $\bs{\beta}_1^t$ equals to the SA-Enet(1) with $\lambda_{11} = \theta_{11}^t \left(1- \phi_1^t \right)$ and $\lambda_{21} = \theta_{21}^t \left(1- \phi_1^t \right)$, where
\begin{equation}
    \phi_1^t = \frac{1}{\delta} \, \mean{\eta^\prime \left( \xmat^\T \e_1^t + \bs{\beta}_1^t \, ; \theta_{11}^t \, \bs{\omega}_1 , \theta_{21}^t \right)} .
\end{equation}
In the large system limit as the AMP iteration $t\uparrow\infty$, the correspondence is given by the functions $\lambda_{11} = \theta_{11}^* \left(1- \phi_1^* \right)$ and $\lambda_{21} = \theta_{21}^* \left(1- \phi_1^* \right)$, where for positive quantities $\alpha_{11}$ and $\alpha_{21}$,
\begin{equation}\label{correspondence saenet cov tau, theta_1, theta_2}
    \tau_1^* \equiv \tau_1^* (\alpha_{11}, \alpha_{21}) = \lim_{t\uparrow\infty} \tau_1^t, \quad \theta_{11}^* = \alpha_{11} \tau_1^*, \quad \theta_{21}^* = \alpha_{21} \tau_1^*, \quad \text{and}
\end{equation}
\begin{equation}\label{correspondence saenet cov phi}
    \phi_1^* = \frac{1}{\delta} \, \E \left[ \eta^\prime \left( B_{0} + \tau_1^* Z ; \theta_{11}^* \, \Omega_1, \theta_{21}^* \right) \right] .
\end{equation}
with $\Omega_1$ as in (\ref{state evolution saenet cov}).

{\color{black}
\begin{prop}\label{ampprop saenet cov}
Suppose model (\ref{model}) holds true for the observed data $(\y, \xmat)$. Let the true signal be $\bs{\beta}_0$ and a $q$-variate auxiliary covariate information $\bs{u} (p) = \left( \bs{u}_1 , \dots, \bs{u}_p \right)^\T$ be available corresponding to each component. Consider the recursion (\ref{amp recursion saenet cov}) at $k=1$ and let $\psi : \R^2 \mapsto \R$ be pseudo-Lipschitz of order $\nu$. Assume the following conditions hold:
\begin{itemize}

\item[(A1)] \textbf{``iid design''.} ${\xmat (p)}_{p \geq 0}$ is a sequence of design matrices $\xmat \in \R^{n \times p} $ indexed by $p$ with iid entries $\xmat_{ij} \sim \bs{N} (0, 1/n)$.

\item[(A2)]\hypertarget{ampcov-a2}{} \textbf{ Large system limit.} $p$ and $n \equiv n(p)$ increase to $\infty$ such that $n/p \rightarrow \delta \in (0,\infty)$.

\item[(A3)] \textbf{Weak convergence of signals and auxiliary covariates.} The joint empirical distribution of $\{ (\bs{\beta}_{01} , \bs{u}_1 ), \dots,  (\bs{\beta}_{0p}, \bs{u}_{p} ) \}_{p \geq 0}$ and the empirical distribution of the sequence of signals $\{\bs{\beta}_{0} (p)\}_{p \geq 0}$ converge weakly to probability measures $(B_0 , U) \sim \prob_{B_0 , U}$ and $B_{0} \sim \prob_{B_{0}}$, respectively, with $\prob_{B_{0}}$ having bounded ${(2\nu-2)}^{th}$ moment.

\item[(A4)] \textbf{Weak convergence of noise.} The noise $\bs{\varepsilon}$ has iid entries and its empirical distribution weakly converges to a probability measure $\prob_W$ with bounded ${(2\nu-2)}^{th}$ moment.

\item[(A5)] \textbf{Conditions on $f$ and $U$.}
\begin{equation}
    \E \Bigg[ \sup_{\bs{\rho}\in\mathcal{B}} \abs{f( U ; \bs{\rho}^*_1 )}^2 \Bigg]<\infty, \quad \mbox{and} \quad \E \Big[ \abs{g( B_0 + \tau_1^t Z , U )} \Big]<\infty,
\end{equation}
where $\bs{\rho}^*_1$ is as defined in (\ref{opt rho amp}).
\end{itemize}
Then for $t\geq0$, we expect
\begin{equation}\label{amprisk saenet cov}
    \lim_{p \rightarrow \infty} \, \frac{1}{p} \sum_{j=1}^p \psi \left( \beta_{1j}^{t+1} , \beta_{0j} \right) \overset{a.s.}{=} \E \left[ \psi \left( \eta \left( B_{0} + \tau_1^t Z \, ; \theta_{11}^t \, \Omega_{1}, \theta_{21}^t \right), B_0 \right) \right],
\end{equation}
where $Z \sim N (0,1)$ is independent of $B_0$, and $\left\{ \tau_1^t \right\}_{t \geq 0}$ is defined by the state evolution in (\ref{state evolution saenet cov}).
\end{prop}

\begin{rem}
    The proof of Proposition~\ref{ampprop saenet cov} follows from the general result presented in Section III(B) in \cite{bayati11}.
\end{rem}

{\color{black}
\begin{rem}
{\rm
Assume the conditions of Proposition~\ref{ampprop saenet cov} and the same notations as in Remark 3.2. As in the group structure, the AMP estimate, in this case, satisfies the KKT conditions for the SA-Enet as $t\uparrow\infty$. Following a similar adaptation with some modifications for the covariate-dependent structure (\ref{amp saenet characterization}) holds true. The interpretation of the result is the same as in the group structure.
}
\end{rem}
}

\begin{rem}
Under the squared error loss, following (\ref{state evolution saenet cov}) and (\ref{amprisk saenet cov}), at any AMP iteration $t\geq0$ the asymptotic squared error risk of the AMP estimate $\bs{\beta}_{1}^{t+1}$ simplifies to $\delta \left( ({\tau^{t+1}_1})^2 - \sigma^2 \right)$. As AMP iteration $t\uparrow\infty$, the risk of the limiting AMP estimate equals to $\delta \left( ({\tau^*_1})^2 - \sigma^2 \right)$.
To our interest, the correspondence (\ref{correspondence saenet cov tau, theta_1, theta_2})--(\ref{correspondence saenet cov phi}) implies that this is also the asymptotic risk of the SA-Enet(1) estimate corresponding to $\lambda_{11} = \theta_{11}^* \left(1- \phi_1^* \right)$ and $\lambda_{21} = \theta_{21}^* \left(1- \phi_1^* \right)$. Under a covariate-dependent structure, we expect this to provide a good approximation to its actual squared error risk.
\end{rem}

\begin{rem}
Similar to the group structure, the AMP arguments for the SA-Enet(1) can be recursively applied to develop the AMP framework in theoretically analyzing the risk of the SA-Enet($T$). For $k = 2, \dots, T$, the key components in the analyses are summarized below.
\begin{itemize}
    \item \textbf{Asymptotic Adaptive Weights.} Based on the limiting AMP SA-Enet($k-1$) estimates, the vector of adaptive weights is defined as $\bs{\omega}_k = {( \omega_{k1},\dots,\omega_{kp})}^\T$ with $\omega_{kj} = f( \bs{u}_j ; {\bs{\rho}}^*_k )$ $\forall \, j=1,\dots,p$. Here $\theta_{1,k-1}^* = \alpha_{1,k-1} \tau_{k-1}^*$, $\theta_{2,k-1}^* = \alpha_{2,k-1} \tau_{k-1}^*$, $$\bs{\rho}^*_k = \argmin_{\bs{\rho}\in\mathcal{B}} L^*_k(\bs{\rho}; \gamma),$$ and 
    \begin{align*}
        L^*_k ( \bs{\rho} ; \gamma) = & \mathbb{E} \Big[ f( U ; \bs{\rho} ) \, \abs{\eta \left( B_0 + \tau_{k-1}^* Z \, ; \theta_{1,k-1}^* \Omega_{k-1}, \theta_{2,k-1}^* \right)} \Big] \\
        & - \mathbb{E} \Big[ \log g \Big( f( U ; \bs{\rho} ) ; \gamma \Big) \Big].
    \end{align*}
    As in the group structure, we note that $\bs{\omega}_k$ is the asymptote of the finite-sample adaptive weights $\hat{\bs{w}}_{k}$ from (\ref{optimization step}) in Algorithm~\ref{algorithm saenet} as $p \uparrow \infty$. Here $\bs{\omega}_k$ is only used to define the AMP algorithm in theoretically analyzing the SA-Enet estimator. For observed data, Algorithm~\ref{algorithm saenet} is used to obtain the SA-Enet estimates.
    
    \item \textbf{AMP Recursions.} The AMP estimates are defined through recursions (\ref{amp recursion saenet cov}), where for feature $j$ the sequence of thresholds is $\left\{\theta_{1k}^t \, \omega_{kj}, \theta_{2k}^t\right\}_{t\geq0}$.
    
    \item \textbf{State Evolution.} The state evolution characterizing the behavior of AMP estimate is given by
    \begin{equation}\label{state evolution saenet_k cov}
    \begin{split}
    {(\tau_k^0)}^2 & = \sigma^2 + \frac{1}{\delta} \E \left(B_0^2\right), \quad \mbox{and}\\
    {(\tau_k^{t+1})}^2 &= \sigma^2 + \frac{1}{\delta} \, \E \left[ \eta \left( \, B_0 + \tau_k^t Z \, ; \theta_{1k}^t \, \Omega_k, \theta_{2k}^t \right) - B_0 \right]^2, \mbox{ for } t\geq 0 ,
    \end{split}
    \end{equation}
    where $\Omega_k = f( U ; {\bs{\rho}}^*_k )$.
    
    \item \textbf{Correspondence between the AMP and the SA-Enet($k$).} Define, $\theta_{1k}^t = \alpha_{1k} \tau_k^t$ and $\theta_{2k}^t = \alpha_{2k} \tau_k^t$. Then for prefixed positive constants $\left(\alpha_{1k} , \alpha_{2k} \right)$, the AMP estimate $\bs{\beta}_k^t$ at any AMP iteration $t$ equals to the $\mbox{SA-Enet}(k)$ estimate corresponding to $\lambda_{1k} = \theta_{1k}^t \left(1- \phi_k^t \right)$ and $\lambda_{2k} = \theta_{2k}^t \left(1- \phi_k^t \right)$, where
    \begin{equation}
        \phi_k^t = \frac{1}{\delta} \, \mean{\eta^\prime \left( \xmat^\T \e_k^t + \bs{\beta}_k^t \, ; \theta_{1k}^t \, \bs{\omega}_k, \theta_{2k}^t \right)} .
    \end{equation}
    In the large system limit, the limiting AMP estimate $\bs{\beta}_k^*$ matches with the $\mbox{SA-Enet}(k)$ estimate corresponding to $\lambda_{1k} = \theta_{1k}^* \left(1- \phi_k^* \right)$ and $\lambda_{2k} = \theta_{2k}^* \left(1- \phi_k^* \right)$ where
\begin{equation}\label{correspondence saenet_k cov tau, theta_1, theta_2}
    \tau_k^* \equiv \tau_k^* (\alpha_{1k}, \alpha_{2k}) = \lim_{t\uparrow\infty} \tau_k^t, \quad \theta_{1k}^* = \alpha_{1k} \tau_k^*, \quad \theta_{2k}^* = \alpha_{2k} \tau_k^*, \quad \text{and}
\end{equation}
\begin{equation}\label{correspondence saenet_k cov phi}
    \phi_k^* = \frac{1}{\delta} \, \E \left[ \eta^\prime \left( B_0 + \tau_k^* Z ; \theta_{1k}^* \, \Omega_k, \theta_{2k}^* \right) \right] .
\end{equation}
    
    \item \textbf{Asymptotic Risk of the SA-Enet($k$) Estimator.} Suppose, the assumptions in Proposition~\ref{ampprop saenet cov} holds true. Following (\ref{amprisk saenet cov}) for a pseudo-Lipschitz function $\psi : \R^2 \mapsto \R$ of order $\nu$, the asymptotic risk of the AMP estimate $\bs{\beta}^{t+1}_k$ at any AMP iteration $t\geq0$ is given by
\begin{equation}\label{amprisk saenet_k cov}
    \lim_{p \rightarrow \infty} \, \frac{1}{p} \sum_{j=1}^p \psi \left( \beta_{kj}^{t+1} , \beta_{0j} \right) \overset{a.s.}{=} \E \left[ \psi \left( \eta \left( B_0 + \tau_k^t Z \, ; \theta_{1k}^t \, \Omega_k, \theta_{2k}^t \right), B_0 \right) \right] ,
\end{equation}
where $Z \sim N (0,1)$ is independent of $B_0$, and $\left\{ \tau_k^t \right\}_{t \geq 0}$ is defined by the state evolution in (\ref{state evolution saenet_k cov}).
Assuming the squared error loss, the asymptotic squared error risk simplifies to $\delta \left( ({\tau^{t+1}_k})^2 - \sigma^2 \right)$, which in the limit equals to $\delta \left( ({\tau^*_k})^2 - \sigma^2 \right)$ as $t \uparrow \infty$. Because of the correspondence, under a covariate-dependent structure, we expect this to provide an accurate approximation of the actual squared error risk of the SA-Enet($k$) estimator. This is empirically confirmed in Section~\ref{subsec: State evolution prediction as a finite sample approximation} through simulation studies.
\end{itemize}
\end{rem}
}

\section{Simulation study}\label{sec: Simulation Study}

In this section, we analyze the performance of the SA-Enet through simulation studies. We assume that the true regression coefficient $\bs{\beta}_0$ is sparse and is generated from a \textit{two-component discrete mixture distribution} discussed below. We consider both group and covariate-dependent structural information that is externally available. Details on the simulation studies are presented in Section~\ref{subsec: Simulation setup}. Section~\ref{subsec: State evolution prediction as a finite sample approximation} validates the asymptotic risk of the SA-Enet predicted from the state evolutions in Section~\ref{sec: Approximate message passing algorithm and state evolution}. In Section~\ref{subsec: Performance comparison}, we compare the performance of the SA-Enet with Lasso \citep{tib96}, A-Lasso \citep{zou06}, Sparse Group Lasso (SGL) \citep{simon2013}, Feature-weighted Elastic-Net (Fwelnet) \citep{tay2020}, Graper \citep{velten19}. We also compare with Structure Adaptive Lasso (SA-Lasso) where we set $\lambda_2 = 0$ in the SA-Enet. We further break the comparison down into two types. Section~\ref{subsubsec: Informative Structural Information} focuses on the case where $\bs{\beta}_0$ is very sparse and the externally available structural information is highly informative. Section~\ref{subsubsec: Robustness with respect to structural information}, on the other hand, investigates the robustness of SA-Enet in terms of collinearity among predictors, the sparsity level, and the informativeness of the available structural information.

\subsection{Simulation setup}\label{subsec: Simulation setup}

We assume that $\bs{\beta}_0$ is sparse with the proportion of signal $\delta_s$, often referred to as the \textit{sparsity level}. Under the group structure, we assume there are $D$ true groups, where in group $d$ the true signal strength equals $\mu_d$. Without loss of generality, group 1 is assumed to be the \textit{null} and we set $\mu_1=0$. The rest of the groups are \textit{non-null} with $\mu_d \neq 0$. Given $\delta_s$, there are $\delta_s p$ signals in $\bs{\beta}_0$ (it is rounded to the nearest integer whenever mentioned). Among them, we assume that a proportion of $p_0$ signals arise in the null group and the rest are divided equally among the $D-1$ non-null groups. So, there are $p_0 \delta_s p$ signals in the null group and $(1-p_0) \delta_s p/(D-1)$ signals in each non-null group where we further assume that the signal strength in group $d$ is $\mu_d$. {\color{black} Heterogeneity in the structure arises either from a difference in the number of signals between groups or from a difference in the signal strengths $\mu_1,\dots,\mu_D$. For brevity, we fix the signal strengths. This lets us control the heterogeneity by appropriately distributing the total signals between the groups. We set each non-null group size to $(1-p_0) p_s/(D-1) p_w$, where $p_w$ denotes the proportion of signals within each non-null group. Given this, the null group is constructed such that the total number of signals in $\bs{\beta}_0$ is $\delta_s p$ and the sum of sizes of all the $D$ groups is $p$. The rest of the $(1-\delta_s)p$ elements in $\bs{\beta}_0$ are set to 0 and within each group, the signals are randomly assigned to the elements. For the simulation we fix $D=3$, $\mu_2=1$, and $\mu_3=2$. In the following sections, we analyze the performance of methods as we vary the sparsity level and the heterogeneity in the structure. To vary the sparsity level, we set $\delta_s$ to 0.1 for a \textit{sparse signal},  0.3 for a \textit{medium signal}, and 0.5 for a \textit{dense signal}. Similarly, we vary heterogeneity in the structure by setting $(p_0,p_w)$ to $(0.05,0.9)$ for \textit{highly informative}, $(0.2,0.8)$ for \textit{moderately informative}, and $(0.3,0.7)$ for \textit{weakly informative}.}

Under the covariate dependent structure, we generate real-valued covariates $( u_1, \cdots , u_p )^\T$ for the $p$ features where $u_j~\overset{iid}{\sim}~\mbox{Unif}(-3,3)$. Given $u_j$'s the features are independently generated as $\beta_{0j} \overset{ind}{\sim} \mbox{Bernoulli} \left( {\left(1+\exp\left(a-b u_j\right)\right)}^{-1} \right)$. {\color{black}Heterogeneity in the structure is reflected by a change in the success probability with increasing $u_j$. Its informativeness increases if there is a large change and the rate of change is high, and vice versa. We tune $(a,b)$ accordingly to achieve a desired level of heterogeneity in the structure and a prespecified sparsity level with similar $\delta_s$ as in the group structure. For the three sparsity levels, the choices $(i)$ in case of a sparse signal are $(7,3)$ for highly informative, $(2.7,0.5)$ for moderately informative, and $(2.6,0.05)$ for weakly informative; $(ii)$ in case of a medium signal are $(3.6,3)$ for highly informative, $(1,0.5)$ for moderately informative, and $(0.9,0.05)$ for weakly informative; and $(iii)$ in case of a dense signal are $(-0.3,3)$ for highly informative, $(0,0.5)$ for moderately informative, and $(0,0.05)$ for weakly informative.}

We consider different types of design matrices $\xmat$ and they are discussed below in respective subsections. Given $\bs{\beta}_0$ and a design matrix $\xmat$, we generate the response $\bs{y}$ from (\ref{model}) with error variance $\sigma^2$. We set $\sigma^2 = 0.2$.

\begin{figure}[h!]
	\centering
    \includegraphics[width=\columnwidth]{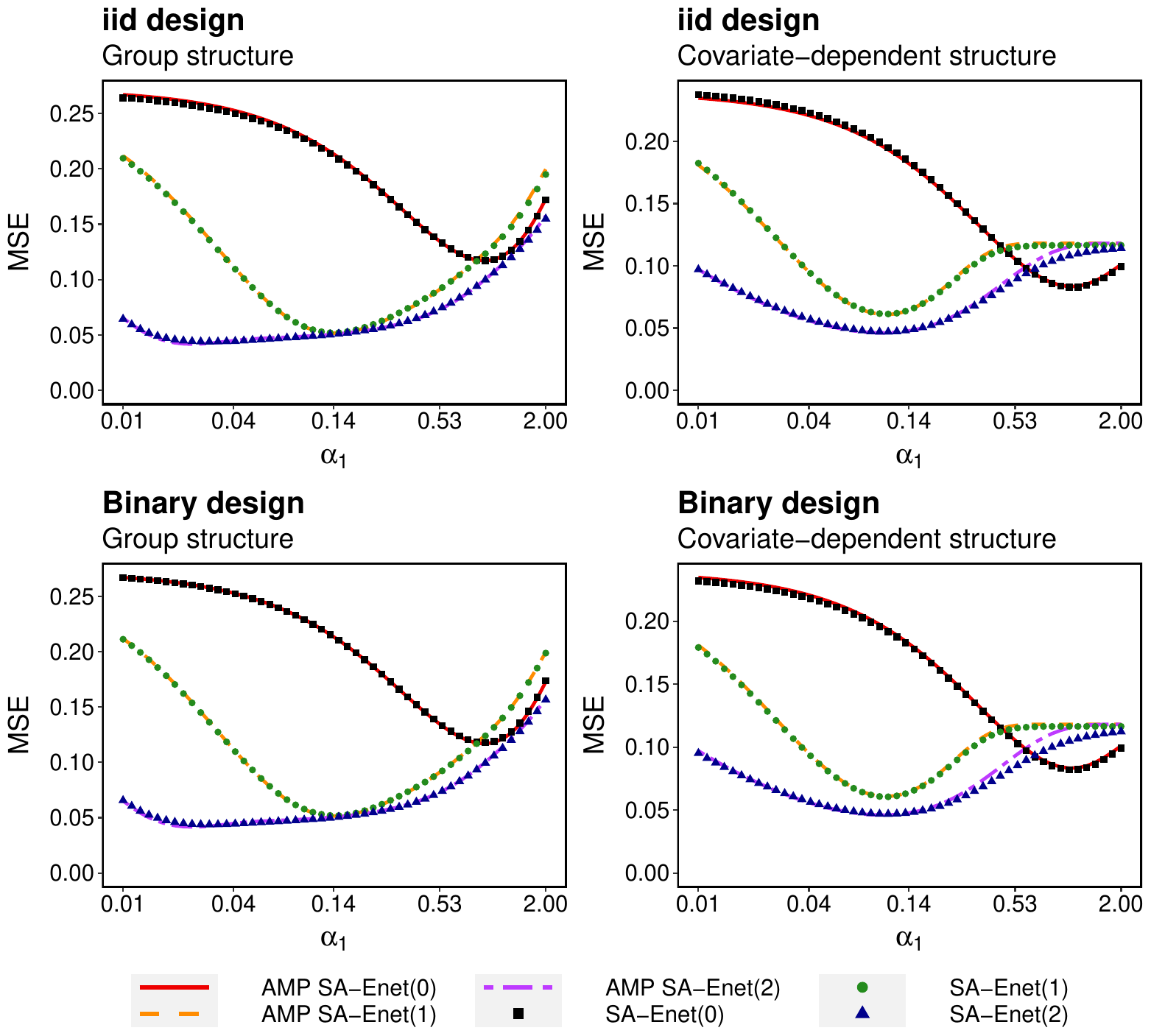}
	\caption{MSE of the SA-Enet estimator as a function of the AMP parameter $\alpha_1$. The number of parameters $p$ is fixed to 500 and $\delta$ equals 0.64. The number of iterations $T$ equals 2 and we compare the MSE of SA-Enet($k$) at $k=0,1,2$. The finite-sample MSEs are overlayed with the risk predicted by the AMP algorithm according to (\ref{amprisk saenet_k group}) and (\ref{amprisk saenet_k cov}).}\label{amp-prediction}
\end{figure}

\subsection{State evolution prediction as a finite sample approximation}\label{subsec: State evolution prediction as a finite sample approximation}

In this section, we validate the asymptotic risk of the SA-Enet estimator predicted by the state evolution in the AMP framework. We compare the predicted risk to the MSE which is the estimated risk based on finite samples from replicated studies. For illustration, we focus on a sparse $\bs{\beta}_0$ with highly informative structural information. We set $p=500$ and $\delta=n/p=0.64$ to determine the sample size. The risk of the SA-Enet estimator based on finite samples is estimated from 100 replications. The AMP predictions are obtained assuming an iid-design. AMP theory predicts that the risk of SA-Enet($k$) is $\delta \left( ({\tau^*_k})^2 - \sigma^2 \right)$ for $k\geq1$. Here ${\tau^*_k}$ is the fixed point of the state evolution SA-Enet($k$) which is defined as (\ref{state evolution saenet group}) and (\ref{state evolution saenet_k group}) for group structure and (\ref{state evolution saenet cov}) and (\ref{state evolution saenet_k cov}) for covariate-dependent structure. To check the validity of the AMP prediction, we consider two choices of the design matrix $\xmat$. In the \textit{iid-design}, each entry of the matrix is independently generated from the normal distribution with mean 0 and variance $1/n$. In \textit{binary design}, each entry of the matrix independently equals $+1/\sqrt{n}$ or $-1/\sqrt{n}$ with equal probability.

For brevity, we fix $\alpha_2 = 0.4$ and compare the risk as a function of $\alpha_1$. To check the validity for multiple iterations, we consider Algorithm~\ref{algorithm saenet} with $T=2$. For $k>1$, the adaptive weights $(\hat{\bs{w}}_k)$ in SA-Enet($k$) is estimated using the SA-Enet($k-1$) estimate corresponding to the minimum risk. Figure~\ref{amp-prediction} presents the comparison for each pair of the design matrices and the types of structures. It suggests that the risks predicted by the state evolutions under both structures match spectacularly with the finite sample risk with $p$ as low as $500$. This result numerically justifies the AMP theory. Although the AMP framework assumes the iid-design, the theoretical prediction from it seems to match very well with the finite sample risk estimated under the binary design. The curves also show that the minimum MSE decreases with the increase in iterations. 
This result shows the potential gain of using the SA-Enet with multiple iterations over the Enet (red curves).

{\color{black}
\subsection{Performance comparison}\label{subsec: Performance comparison}

In this section, we compare the finite sample performance of the SA-Enet with some of the existing methods across a wide range of simulated scenarios. To evaluate the effect of multicollinearity among predictors, we consider three design matrices: the iid-design, AR(1) design with $\rho=0.5$, and equicorrelated design with $\rho=0.5$. For each design matrix, we analyze the robustness of each method with respect to the sparsity level and the structural information. For comparing different methods, we use MSE for checking the quality in estimating the signal strength and the Matthews Correlation Coefficient (MCC) for checking the model selection performance of detecting signals. MCC quantifies the accuracy of classifying true signals. By definition, the MCC is a correlation coefficient between the observed and predicted binary classifications and so it takes values between --1 and +1. A coefficient of +1 indicates a perfect classification, 0 indicates no better than a random classification, and --1 indicates a total disagreement between prediction and observation. When implementing the Graper, we use the version where the posterior mean is used as the estimate. Thus we do not include Graper in the MCC comparison. To set the sample size, we define $\delta_e = \delta_s p/n$ as the ratio of the total number of signals to the sample size. The larger the ratio the higher the difficulty in the estimation problem and vice versa. For the simulation, we vary $\delta_e$ as 0.5, 0.75, and 1, set $p = 300$, and summarize the performance over 100 replications. Five iterations are performed for the SA-Lasso and SA-Enet as per the default suggestion in Section~\ref{subsubsec: Estimator and algorithm}, and the results are presented for the first and fifth iterations. Similar to SA-Enet($T$), SA-Lasso($T$) denotes the SA-Lasso estimator at iteration $T$. Following \cite{zhou09} we consider $\hat{w}_j = \abs{\hat{\beta}^\tL_j}^{-\gamma}$ for $\gamma>0$ as adaptive weights in the A-Lasso, where $\hat{\bs{\beta}}^{\tL}$ is the Lasso estimate. For implementation, $(\lambda,\gamma)$ are considered as tuning parameters of the A-Lasso. When implementing SA-Enet and SA-Lasso using Algorithm~\ref{algorithm saenet}, $(\lambda_{10},\lambda_{20})$ at iteration 0 and $(\lambda_{1k},\lambda_{2k},\gamma_k)$ at iteration $k$ are also considered as tuning parameters. We use 10-fold cross-validations to find optimal values of the tuning parameters. We also use the true group structure and the external covariates $( u_1, \dots , u_p )^\T$, which were used to generate the true $\bs{\beta}_0$, when implementing SGL, Fwelnet, Graper, SA-Lasso, and SA-Enet. For each structure, the same true $\bs{\beta}_0$ is used to replicate data for each design matrix.

\begin{figure}[h]
	\centering
    \includegraphics[width=\columnwidth]{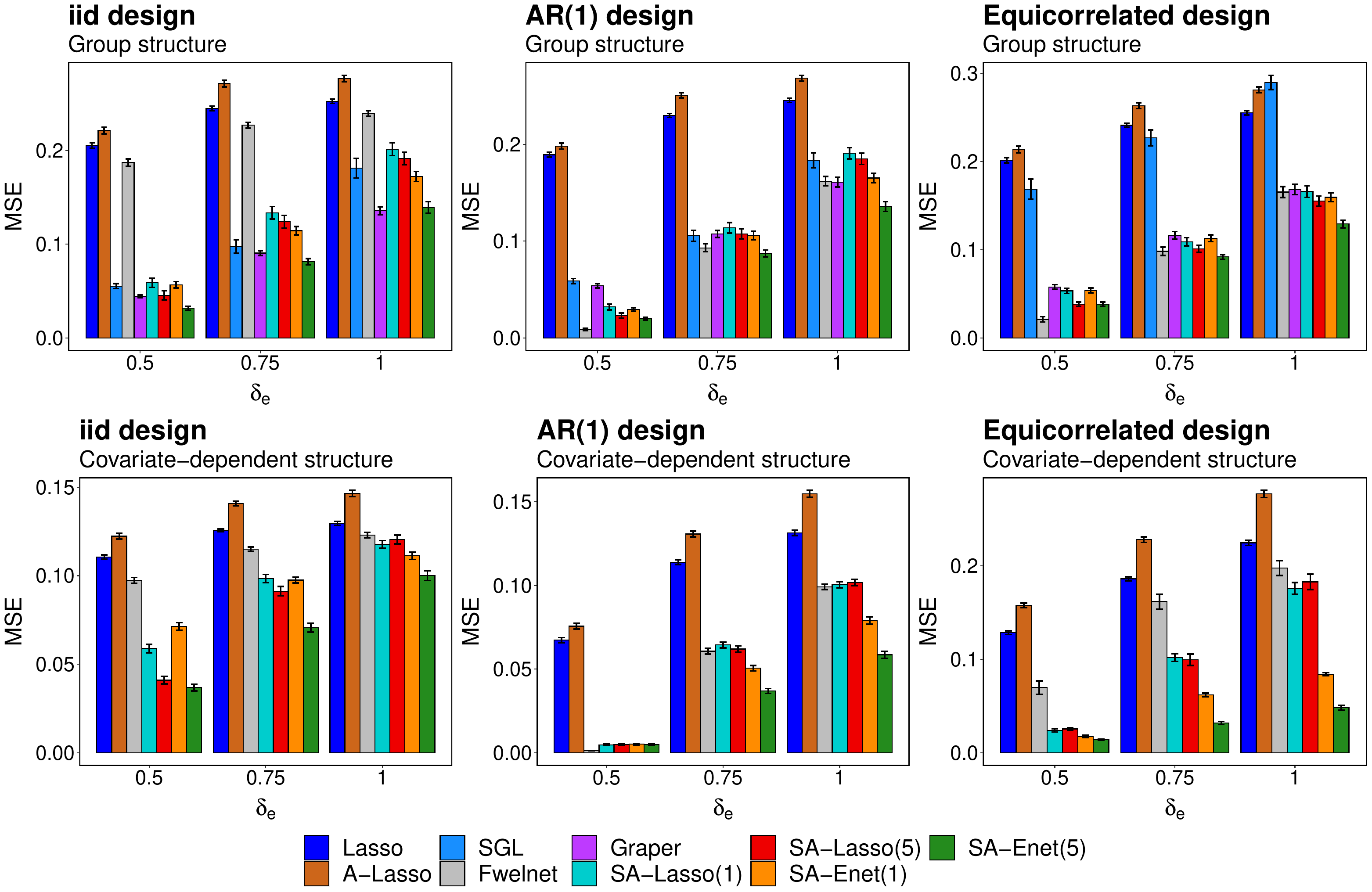}
	\caption{MSE ($\pm1$ standard error) of regression coefficient estimates from different methods when the structural information is highly informative and the true signal is sparse.}\label{fig: mse-hhet-lsp}
\end{figure}

\subsubsection{Informative structural information}\label{subsubsec: Informative Structural Information}

In this section, we focus on the scenario where $\bs{\beta}_0$ is sparse and the structural information is highly informative. Figure~\ref{fig: mse-hhet-lsp} compares MSEs of different methods in this setup. There are two key findings: (1) change in performance as the correlation among predictors increases, and (2) change in performance as $\delta_e$ increases. Figure~\ref{fig: mse-hhet-lsp} shows that in an iid design when $\delta_e = 0.5$, the MSEs of the SGL, Graper SA-Lasso(1), SA-Lasso(5), SA-Enet(1) are similar to each other. Lasso, A-Lasso, and Fwelnet have larger MSEs compared to them. SA-Enet(5) has the lowest MSE among them. As $\delta_e$  increases and the estimation gets more difficult, MSE increases for all methods. SA-Enet(5) and Graper perform similarly to each other and have the lowest MSEs. As we move from iid design to equicorrelated design through AR(1) design, the correlation among the predictors increases. In this case for $\delta_e = 0.5$, Fwelnet performs the best and is slightly better than SA-Lasso(5) and SA-Enet(5). As $\delta_e$ increases to 0.75 and 1, SA-Enet(5) performs better than the others. It also shows that SA-Enet(5) offers a significant improvement over SGL.

\begin{figure}[h!]
	\centering
    \includegraphics[width=\columnwidth]{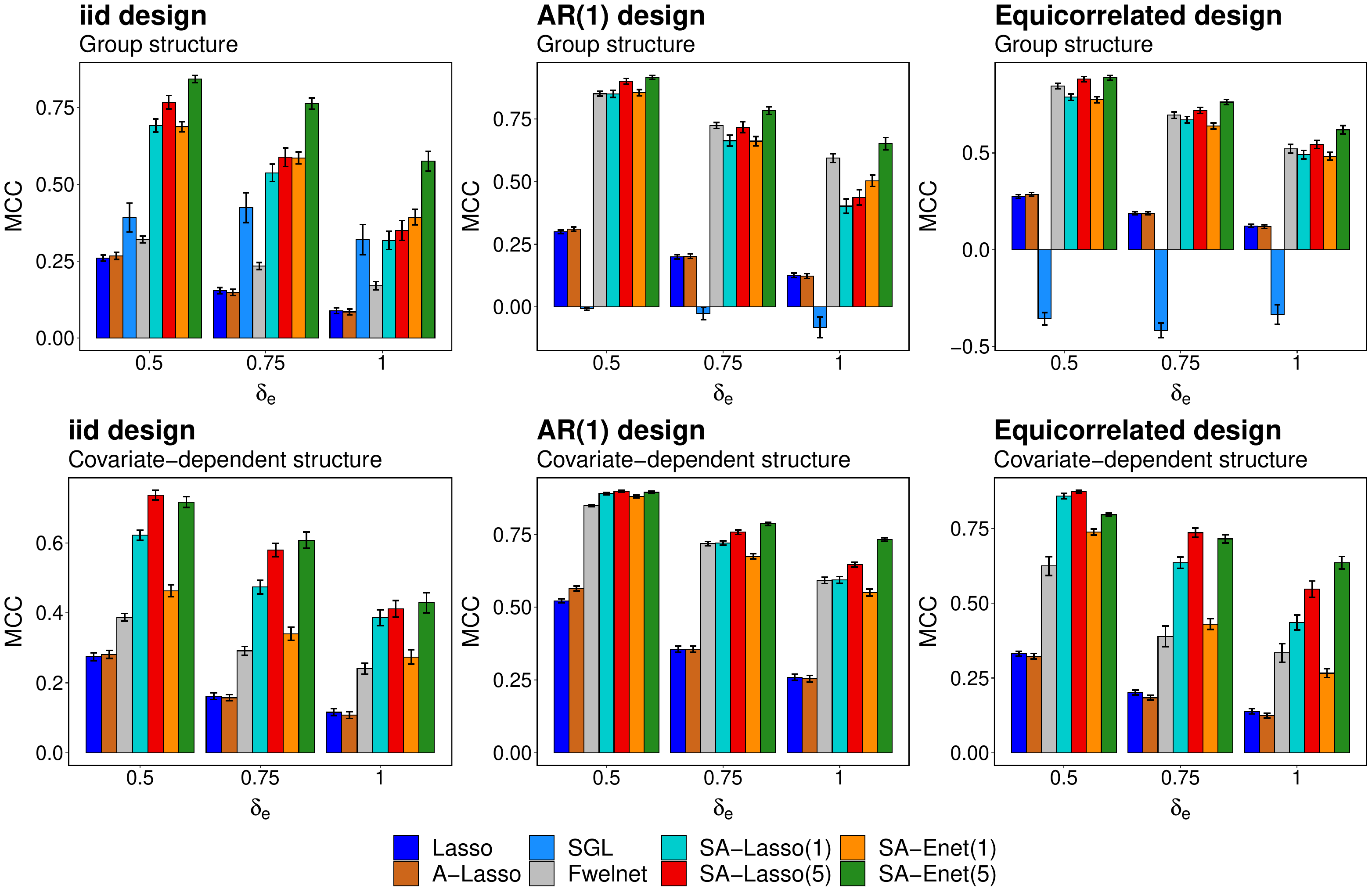}
	\caption{MCC ($\pm1$ standard error) of regression coefficient estimates from different methods when the structural information is highly informative and the true signal is sparse.}\label{fig: mcc-hhet-lsp}
\end{figure}
We also compare their model selection performance and they are presented in Figure~\ref{fig: mcc-hhet-lsp}. Graper R-package provides non-sparse estimates and hence is not included in this comparison. The figure shows that SA-Enet chooses the true signals and noises more often or at least as well as others for all $\delta_e$ that we consider here.

The performance improvements in terms of MSE and MCC by SA-Enet is due to the fact that it imposes the structures through a soft constraint, unlike SGL which has a hard constraint and suffers from it. The MSE of Graper increases with an increase in the correlation because it assumes the \textit{mean-field} approximation on the variational family. The difference in performance between Fwelnet and SA-Enet can be attributed to the fact the penalty on the regression coefficients is in accordance with that proposed in \cite{zou2009}, which is known to have improved performance when the number of predictors diverges with sample size. Lasso and A-Lasso have large MSEs because they are agnostic to structural information.

\subsubsection{Robustness with respect to structural information}\label{subsubsec: Robustness with respect to structural information}

In many real-life applications, it is possible to identify external groups or other structural information on the predictors. SA-Enet provides a framework to make use of such auxiliary information in estimating $\bs{\beta}$ for improved estimation, model selection, and prediction.
\begin{figure}[h!]
	\centering
    \includegraphics[width=\columnwidth]{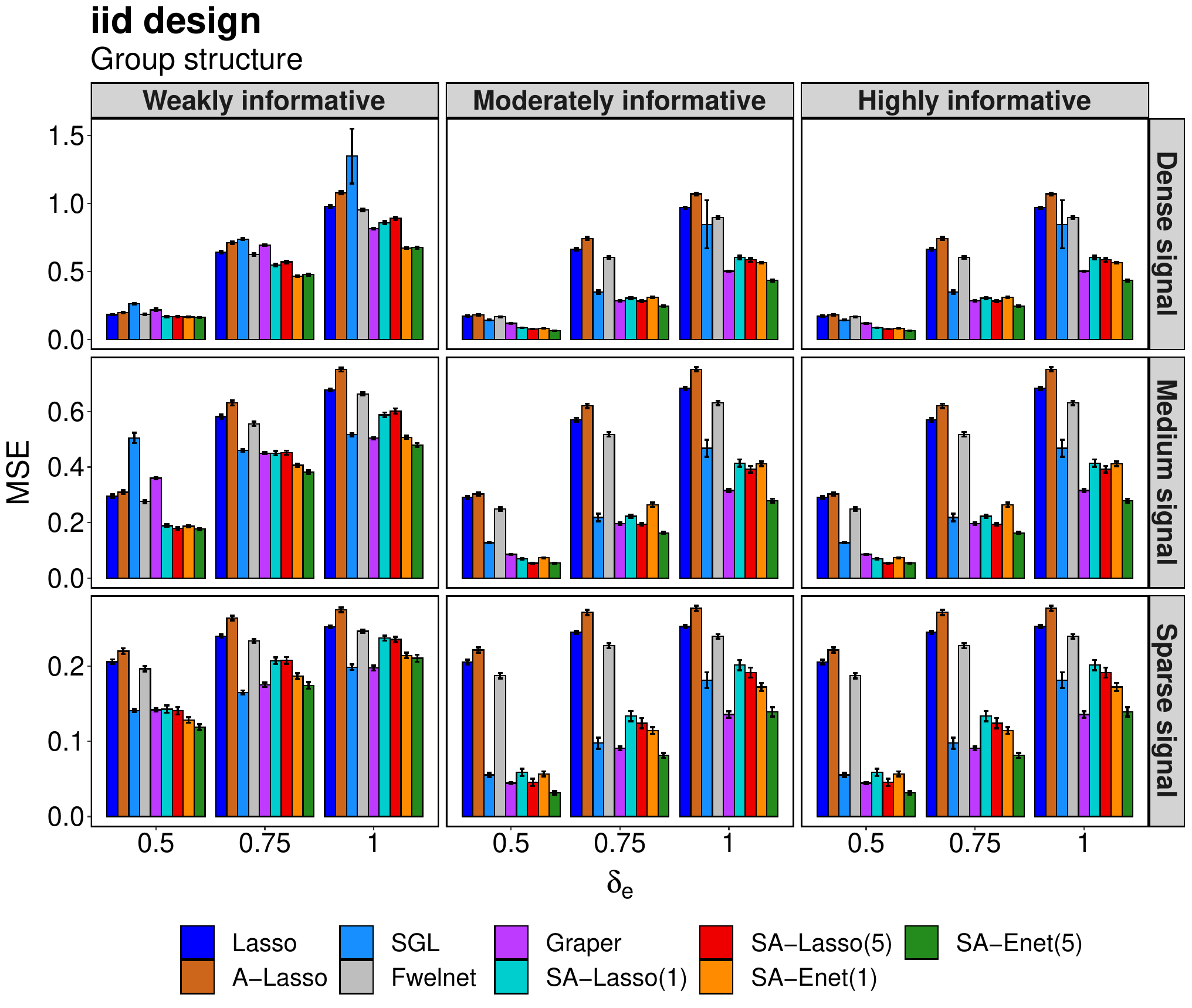}
	\caption{Comparison of the MSEs ($\pm1$ standard deviation) from different methods for a varying proportion of signals and the heterogeneity of a group structure in an iid-design.}\label{fig: iid-mse-grp}
\end{figure}
\begin{figure}[h!]
	\centering
    \includegraphics[width=\columnwidth]{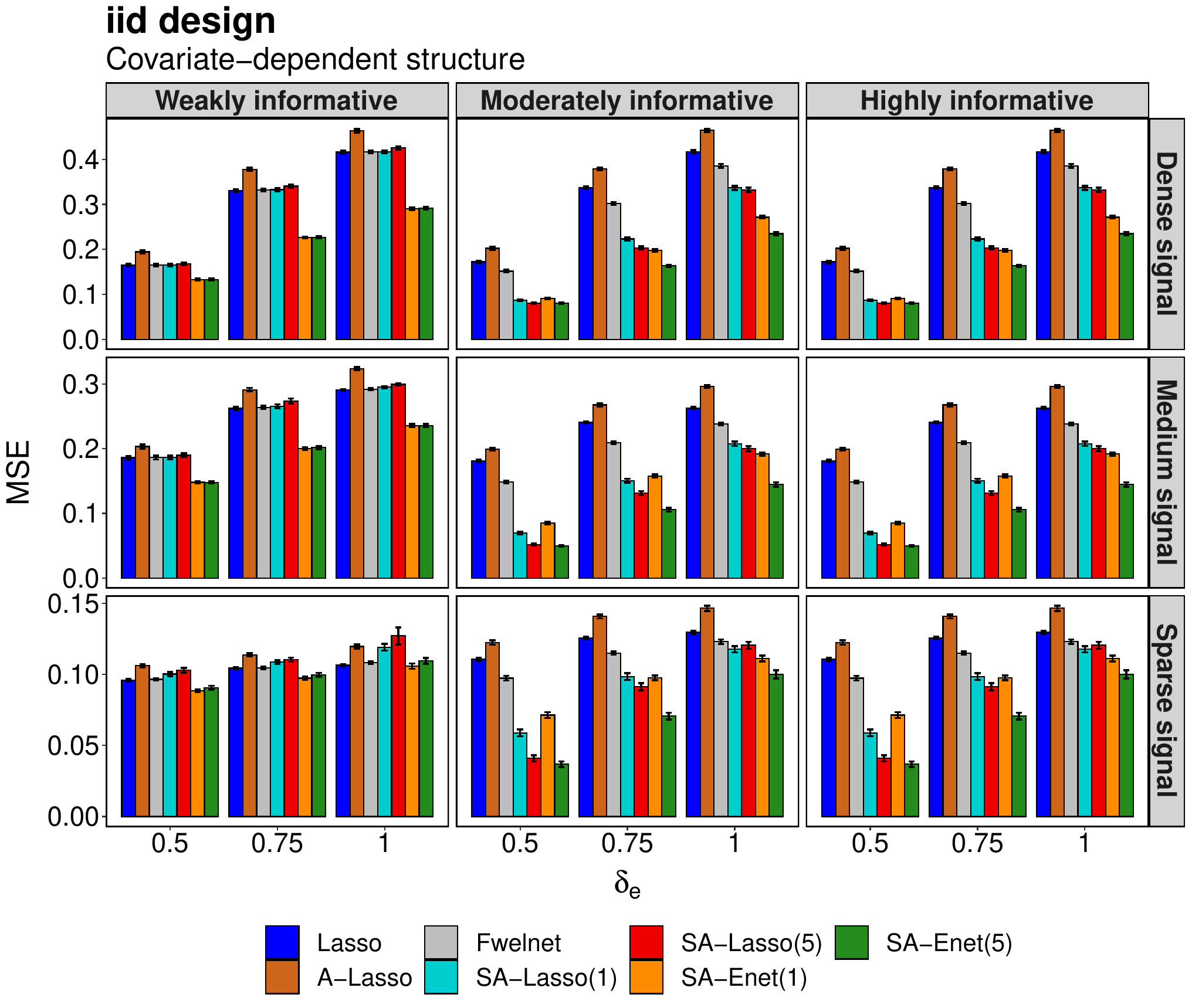}
	\caption{Comparison of the MSEs from different methods for a varying proportion of signals and the heterogeneity of a covariate-dependent structure in an iid-design.}\label{fig: iid-mse-cov}
\end{figure}
\begin{figure}[h!]
	\centering
    \includegraphics[width=\columnwidth]{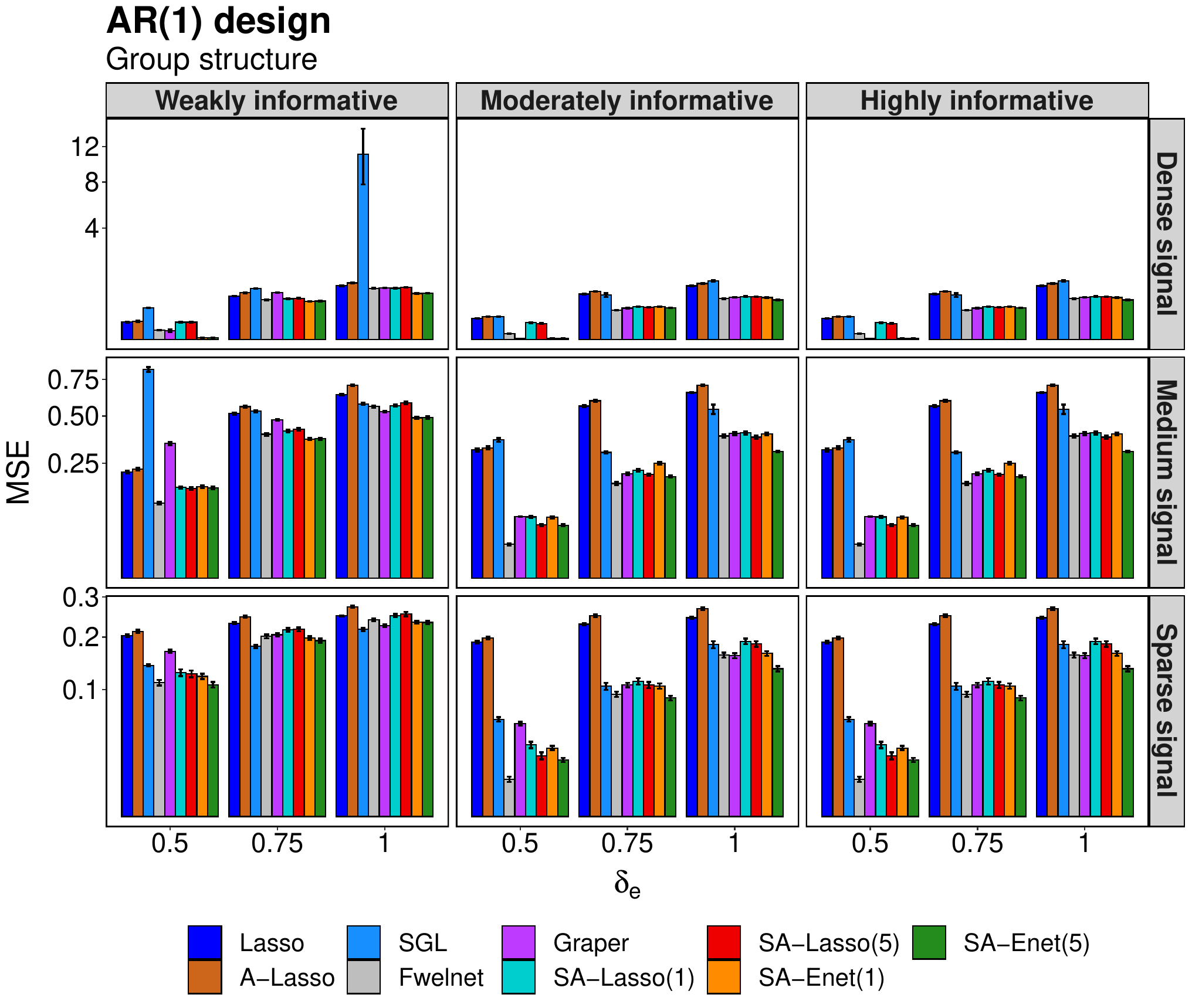}
	\caption{Comparison of the MSEs from different methods for a varying proportions of signals and the heterogeneity of a group structure in an AR(1) design.}\label{fig: ar-mse-grp}
\end{figure}
\begin{figure}[h!]
	\centering
    \includegraphics[width=\columnwidth]{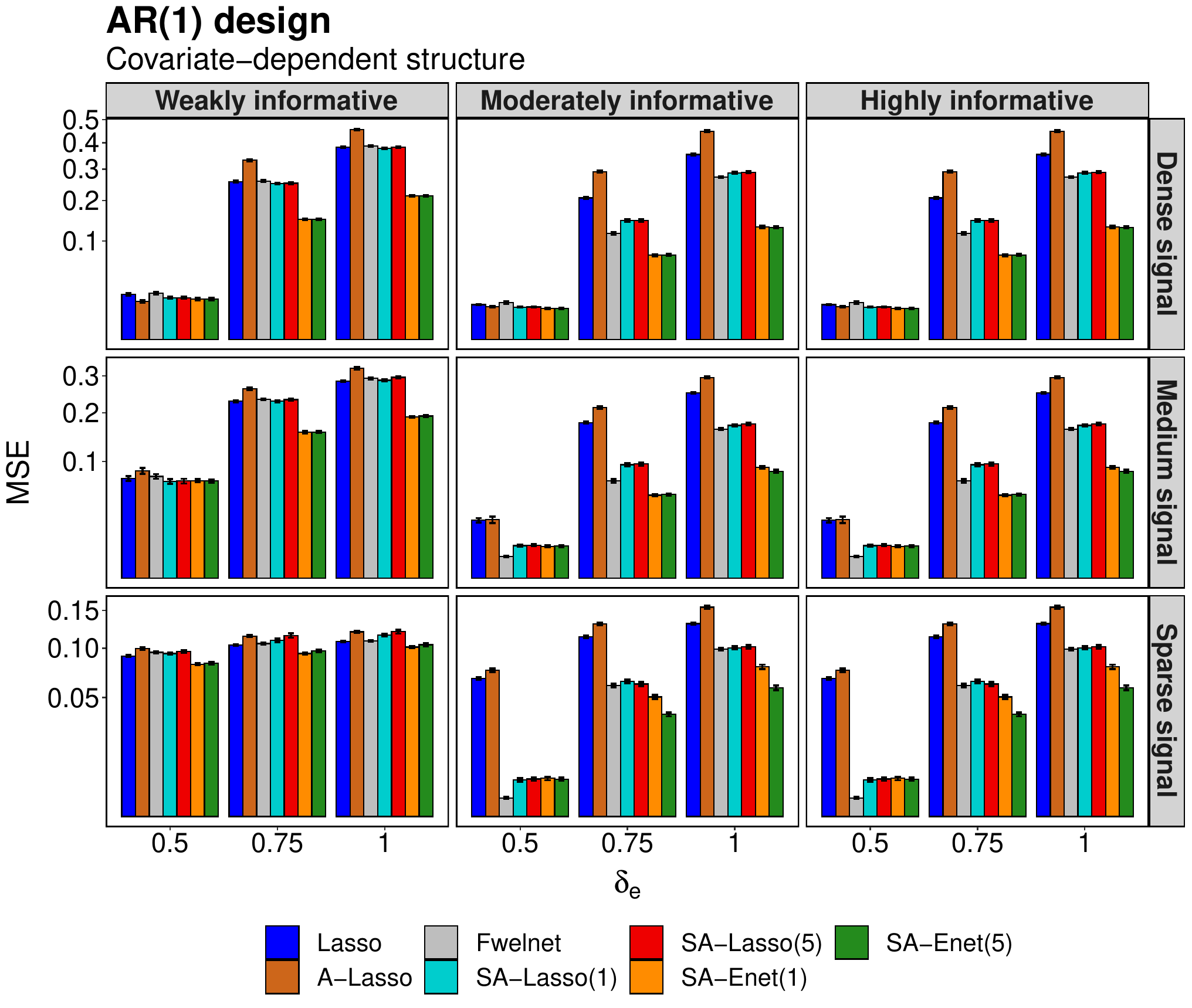}
	\caption{Comparison of the MSEs from different methods for a varying proportions of signals and the heterogeneity of a covariate-dependent structure in an AR(1) design.}\label{fig: ar-mse-cov}
\end{figure}
\begin{figure}[h!]
	\centering
    \includegraphics[width=\columnwidth]{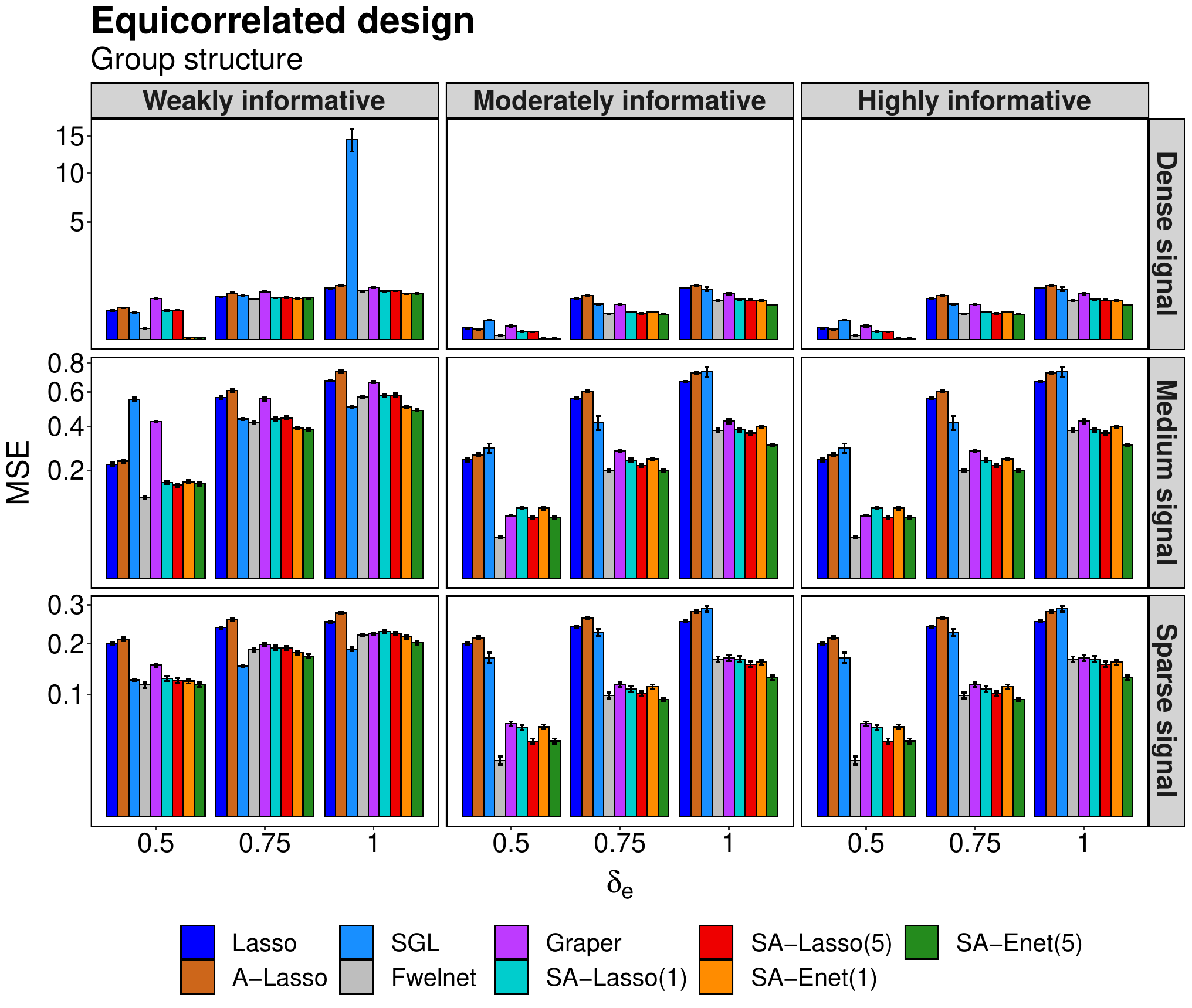}
	\caption{Comparison of the MSEs from different methods for a varying proportions of signals and the heterogeneity of a group structure in an equicorrelated design.}\label{fig: eq-mse-grp}
\end{figure}
\begin{figure}[h!]
	\centering
    \includegraphics[width=\columnwidth]{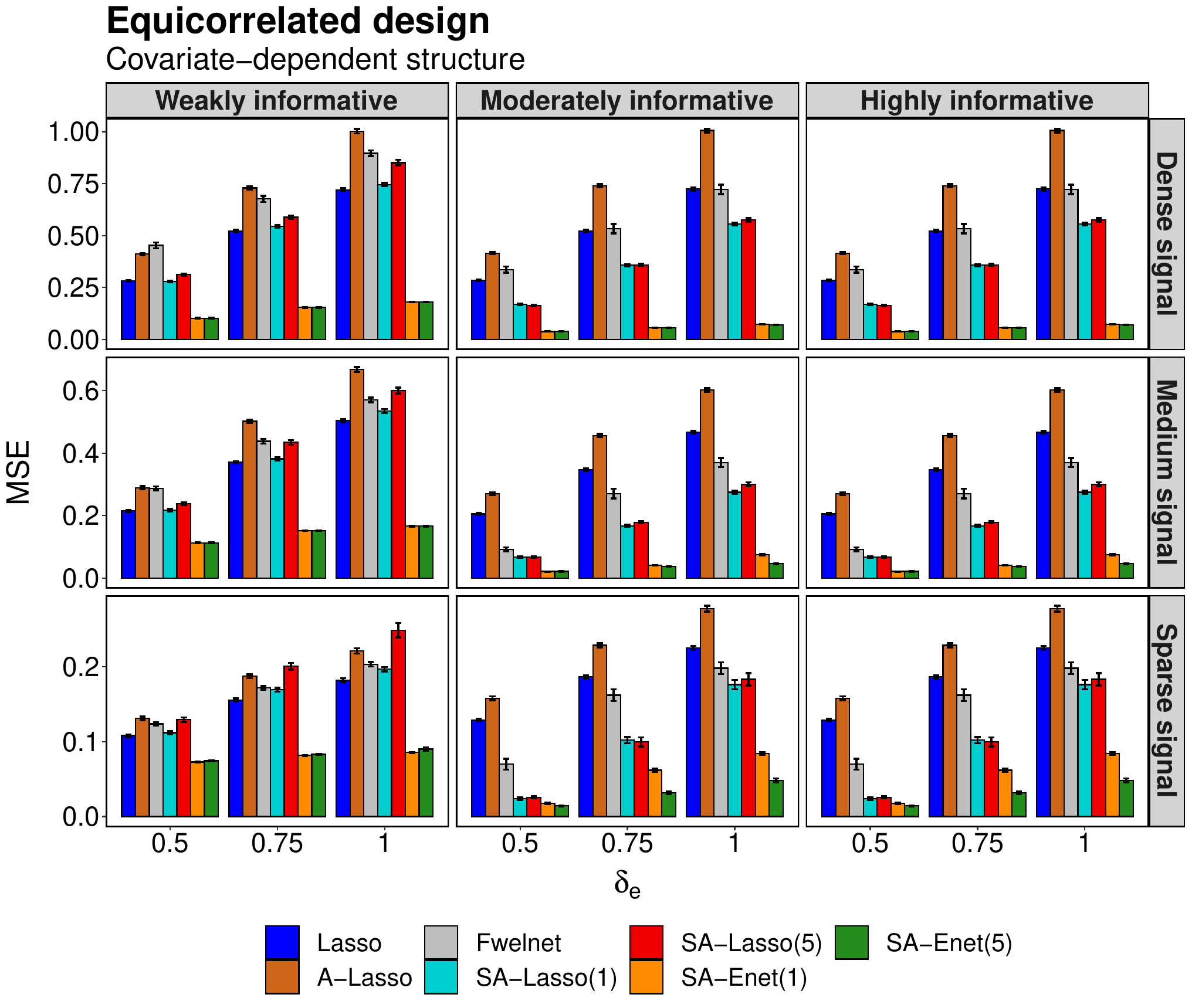}
	\caption{Comparison of the MSEs from different methods for a varying proportions of signals and the heterogeneity of a covariate-dependent structure in an equicorrelated design.}\label{fig: eq-mse-cov}
\end{figure}
Figures~\ref{fig: mse-hhet-lsp} and \ref{fig: mcc-hhet-lsp} show that SA-Enet provides a significant improvement in performance over the other methods when the structure is informative. But in real-life applications, the strength of any auxiliary information is often unknown. In this section, we analyze the robustness of different methods with respect to heterogeneity in structural information and the true proportion of signals. We consider the same simulation setup as above and vary the proportion of signals and structural information as described in Section~\ref{subsec: Simulation setup}.

Figures~\ref{fig: iid-mse-grp}--\ref{fig: eq-mse-cov} compare MSEs from different methods for different design matrices for group and covariate-dependent structures for a varied range of $\delta_e$, proportion of signals, and heterogeneity in structural information. We also compare model selection performance by comparing MCC from different methods in the same simulation settings and they are presented in Figures~\ref{iid-mcc}--\ref{eq-mcc}.

The findings from the figures extend the observations from Section~\ref{subsubsec: Informative Structural Information} and they can be summarized as follows. When the signal is sparse and the structure is weakly informative, the SA-Enet(5) performs as well as the other methods for all $\delta_e$, except in equicorrelated design where SA-Enet performs substantially better than the other methods for covariate-dependent structure. As heterogeneity in the structure increases, SA-Enet outperforms other methods in iid design for all $\delta_e$. In AR(1) and equicorrelated designs, Fwelnet performs better than others when $\delta_e=0.5$. But as $\delta_e$ increases to 0.75, SA-Enet performs as well as Fwelnet and then outperforms it for $\delta_e=1$. We observe similar findings as the proportion of signals increases. This highlights that the SA-Enet is able to take advantage of the structure whenever it can, and thus provides a significant improvement in MSE. In most other cases, it performs as well as the other methods. For informative group structure, the performance of SGL depends on the correlation among the predictors. Under the iid-design, for a varied proportion of signals and $\delta_e$, the SGL performs similarly to Graper or SA-Lasso. Under AR(1) and equicorrelated designs, it starts to perform poorly and performs as well as the Lasso and the A-Lasso. This is because SGL uses the group structure by imposing a hard constraint, unlike the SA-Enet which imposes a soft constraint. In the same setting, Graper performs similarly to SA-Enet under iid design. But for AR(1) and equicorrelated designs, Graper's performance worsens and it performs as well as Fwelnet. This is because Graper takes a variational Bayes approach which assumes the mean-field approximation on the variational family. This fails to account for the correlation among the predictors. We also observe there is a difference in performance between Fwelnet and SA-Enet. This is because the adaptive Elastic-Net penalties used by the methods are different. In this regard, SA-Enet uses the penalty proposed in \cite{zou2009}, which is known to have improved performance when the number of predictors diverges with the sample size. In most cases, Lasso and A-Lasso have large MSEs because they do not take any structural information into account.

\begin{figure}[h!]
	\centering
    \includegraphics[width=\linewidth]{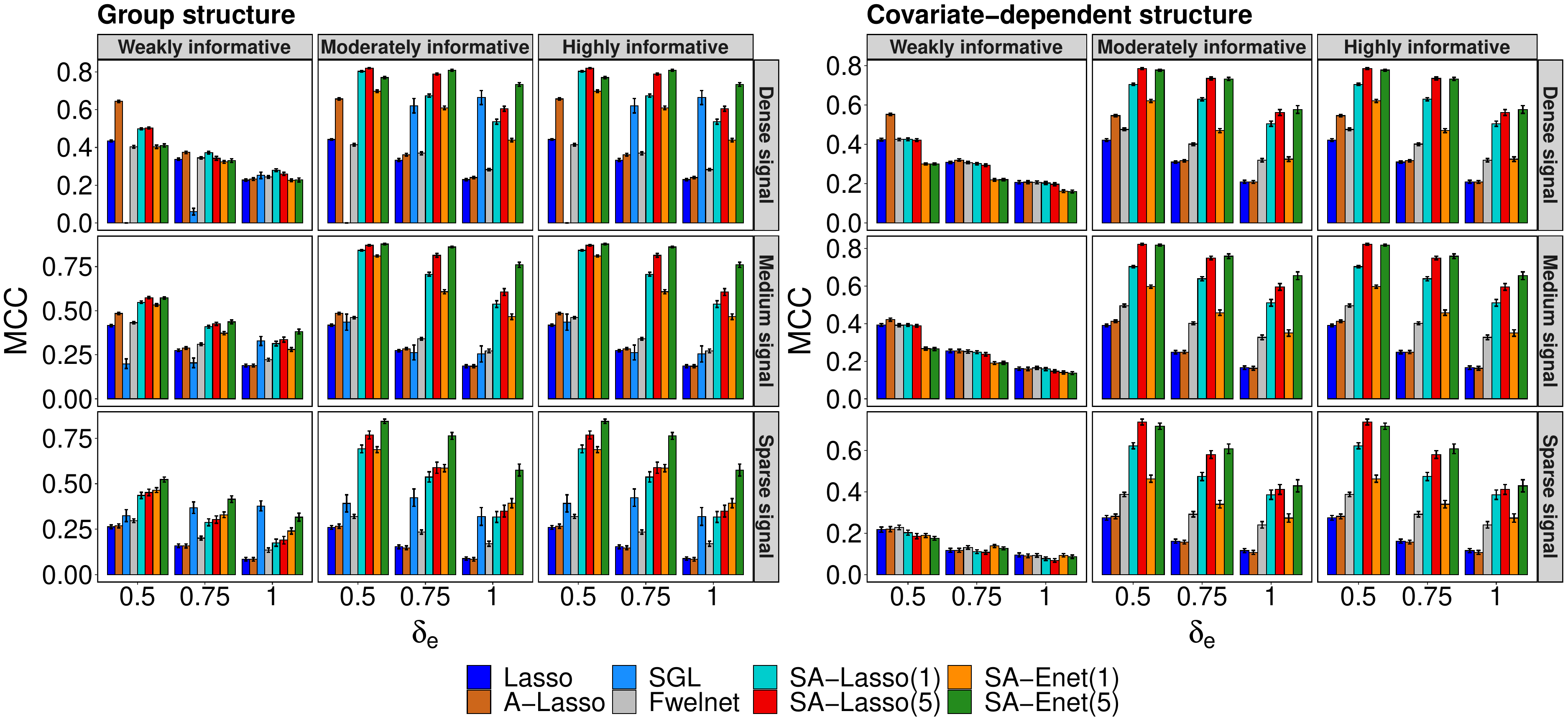}
	\caption{Comparison of the MCCs from different methods for a varying proportions of signals and the heterogeneity of group and covariate-dependent structure in an iid design.}\label{iid-mcc}
\end{figure}
\begin{figure}[h!]
	\centering
    \includegraphics[width=\linewidth]{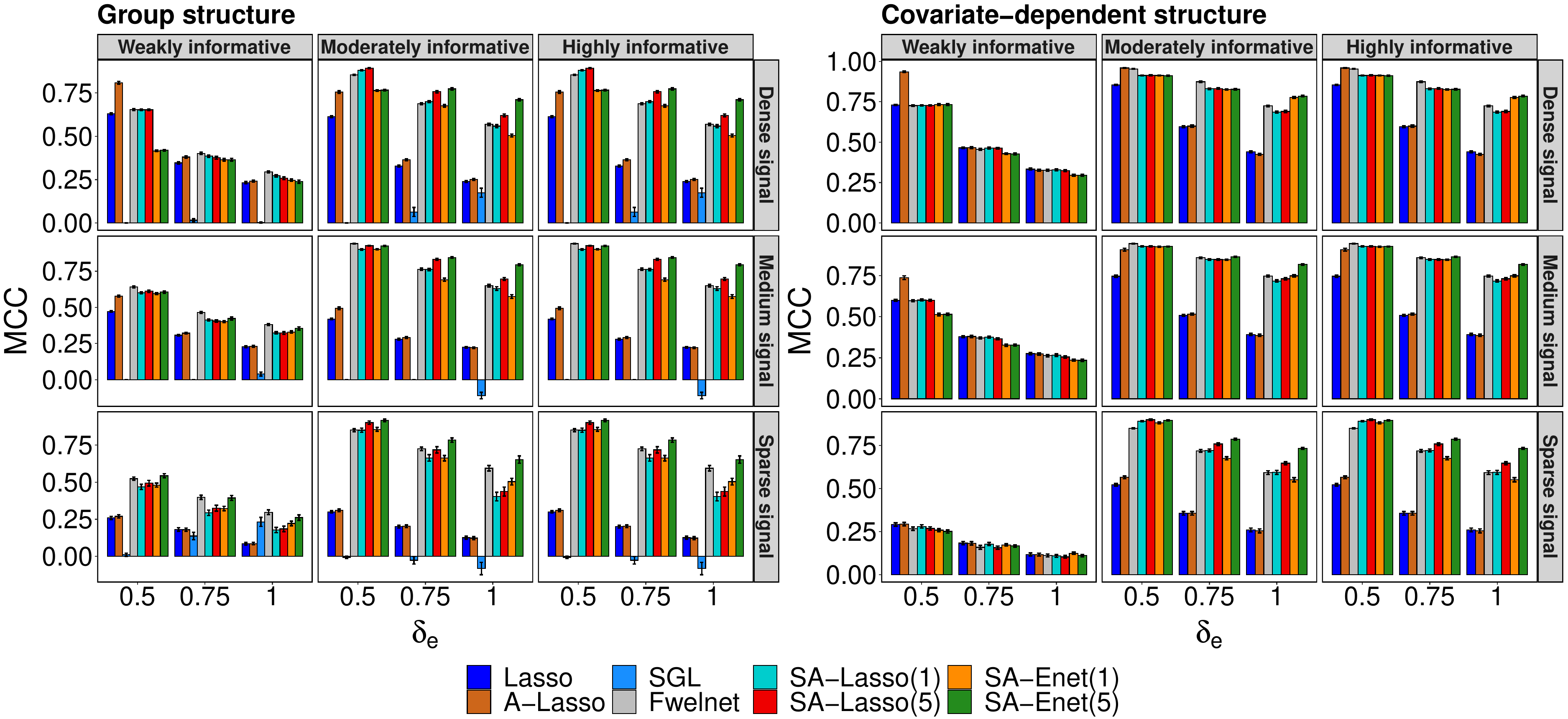}
	\caption{Comparison of the MCCs from different methods for a varying proportions of signals and the heterogeneity of group and covariate-dependent structure in an AR(1) design.}\label{ar-mcc}
\end{figure}
\begin{figure}[h!]
	\centering
    \includegraphics[width=\linewidth]{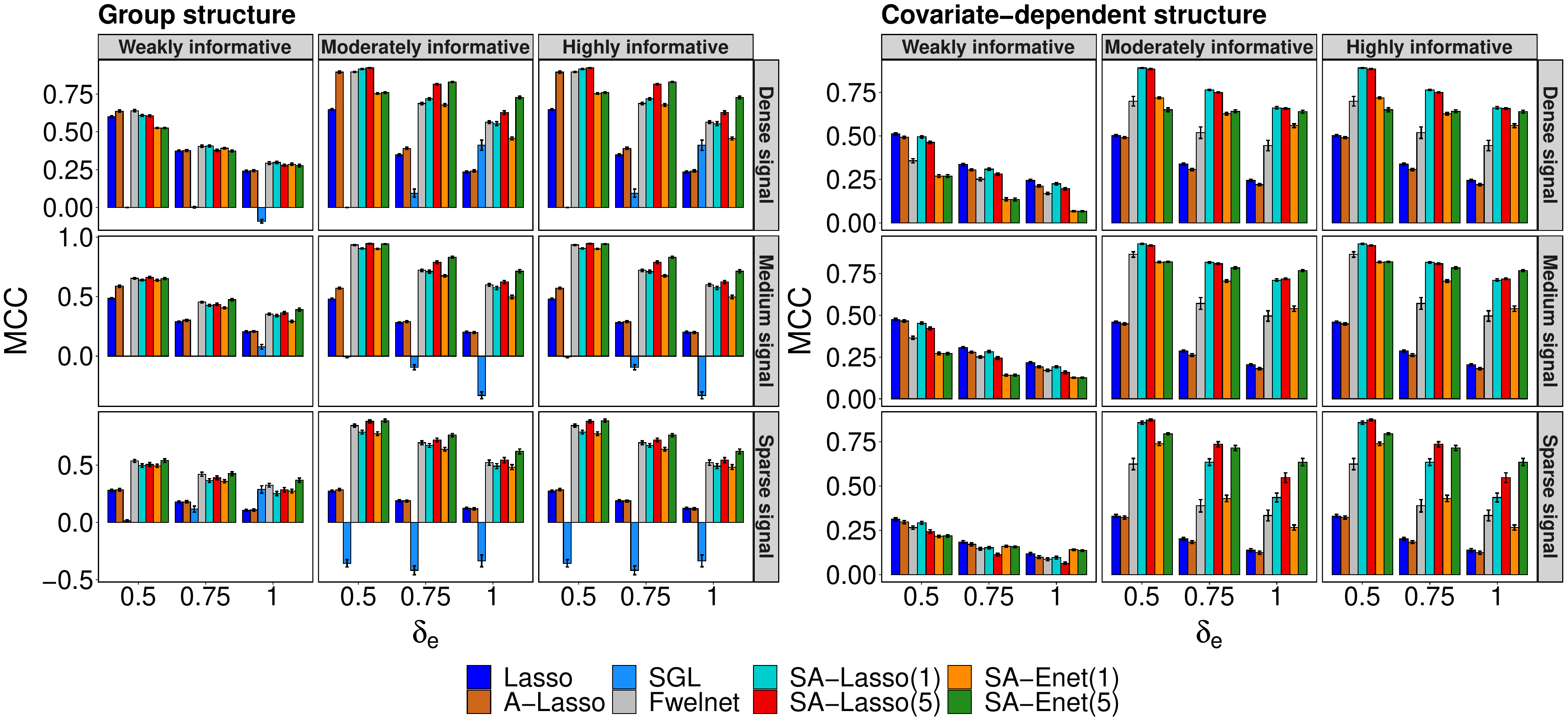}
	\caption{Comparison of the MCCs from different methods for a varying proportions of signals and the heterogeneity of group and covariate-dependent structure in an equicorrelated design.}\label{eq-mcc}
\end{figure}
}

\section{Drug response prediction in leukemia samples}\label{sec: Drug response prediction in leukemia samples}

To illustrate the efficacy of SA-Enet on real data, we apply the method in drug response prediction based on several molecular predictors. Nowadays, a large number of molecular features from different biological layers can be jointly measured using high-throughput technologies \citep{hasin17,ritchie15}.
The CLL data that we consider here consists of several omic measurements from 121 patients and are obtained from the Bioconductor package \textit{MOFAdata 1.0.0} \citep{mofadata, dietrich18}.
There are 3 omic types: $(a)$ expression values for the 5000 most variable genes (mRNA), $(b)$ methylation M-values for the 4248 most variable CpG sites (Methylation), and $(c)$ viability values in response to 310 different drugs and concentrations (Drugs) (5 different doses for each of 61 drugs).
{\color{black}Following \cite{velten19} we consider the problem of viability value prediction for each dose of the Ibrutinib drug. Since there are 5 doses of each drug, we carry out five separate regressions for each dose of Ibrutinib. So in each regression, the response $\bs{y}$ is the viability value of Ibrutinib for a particular dose. Everything other than Ibrutinib's five responses is considered a predictor in the design matrix $\xmat$. This leads to $n=121$ observations and $p=9553$ predictors consisting of 5000 mRNA expression values, 4248 Methylation M-values, and 305 viability values from other drugs.} For SGL, Fwelnet, Graper, SA-Lasso, and SA-Enet we use a 3-group structure representing the omic types as structural information. 
For each regression, we randomly partition the data into 81 training and 40 test samples, fit the model on the 81 training observations, and then compute the root mean squared prediction error (RMSPE) on the 40 test observations to quantify an overall quality of predictions.

\begin{figure}[h!]
	\centering
	\includegraphics[width =\columnwidth]{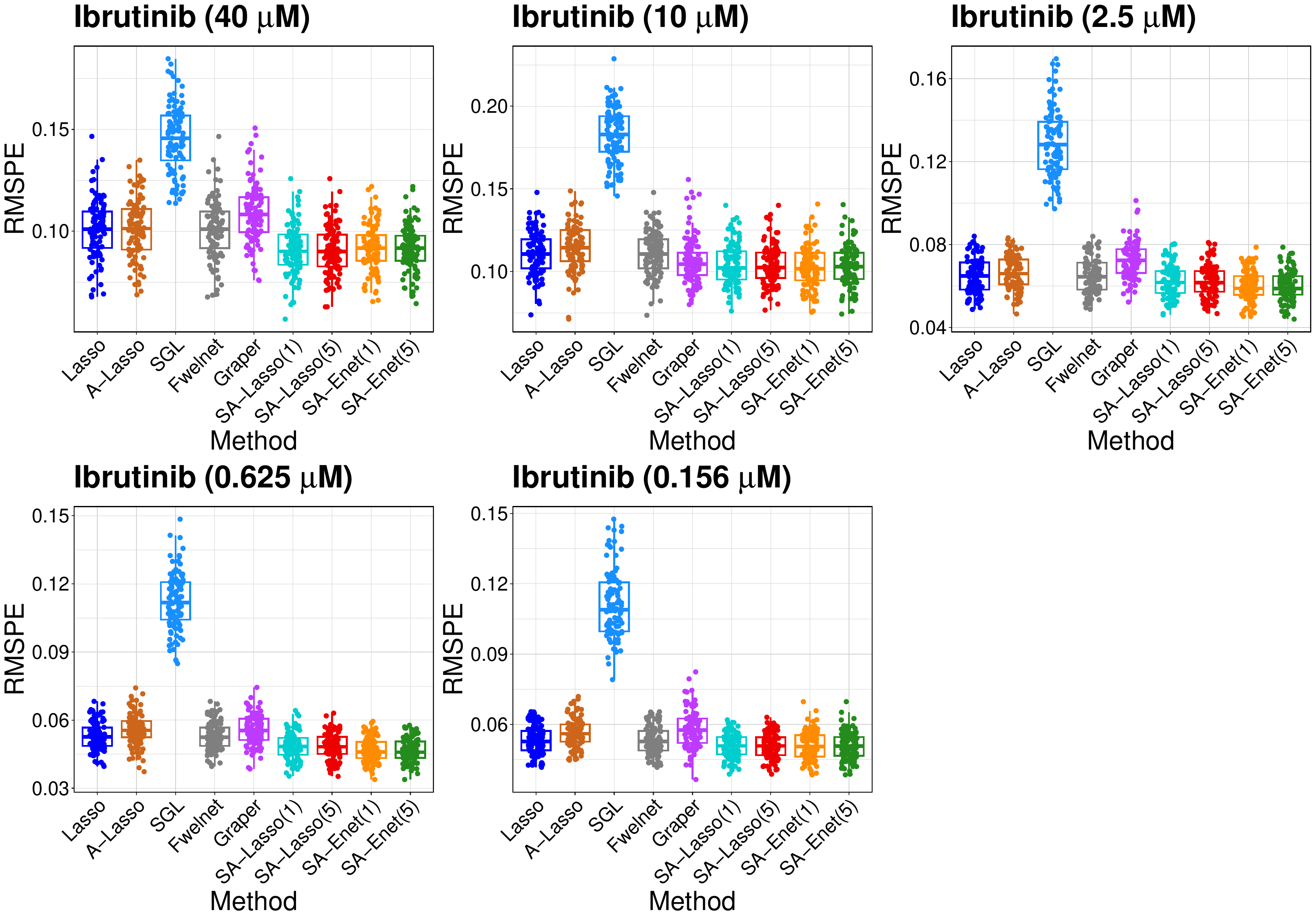}
	\caption{Boxplots of RMSPEs for different methods over 100 random training and test set partitions of the CLL data.}\label{CLL-boxplot}
\end{figure}

Figure~\ref{CLL-boxplot} presents boxplots of RMSPEs for different methods for 100 random partitions. Aggregated over the five regressions, SA-Enet(5) reduces average RMSPE by 7--22\% over Lasso, 18--29\% over A-Lasso, 60--82\% over SGL, 7--21\% over fwelnet, and 6--30\% over Lasso. When compared to SA-Lasso, the SA-Enet(5) reduces RMSPE by 7\% for doses $2.5 \mu M$ and $0.625 \mu M$ and performs similarly in others.
\begin{figure}[h!]
    \centering
    \includegraphics[width=\columnwidth]{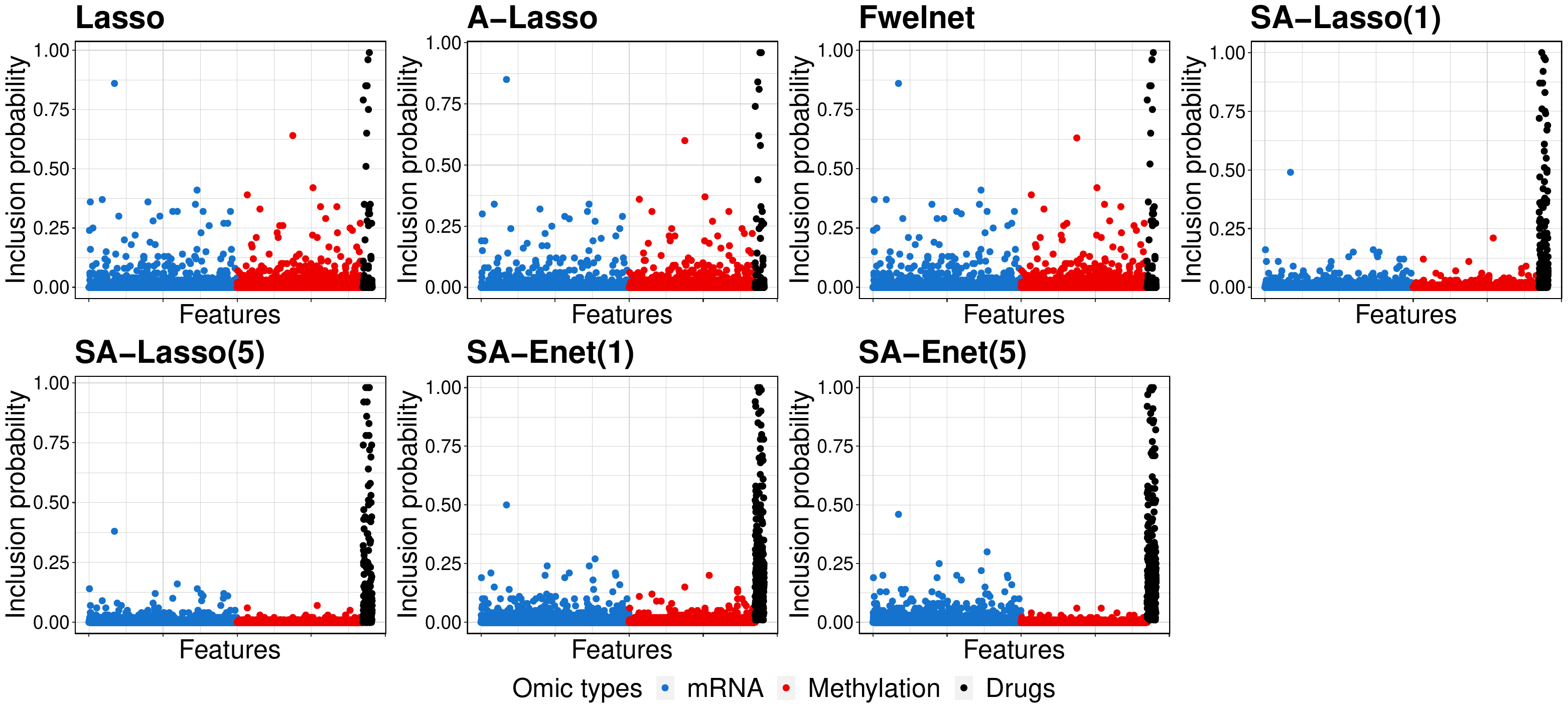}
    \caption{Feature inclusion probability of different methods in Ibrutinib $(40 \, \mu M)$ response prediction. In each panel, the vertical axis shows the proportion of times the method selects each feature in the model. This reflects the consistency of feature selection by the methods. The horizontal axis corresponds to the features color-coded according to the omic types.}
    \label{CLL-feature-inclusion}
\end{figure}
We also compare the feature selection performance of the methods. We calculate the proportion of times out of 100 random partitions each feature was included in the model. Figure~\ref{CLL-feature-inclusion} compares the inclusion probabilities for the Ibrutinib response prediction for $40 \, \mu M$. {\color{black}SGL never selects any feature in the model. We suspect this is because SGL uses the group structure by imposing a hard constraint on the features in each group. The constraint forces the coefficients in the same group to be simultaneously large or small. This type of constraint is also not expected to be robust with respect to heterogeneity in the structure. Contrary to this constraint, the group structure in this example is relatively soft. It only represents the omic types and it is also unknown whether it represents the true underlying structure among the features. This explains its poor predictive performance and hence is removed from the comparison of inclusion probability in Figure~\ref{CLL-feature-inclusion}. As for the other methods, out of the three omic types all methods select other drug responses more frequently in the model. This is a desirable property for a method since ideally, the goal is to select important features in the model more frequently.} We also find that, on average, SA-Enet(5) has higher inclusion probabilities for the drug responses and lower probabilities for other omic types. This is because of adaptive penalization which uses the group structure by imposing a soft constraint. This results in a more consistent feature inclusion and improved predictive performance. Similar conclusions can be drawn from the four other regressions. In this case, we have also studied the robustness of SA-Enet with respect to the number of iterations $T$. Similar to the simulation studies, here also the method shows robustness for increased iterations. Numerical results supporting them are deferred to Section \ref{subsec app: Robustness across iterations} in the appendix.

\section{Conclusion}\label{sec: Conclusion}

In this research, we have proposed a structure-adaptive framework to incorporate auxiliary information on features for estimating the sparse regression coefficients in high-dimensional linear regression. The SA-Enet estimator is intuitive, practical to implement, and effective in real-life applications as demonstrated here. Our framework is flexible enough to incorporate various types of structural information that can arise in many genomics applications. Examples of structures range from something as intuitive as groups to something as general as covariate information corresponding to each coefficient.

Compared to the group Lasso and the fused Lasso, we do not directly impose constraints on the regression coefficients. Instead, we use the external information together with the data to jointly determine the penalization strength for each regression coefficient. In this sense, we have translated the external information into a soft constraint on the regression coefficients compared to the hard constraints imposed by the group Lasso and fused Lasso. Therefore, our method is expected to be more robust to misspecified or non-informative external information. For a general purpose, we recommend the use of the SA-Enet with 5 iterations.

Under the iid-design when $p$ grows in the same order as $n$, we introduce an AMP algorithm to analyze the SA-Enet estimator. This helps us study the asymptotic risk of the SA-Enet estimator through a one-dimensional recursion, known as the state evolution. A numerical study confirms the practical relevance of our theory in predicting the finite sample risk of the SA-Enet. Although the risk is obtained under an asymptotic setting, the prediction offered by the AMP theory seems to hold even when $p$ is as small as 500. This justifies the finite sample validity of the predicted asymptotic risk and also confirms the practical relevance of our theory in predicting the finite sample risk of the SA-Enet estimator. In conclusion, the promising finite sample performances demonstrated via simulations and a real data illustration suggest that the framework might be useful in a variety of statistical problems.

\section{Supplementary matrials}
Software scripts for implementing SA-Enet in \texttt{R} \citep{R} are available at \href{https://github.com/sandy-pramanik/saenet.git}{GitHub}. R-scripts used to obtain the results presented here are available in the supplement.

\section{Appendix}

Below we provide some technical details and additional materials. The appendix is organized as follows. Section~\ref{sec app: Theoretical analysis in the location model under a group structure} provides some analyses of a location model under the group structure. This provides a theoretical motivation for the framework proposed here. Section~\ref{sec app: Proof of Theorem 3.1} presents a general version of Theorem~\ref{ampthm saenet group} and a technical lemma required for the proof. In the same section, we also provide a sketch for the proof of Proposition~\ref{ampprop saenet cov}. Sections \ref{sec app: Robustness across iterations} and \ref{sec app: Drug response prediction in leukemia samples} conclude with additional simulation results and real data analysis that are omitted above for brevity.

\appendix

\section{Theoretical analysis in the location model under a group structure}\label{sec app: Theoretical analysis in the location model under a group structure}

To get some insights on how SA-Enet can become superior by making use of the auxiliary information, we study the $ L^2$-risk of the estimator under the group structure in this section. For this, we particularly consider SA-Lasso with a single iteration.

Consider the location model $\bs{Y} = \bs{\mu} + \bs{\varepsilon}$, where $ \bs{\mu}_{n \times 1} = {( \mu_1, \mu_2, \cdots, \mu_n )}^\T $ and everything else is as in Section 2.1 in the main article. Also, assume that $\bs{\mu}$ has an underlying group structure as described in Section 2.3.4 in the main article with $\abs{S_d}=n_d$ for $1\leq d\leq D$. For all $i\in S_d$, define the thresholding estimator $\hat{\mu}_i$ of $\mu_i$ as follows:
\begin{equation}\label{thresh-est}
    \hat{\mu}_i = \text{sgn}(Y_i) \, \left( \abs{Y_i} - \frac{\lambda_d}{M_d} \right)_+  \quad \text{for some $\lambda_d >0$},
\end{equation}
where $M_d = \mean{\abs{\hat{\bs{\theta}}}_{S_d}}$ with $\hat{\theta}_i = \text{sgn}(Y_i) \, \left( \abs{Y_i} - \tau \right)_+$ for some properly chosen $\tau >0$, $\text{sgn}(x) = x/\abs{x}$ and $(x)_+ = x \bs{1}\{x \geq 0\}$. Denote by $\phi(x) $ the density function for the standard normal random variable.

\begin{theorem}\label{thm1}
Consider the above setup and in addition assume that $\varepsilon_i$'s are independent and identically distributed as $N(0, \sigma^2)$. Further for simplicity, let $\mu_i=0$ if $i\in S_1$ (corresponds to the null group), and $\mu_i=a_d$ if $i\in S_d$ for all $d>1$. Denote by $M_0 = \E \abs{Y_1 - \mu_1}$ and let $\lambda_d=\sigma M_0\sqrt{2\log n_d}$ for $1\leq d\leq D$ in (\ref{thresh-est}). Under these assumptions, if
\begin{equation}\label{thm1-assmp}
    \min_{2 \leq d \leq D} \, \left[ \big( 2 \phi(1)+1 \big) a_d^2 - 2 \sigma^2 \, \log n_d \, \E \bigg( \frac{M_0}{M_d} \bigg)^2 - \sigma^2 \right] >0,
\end{equation}
then an upper bound for the $L^2$-risk of $\hat{\bs{\mu}}$ (as defined in (\ref{thresh-est})) is given by
\begin{equation}\label{risk-bound}
\begin{split}
    \E \norm{\hat{\bs{\mu}} - \bs{\mu}}^2_2 \leq & \sum_{d=2}^D \, \frac{2 \sigma (\sigma^2 + a_d^2)}{M_0 \sqrt{2 \log n_d}} + \sum_{d=2}^D \, n_d \left[ 2 \sigma^2 \, \log n_d \, \E \left( \frac{M_0}{M_d} \right)^2 + \sigma^2 \right] + \\ 
    & O \left( \frac{1}{\sqrt{\log n_1}} \right) .
\end{split}
\end{equation}
\end{theorem}

\begin{rem}
The leading term in (\ref{risk-bound}) is $ 2 \sigma^2 \sum_{d=2}^D \, n_d \, \log n_d \, \E ( M_0 / M_d )^2$.
\end{rem}

\begin{rem}
Recall that the risk for the soft thresholding estimator with the universal threshold $\lambda = \sigma \sqrt{2 \log n}$ is upper bounded by $(2 \log n +1) \big[ \sigma^2 + \sum_{i=1}^n \min(\sigma^2 , \mu_i) \big]$. With $\min_{2 \leq d \leq D} \, a_d^2 > \sigma^2$, the bound becomes $(2 \log n +1) \sigma^2 \big[ 1+ \sum_{d=2}^D n_d ] \big]$. When $\E ( M_0 / M_d )^2 <1$, this is larger than the leading term of the upper bound in (\ref{risk-bound}).
\end{rem}

\noindent {\bf Proof of Theorem \ref{thm1}.} For all $i=1,\ldots,n$ and $d=1,\ldots,D$, if $ i \in S_d$ for some $d$, then based on the definition (\ref{thresh-est}) define
\begin{align*}
g_i(\bs{Y}):=\hat{\mu}_i - Y_i =\begin{cases}
                         -\frac{\lambda_d}{M_d}, & \mbox{if } Y_i >\frac{\lambda_d}{M_d}, \\
                         -Y_i, & \mbox{if } |Y_i|\leq \frac{\lambda_d}{M_d},\\
                         \frac{\lambda_d}{M_d}, & \mbox{if } Y_i <-\frac{\lambda_d}{M_d}.
                       \end{cases}
\end{align*}
Below we study the risk of the estimator in the $d^{th}$ group, and the result will follow from the summation over all groups. For the ease of notation, write $s=n_d$ and $\lambda=\lambda_d$. By the Stein's lemma, we have
\begin{align*}
\E \norm{\bs{\hat{\mu}}_{S_d}-\bs{\mu}_{S_d}}^2=\sum_{i \in S_d} \E \, h_i(\bs{Y}),
\end{align*}
where
\begin{align*}
h_i(\bs{Y})=2\sigma^2 \,\, \nabla_i \, g_i(\bs{Y}) +\left(\frac{\lambda}{M_d}\right)^2\wedge Y_i^2 + \sigma^2.
\end{align*}
Direct calculation shows that
\begin{align*}
\nabla_i \, g_i(\bs{Y}) =\frac{\lambda}{s M_d^2} \, \mathbf{1} \bigg\{ |Y_i| >\left(\frac{\lambda}{M_d}\right) \vee \tau \bigg\} - \mathbf{1} \{ M_d|Y_i| \leq \lambda \}.
\end{align*}
On the one hand,
\begin{align}\label{stein-bound}
\E [h_i(\bs{Y})] \leq 2\sigma^2 \, \E \Bigg[ \frac{\lambda}{s M_d^2} \mathbf{1} \bigg\{ |Y_i| >\left(\frac{\lambda}{M_d}\right) \vee \tau \bigg\} \Bigg] + E\left(\frac{\lambda}{M_d}\right)^2 +\sigma^2.
\end{align}
In (\ref{stein-bound}), we have used the fact that $\left(\frac{\lambda}{M_d}\right)^2\wedge Y_i^2\leq \left(\frac{\lambda}{M_d}\right)^2$. Using the other bound,
we get
\begin{align*}
\E [h_i(\bs{Y})] \leq 2\sigma^2 \, \E \Bigg[ \frac{\lambda}{s M_d^2} \mathbf{1} \bigg\{ |Y_i| >\left(\frac{\lambda}{M_d}\right) \vee \tau \bigg\} \Bigg] + \mu^2_i + 2\sigma^2 \, P(M_d|Y_i|>\lambda).
\end{align*}
We divide the rest of the arguments into the following 5 steps.

\vskip 0.5cm
\paragraph{Step 1:} Notice that $[ \abs{Y_i} - \tau ]_+$'s are sub-Gaussian, i.e., for all $i$
\begin{equation*}
    \E \Big[e^{t \big([ \abs{Y_i} - \tau ]_+ - \E[ \abs{Y_i} - \tau ]_+ \big)} \Big] \leq e^{t^2\eta_i^2/2},
\end{equation*}
for all $t\in\mathbb{R}$ and some $\eta_i>0$. By the concentration inequality for sub-Gaussian variable, we obtain
\begin{align*}
P(M_d- \E M_d>\xi) & = P\left[ \frac{1}{s}\sum_{i \in S_d} \, \Big([ \abs{Y_i} - \tau ]_+ - \E[ \abs{Y_i} - \tau ]_+\Big) >\xi\right]
\leq \exp \left(-\frac{s\xi^2}{2\eta}\right),
\end{align*}
with $\eta=s^{-1}\sum_{i \in S_d} \eta_i^2$.

\vskip 0.5cm
\paragraph{Step 2:} We study the term $P(M_d|Y_i|>\lambda)$. Using the result in Step 1, we have
\begin{align*}
P(M_d|Y_i|>\lambda)\leq& P(M_d|Y_i|>\lambda, M_d - \E M_d\leq \xi)+P(M_d - \E M_d>\xi)
\\ \leq & P \left( (\E M_d+\xi)|Y_i|>\lambda \right)+\exp \left(-\frac{s\xi^2}{2\eta}\right).
\end{align*}
Let $g(\mu_i)=P(|Y_i|>\lambda(\E M_d+\xi)^{-1})$. As $g$ is symmetric about zero,
$$g(\mu_i)\leq g(0)+(1/2)(\sup_x|g''(x)|)\mu_i^2.$$

By the Mill's ratio,
\begin{equation*}
g(0) = 2P_{\mu_i=0}(Y_i>\lambda(\E M_d+\xi)^{-1}) \leq \frac{2\phi(\lambda(\E M_d+\xi)^{-1}/\sigma)}{\lambda(\E M_d+\xi)^{-1}/\sigma} = \frac{2s^{-\gamma^2}}{\gamma\sqrt{4\pi \log s}}
\end{equation*}
where $\gamma=M_0/(\E M_d+\xi)$. Some calculus shows that
\begin{align*}
\sup_x|g''(x)|\leq 2\sigma^{-2}\sup_x|x\phi(x)| =2\phi(1)\sigma^{-2},
\end{align*}
which then implies that
\begin{align*}
P(M_d|Y_i|>\lambda) \leq  \frac{2s^{-\gamma^2}}{\gamma\sqrt{4\pi \log s}}+\phi(1) \left( \frac{\mu_i}{\sigma} \right)^2+\exp \left(-\frac{s\xi^2}{2\eta}\right).
\end{align*}

\vskip 0.5cm
\paragraph{Step 3:} Next we study the term $\E \Big[ \frac{\lambda}{s M_d^2} \mathbf{1} \Big\{ |Y_i| >\left(\frac{\lambda}{M_d}\right) \vee \tau \Big\} \Big]$. It is not hard to see that.
\begin{align}
\E \Bigg[ \frac{\lambda}{s M_d^2} \mathbf{1} \Bigg\{ |Y_i| >\left(\frac{\lambda}{M_d}\right) \vee \tau \Bigg\} \Bigg] \leq & \frac{1}{s\lambda} \E \Bigg[ Y_i^2 \, \mathbf{1} \Bigg\{ |Y_i| >\left(\frac{\lambda}{M_d}\right) \vee \tau \Bigg\} \Bigg] \leq \frac{\sigma^2+\mu_i^2}{s\lambda}.
\end{align}

\vskip 0.5cm
\paragraph{Step 4:} Using the above results, we have
\begin{align*}
& \E [h_i(\bs{Y})]\\ 
\leq & \frac{2\sigma(\sigma^2+\mu_i^2)}{s M_0\sqrt{2\log s}}+\\
& \left\{ \E \left(\frac{\lambda}{M_d}\right)^2 +\sigma^2\right\}\wedge \left\{\frac{4\sigma^2 s^{-\gamma^2}}{\gamma\sqrt{4\pi \log s}}+(2\phi(1)+1)\mu_i^2+2\sigma^2\exp \left(-\frac{s\xi^2}{2\eta}\right)\right\}.
\end{align*}
Thus we get
\begin{align*}
& \E\|\hat{\bs{\mu}}_{S_d} - \bs{\mu}_{S_d} \|^2_2\\
\leq & \frac{2\sigma( s \sigma^2+\sum_{i \in S_d} \mu_i^2)}{sM_0\sqrt{2\log s}} +\\
& \sum_{i \in S_d} \left\{ \E \left(\frac{\lambda}{M_d}\right)^2 +\sigma^2\right\}\wedge \left\{\frac{4\sigma^2 s^{-\gamma^2}}{\gamma\sqrt{4\pi \log s}}+(2\phi(1)+1)\mu_i^2+2\sigma^2\exp \left(-\frac{s\xi^2}{2\eta}\right)\right\}.
\end{align*}
Under assumption (\ref{thm1-assmp}), $(2\phi(1)+1)a^2> \E \left(\frac{\lambda}{M_d}\right)^2 +\sigma^2$ for all $d>1$. The above bound then becomes
\begin{align*}
\E \|\hat{\bs{\mu}}_{S_d}- \bs{\mu}_{S_d}\|^2_2
\leq & \frac{2\sigma(\sigma^2+a^2)}{M_0\sqrt{2\log s}}+ s\left\{ 2\sigma^2\log (s) \,\, \E \left(\frac{M_0}{M_d}\right)^2 +\sigma^2\right\}.
\end{align*}
For $d=1$, by choosing $\xi=s^{-1/3}$, we have $\E M_1=M_0$ and $\gamma=M_0/(M_0+ s^{-1/3})$. It is not hard to verify that $s^{1-\gamma^2}\rightarrow 1.$ Then we have
\begin{equation*}
    \E\|\hat{\bs{\mu}}_{S_1}- \bs{\mu}_{S_1}\|^2_2 =O\left(\frac{1}{\sqrt{\log s}}\right).
\end{equation*}

\vskip 0.5cm
\paragraph{Step 5:} Under condition (\ref{thm1-assmp}) and combining the results above, we finally obtain
\begin{align*}
\E \| \hat{\bs{\mu}} - \bs{\mu} \|^2_2 \leq & \sum^{D}_{d=2}\frac{2\sigma(\sigma^2+a^2_i)}{M_0\sqrt{2\log n_d}}+\sum^{D}_{d=2} \, n_d\left\{ 2\sigma^2 \log(n_d) \, \, \E\left(\frac{M_0}{M_i}\right)^2 +\sigma^2\right\} +\\
& O\left(\frac{1}{\sqrt{\log n_1}}\right).
\end{align*}

\section{Proof of Theorem~\ref{ampthm saenet group}}\label{sec app: Proof of Theorem 3.1}

The proof is along a similar line as that of Theorem~2 in \cite{bayati11}. We provide a general result similar to this theorem which directly leads to the proof of Theorem 3.1 in the main article as a special case. 

\subsection{A general result under a group structure}\label{subsec app: A general result under a group structure}

\begin{theorem}\label{thm2}
For $t \geq 0$, consider the sequences of functions $\{g_{t}\}_{t \geq 0}$ and $\{f_{dt}\}_{t \geq 0}$ for all $d=1, \ldots,D$, where for each $d$, $f_{dt} : \mathbb{R}^2 \mapsto \mathbb{R}$ and $g_{t} : \mathbb{R}^2 \mapsto \mathbb{R}$ are assumed to be Lipschitz continuous. Now, given $\bs{\varepsilon} \in \mathbb{R}^n$ and $\bs{\beta}_0 \in \mathbb{R}^p$, define the sequence of vectors $\bs{h}^t, \bs{q}^t \in \mathbb{R}^p$ and $\bs{b}^t, \bs{m}^t \in \mathbb{R}^n$, by fixing initial condition $q^0$, and obtaining $\{ b^t \}_{t \geq 0}$, $\{ m^t \}_{t \geq 0}$, $\{ h^t \}_{t \geq 1}$ and $\{ q^t \}_{t \geq 1}$ through the following recursions
\begin{align}
    &   \bs{h}^{t+1} = \xmat^\T \bs{m}^t - \xi_t \bs{q}^t , \label{h-rec}\\
    &   \bs{b}^t = \xmat \bs{q}^t - \lambda_t \bs{m}^{t-1} ,  \label{b-rec}\\
    &   \bs{q}^t = {(q_1^t , \cdots, q_p^t)}^\T \,\, \text{where} \,\, q_j^t = f_{dt} (h_j^t , \beta_{0j}) \,\, \text{if} \,\,  j \in S_d \,\, \text{for all} \,\, j,d ,  \label{q-rec}\\
    &   \bs{m}^t = g_t (\bs{b}^t , \bs{\varepsilon}) , \label{m-rec}
\end{align}
where $\xi_t = \mean{g^{\prime}_t ( \bs{b}^t , \bs{\varepsilon} )}$, $\lambda_t = \delta^{-1} \sum_{d=1}^D c_d \, \mean{f^{\prime}_{dt} ( \bs{h}^t_{S_d} , \bs{\beta}_{0 S_d})}$ and define $m^{-1}=0$. In addition, assume the following conditions hold:

\begin{ass}
$p$ and $\{p_{d} \}_{d=1}^D$ are such that as $n \uparrow \infty$, $n/p \rightarrow \delta \in (0, \infty)$ and $p_{d}/p \rightarrow c_d \in (0,1)$ for all $d$.
\end{ass}

\begin{ass}
$ \displaystyle{ \lim_{p_d \rightarrow \infty} } \, \frac{1}{p_d} \norm{\bs{q}^0_{S_d}}^2 \in (0, \infty)$ for all d. This implies,

    \begin{equation}\label{sigma0}
        \delta^{-1} \displaystyle{ \lim_{p \rightarrow \infty}} \frac{1}{p} \norm{\bs{q}^0}^2 = \delta^{-1} \sum_{d=1}^D c_d \,\, \displaystyle{ \lim_{p_d \rightarrow \infty}} \frac{1}{p_d} \norm{\bs{q}^0_{S_d}}^2 = \left(\sigma^0\right)^2 \in (0, \infty).
    \end{equation}
\end{ass}

\begin{ass}
Empirical distributions of the sequence of vectors $\{\bs{\beta}_{0 S_d} (p_d)\}_{p_d \geq 0}$ and $\{  \bs{\varepsilon} (p) \}_{p \geq 0}$ converge weakly to probability measures $\prob_{B_{0d}}$ and $\prob_W$ for any $d$. We also assume they have bounded $(2\nu-2)^{\text{th}}$ moment, and
    \begin{align}
        &  (i) \,\, \displaystyle{ \lim_{p_d \rightarrow \infty}} \E_{\hat{\prob}_{\bs{\beta}_{0 S_d} (p_d)}} \Big[B_{0d}^{2\nu-2}\Big] = \E_{\prob_{B_{0d}}} \Big[B_{0d}^{2\nu-2}\Big] < \infty , \label{thm2-assmp3-1}\\
        &  (ii) \,\, \displaystyle{ \lim_{p \rightarrow \infty}} \E_{\hat{\prob}_{\varepsilon (p)}} \Big[W^{2\nu-2}\Big] = \E_{\prob_W} \Big[W^{2\nu-2}\Big] < \infty , \label{thm2-assmp3-2}\\
        &  (iii) \,\, \displaystyle{ \lim_{p_d \rightarrow \infty}} \E_{\hat{\prob}_{\bs{q}_{0 S_d} (p_d)}} \Big[B_{0d}^{2\nu-2}\Big] < \infty . \label{thm2-assmp3-3}
    \end{align}
    Therefore, the empirical distribution of $\beta_{0j}$ converges to $B_0 \sim \prob_{B_0} := \sum_{d=1}^D c_d \prob_{B_{0d}}$.
\end{ass}

With $\sigma^0$ as in (\ref{sigma0}), State evolution defines quantities $\{ \tau^t \}_{t \geq 0}$ and $\{ \sigma^t \}_{t \geq 0}$ as follows:
\begin{equation}\label{state-evolution-eqn}
    \left(\tau^t\right)^2 = \E \Big\{ g_t (\sigma^t Z , W)^2 \Big\} \quad \text{and} \quad \left(\sigma^t\right)^2 = \delta^{-1} \sum_{d=1}^D c_d \, \E \Big\{ f_{dt} (\tau^{t-1} Z , B_{0d})^2 \Big\},
\end{equation}
where $B_{0d} \sim \prob_{B_{0d}} $ for all $d$, and $W \sim \prob_W$ are independent of $Z \sim N(0,1)$. Then for all $t \geq 0$, and for any pseudo-Lipschitz functions $\psi_d : \R^2 \mapsto \R $ for all $d$ and $ \psi : \R^2 \mapsto \R $ of orders $\nu$,
\begin{align*}
    &   \displaystyle{ \lim_{p_d \rightarrow \infty} } \, \frac{1}{p_d} \sum_{j \in S_{0d}} \psi_d (h_j^{t+1} , \beta_{0j}) \overset{a.s.}{=} \E \big[ \psi_d (\tau^t Z , B_{0d} ) \big], \quad \forall d = 1, \cdots, D, \\
    &   \displaystyle{ \lim_{n \rightarrow \infty} } \, \frac{1}{n} \sum_{i=1}^n \psi (b_i^t , \varepsilon_i) \overset{a.s.}{=} \E \big[ \psi (\sigma^t Z , W ) \big],
\end{align*}
where $\sigma^t$ and $\tau^t$ are determined by the recursion (\ref{state-evolution-eqn}).
\end{theorem}

\begin{cor}
The fact that the AMP algorithm in Theorem 3.1 in the main article is a special case of the recursions (\ref{h-rec})--(\ref{m-rec}) can be observed by defining
\begin{align*}
    &   \bs{h}^{t+1} = \bs{\beta}_0 - ( \xmat^\T \bs{e}_1^t + \bs{\beta}_1^t) ,\\
    &   \bs{q}^t = \bs{\beta}_1^t - \bs{\beta}_0 , \\
    &   \bs{b}^t = \bs{\varepsilon} - \bs{e}_1^t , \\
    &   \bs{m}^t = - \bs{e}^t.
\end{align*}
The functions $f_{dt}$ and $g_t$ are defined as
\begin{align*}
    &   f_{dt} (r, s) = \eta \left( s-r \, ; \theta_{11}^{t-1} \, \omega_{1d},\theta_{21}^{t-1} \right) -s , \\
    &   g(r,s) = r - s,
\end{align*}
with the same initial condition $ \bs{q}^0 =- \bs{\beta}_0$. Also, $\tau^t \equiv \tau^t_1$ and $\sigma^t \equiv \sigma^t_1$.
\end{cor}

\subsection{A technical lemma concerning the proof of Theorem \ref{thm2}}\label{subsec app: A technical lemma concerning the proof of Theorem B.1}

In this subsection, similar to Lemma 1 in \cite{bayati11} we provide a more general result that will lead to the proof of Theorem \ref{thm2}.

Denote by $\mathcal{G}_{t_1 , t_2}$ the $\sigma$-algebra generated by $ \bs{b}^0 , \cdots, \bs{b}^{t_1 -1}$, $ \bs{m}^0 , \cdots, \bs{m}^{t_1 -1}$, $ \bs{h}^1 , \cdots, \bs{h}^{t_2}$, $ \bs{q}^1 , \cdots, \bs{q}^{t_2}$, $\bs{\beta}_0$ and $\varepsilon$.
\begin{lemma}
Let $\{\xmat(p) \}_p$, $\{ \bs{q}^0 (p) \}_p$, $\{\bs{\beta}_0 (p) \}_p$ and $\{ \bs{\varepsilon} (p) \}_p$ be sequences as in Theorem \ref{thm2}, with $n/p \, \rightarrow \delta \in (0, \infty)$ and let $\{ \sigma^t , \tau^t \}_{t \geq 0}$ be defined by the recursion (\ref{state-evolution-eqn}) with initialization $\left(\sigma_0\right)^2 =\delta^{-1} \lim_{n \rightarrow \infty} \mean{\bs{q}^0 , \bs{q}^0}$, where $\mean{a,b} = m^{-1} \sum_{i=1}^m a_i b_i$ for $\bs{a}, \bs{b} \in \R^m$. Then for all $t \geq 0$, the followings hold:
\paragraph{(a)} \begin{align*}
    &   \bs{h}^{t+1} |_{\mathcal{G}_{t+1 , t}} \overset{d}{=} \sum_{i=0}^{t-1} \, \alpha_i \bs{h}^{i+1} + \Tilde{\xmat}^\T \bs{m}^t_\perp + \Tilde{Q}_{t+1} \overrightarrow{\bs{o}}_{t+1} (1) ,\\
    &   \bs{b}^t |_{\mathcal{G}_{t , t}} \overset{d}{=} \sum_{i=0}^{t-1} \, \beta_i \bs{b}^i + \Tilde{\xmat} \bs{q}^t_\perp + \Tilde{M}_t \overrightarrow{\bs{o}}_t (1),
\end{align*}
where $\Tilde{\xmat}$ is an independent copy of $\xmat$, $\Tilde{Q}_t$ ($\Tilde{M}_t$) is such that their columns form an orthogonal basis for the column space of $Q_t$ ($M_t$), and $\Tilde{Q}_t^\T \, \Tilde{Q}_t = p \mathbf{I}_{t \times t}$ ($\Tilde{M}_t^\T \, \Tilde{M}_t = n \mathbf{I}_{t \times t}$).\\

\paragraph{(b)} For all pseudo-Lipschitz functions $\phi_{hd}, \, \phi_b \mapsto \R^{t+2} \rightarrow R$ of order $\nu$, we have
\begin{align*}
    &   \lim_{p_d \rightarrow \infty} \, \frac{1}{p_d} \sum_{j \in S_d} \, \phi_{hd} \left( h_j^1, \cdots, h_j^{t+1}, \beta_{0j} \right) \overset{a.s.}{=} \E \left[ \phi_{hd} \left( \tau^0 Z_{0d}, \cdots, \tau^t Z_{td}, B_{0d} \right) \right]  \quad \forall \, d, \\
    &   \lim_{n \rightarrow \infty} \, \frac{1}{n} \sum_{i=1}^n \, \phi_b \left( b_i^0, \cdots, b_i^t, \varepsilon_i \right) \overset{a.s.}{=} \E \left[ \phi_b \left( \sigma^0 \hat{Z}_0, \cdots, \sigma^t \hat{Z}_t, W \right) \right],
\end{align*}
where where $(Z_{0d} , \cdots , Z_{td})$ and $(\hat{Z}_0, \cdots, \hat{Z}_t )$ are two zero-mean Gaussian vectors independent of $B_{0d}$ and $W$ with $Z_{id}, \hat{Z}_i \sim N(0,1)$ for all $d$.\\

\paragraph{(c)} For all $0 \leq r,s \leq t$,
\begin{align*}
    &   \lim_{p_d \rightarrow \infty} \, \mean{\bs{h}_{S_d}^{r+1}, \bs{h}_{S_d}^{s+1}} \overset{a.s.}{=} \lim_{n \rightarrow \infty} \, \mean{\bs{m}^r , \bs{m}^s} < \infty \quad \forall d, \\
    &   \lim_{n \rightarrow \infty} \, \mean{\bs{b}^r , \bs{b}^s} \overset{a.s.}{=} \frac{1}{\delta} \, \lim_{p \rightarrow \infty} \, \mean{\bs{q}^r , \bs{q}^s} < \infty.\\
\end{align*}

\paragraph{(d)} For all $0 \leq r,s \leq t$, and Lipschitz functions $\phi_d , \phi \mapsto \R^2 \rightarrow \R$,
\begin{align*}
    &   \lim_{p_d \rightarrow \infty} \, \mean{\bs{h}_{S_d}^{r+1}, \phi_d ( \bs{h}_{S_d}^{s+1}, \bs{\beta}_{0 S_d} ) } \overset{a.s.}{=} \lim_{p_d \rightarrow \infty} \, \mean{\bs{h}_{S_d}^{r+1}, \bs{h}_{S_d}^{s+1} } \, \mean{ \phi^\prime_d ( \bs{h}_{S_d}^{s+1}, \bs{\beta}_{0 S_d} ) } \quad \forall d, \\
    &   \lim_{n \rightarrow \infty} \, \mean{\bs{b}^r , \phi( \bs{b}^s, \varepsilon )} \overset{a.s.}{=} \lim_{n \rightarrow \infty} \, \mean{\bs{b}^r , \bs{b}^s } \, \mean{ \phi^\prime( \bs{b}^s, \varepsilon )}.
\end{align*}
Here $\phi^\prime$ denotes derivative with respect to the first coordinate of $\phi$.\\

\paragraph{(e)} For $l = \nu-1$,
\begin{align*}
    &   \limsup_{p_d \rightarrow \infty} \, \frac{1}{p_d} \, \sum_{j \in S_d} \, \left( h_j^{t+1} \right)^{2l} < \infty \quad \text{almost surely} \quad \forall d, \\
    &   \limsup_{n \rightarrow \infty} \, \frac{1}{n} \, \sum_{i=1}^n \, \left( b_i^t \right)^{2l} < \infty \quad \text{almost surely}.\\
\end{align*}

\paragraph{(f)} For $0 \leq r \leq t$,
\begin{equation*}
    \lim_{p_d \rightarrow \infty} \, \mean{\bs{h}_{S_d}^{r+1}, \bs{q}_{S_d}^0} \overset{a.s.}{=} 0.\\
\end{equation*}

\paragraph{(g)} For $0 \leq r \leq t$ and $0 \leq r \leq t-1$, there exists strictly positive constants $\rho_r$ and $\zeta_s$ (independent of $n$ and $p$) such that
\begin{align*}
    &   \lim_{p \rightarrow \infty} \, \mean{\bs{q}_{\perp}^r , \bs{q}_{\perp}^r } > \rho_r \quad \text{almost surely}, \\
    &   \lim_{n \rightarrow \infty} \, \mean{\bs{m}_{\perp}^s , \bs{m}_{\perp}^s } > \zeta_s \quad \text{almost surely}.\\
\end{align*}
\end{lemma}

\section{Adaptive weights in AMP algorithm under covariate-dependent structure}\label{sec app: Adaptive weights in AMP algorithm under covariate-dependent structure}

Here we provide a proof sketch to show the convergence of adaptive weights used in the AMP Algorithm~\ref{amp algorithm cov} for covariate-dependent structure. Following the notations in Section~\ref{sec: AMP algorithm under covariate-dependent structure} write
\begin{equation}
    \hat{\bs{\rho}}_1 = \argmin_{\bs{\rho}\in\mathcal{B}} L_{1p} (\bs{\rho}; \gamma), \quad \text{and} 
    \quad \bs{\rho}^*_1 =\argmin_{\bs{\rho}\in\mathcal{B}} L^*_1 (\bs{\rho}; \gamma) ,
\end{equation}
where
\begin{equation}
    L_{1p} ( \bs{\rho} ; \gamma) = \frac{1}{p} \sum_{j=1}^{ p } \Big[ f( \bs{u}_j ; \bs{\rho} )\abs{\beta^*_{0j}} - \log g \Big( f( \bs{u}_j ; \bs{\rho} ) ; \gamma \Big) \Big], \,\, \text{and}
\end{equation}
\begin{equation}
    L^*_1 ( \bs{\rho} ; \gamma) = \mathbb{E} \Big[ f( U ; \bs{\rho} ) \, \abs{\eta \left( B_0 + \tau_0^* Z \, ; \theta_{10}^{*}, \theta_{20}^{*} \right)} \Big]  - \mathbb{E} \Big[ \log g \Big( f( U ; \bs{\rho} ) ; \gamma \Big) \Big] .
\end{equation}
Here $\bs{\beta}_{0}^* = (\beta^*_{01},\dots,\beta^*_{0p})^\top$ is the limiting AMP-Enet estimate as the number of AMP iterations goes to infinity. Under iid design, as $p \uparrow\infty$, this coincides with the Enet estimate defined in (\ref{eqenet}). We want to justify the adaptive weights in Step 2(a) of the AMP Algorithm~\ref{amp algorithm cov} in the main article. So the goal is to show that
\begin{align*}
    \hat{\bs{\rho}}_1 \overset{P}{\rightarrow} {\bs{\rho}}^*_1 .
\end{align*}
We can complete the task in three steps.

\paragraph{Step 1.} For fixed $\bs{\tau}$ and $\gamma$, we show that 
\begin{align*}
    L_{1p} (\bs{\rho}; \gamma) \overset{P}{\rightarrow} L^*_1 (\bs{\rho}; \gamma) .
\end{align*}
According to the Weak Law of Large Numbers,
\begin{align*}
    \frac{1}{p} \sum_{j=1}^{p} \log g \Big( f( \bs{u}_j ; \bs{\rho} ) ; \gamma \Big) \overset{P}{\rightarrow} \mathbb{E} \Big[ \log g \Big( f( U ; \bs{\rho} ) ; \gamma \Big) \Big] .
\end{align*}
After inspecting some of the steps as in (3.34) in \cite{bayati12}, it appears that the $\phi_h$ and $\phi_b$ functions in (3.16) and (3.17) allow additional covariates.
Under the moment assumption $$\mathbb{E}\Bigg[ \sup_{\bs{\rho}\in\mathcal{B}} \abs{f( U ; \tau_0 , \bs{\tau}_1 )}^2 \Bigg]<\infty,$$ we have $$\frac{1}{p} \sum_{j=1}^{ p } f( \bs{u}_j ; \bs{\rho} )\abs{\beta^*_{0j}} \overset{P}{\rightarrow} \mathbb{E} \Big[ f( U ; \bs{\rho} ) \, \abs{\eta \left( B_0 + \tau_0^* Z \, ; \theta_{10}^{*}, \theta_{20}^{*} \right)} \Big].$$ Here it seems $k=2$ is enough for the arguments in and below (3.34) of \cite{bayati12} to go through.

\paragraph{Step 2.} Let $l(\bs{u}_j, \beta^*_{0j}, \bs{\rho}) = f( \bs{u}_j ; \bs{\rho} )\abs{\beta^*_{0j}} - \log g \Big( f( \bs{u}_j ; \bs{\rho} ) ; \gamma \Big)$. Suppose $$|l(\bs{u}_j,\beta_j^*,\bs{\tau})-l(\bs{u}_j,\beta_j^*,\bs{\tau}')|\leq g(\beta_j^*,\bs{u}_j)\|\bs{\tau}-\bs{\tau}'\|$$ where $\mathbb{E}[|g(\beta_j^*,\bs{u}_j)|]<\infty.$ Following similar arguments from Uniform law of large numbers, we can show that $$\sup_{\bs{\rho}\in\mathcal{B}} \abs{L_{1p} (\bs{\rho}; \gamma) - L^*_1 (\bs{\rho}; \gamma)} \overset{P}{\rightarrow} 0.$$

\paragraph{Step 3.} Suppose that for any $\epsilon>0$, $$\min_{\bs{\rho}: \norm{\bs{\rho} - \bs{\rho}^*_1}>\epsilon} L^*_1 (\bs{\rho}; \gamma)>L^*_1 (\bs{\rho}^*_1; \gamma).$$ Then following Theorem~5.7 in \cite{vaart1998}, we have
\begin{align*}
    \hat{\bs{\rho}}_1 \overset{P}{\rightarrow} {\bs{\rho}}^*_1 .
\end{align*}

\section{Robustness across iterations}\label{sec app: Robustness across iterations}

\begin{figure}[H]
	\centering
    \includegraphics[width=\linewidth]{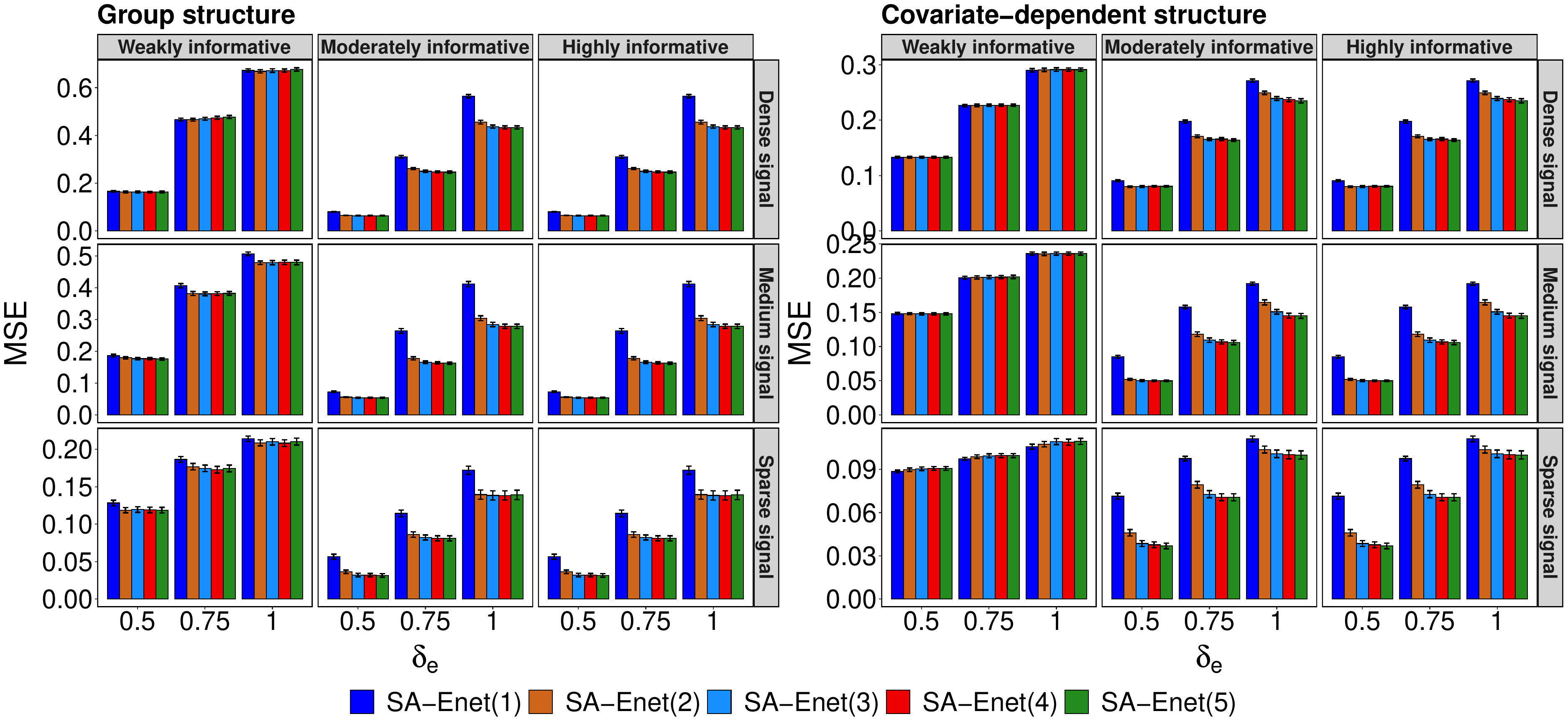}
	\caption{Robustness of the MSE for SA-Enet($T$) in an iid-design.}\label{iid-mse-all}
\end{figure}

\begin{figure}[H]
	\centering
    \includegraphics[width=\linewidth]{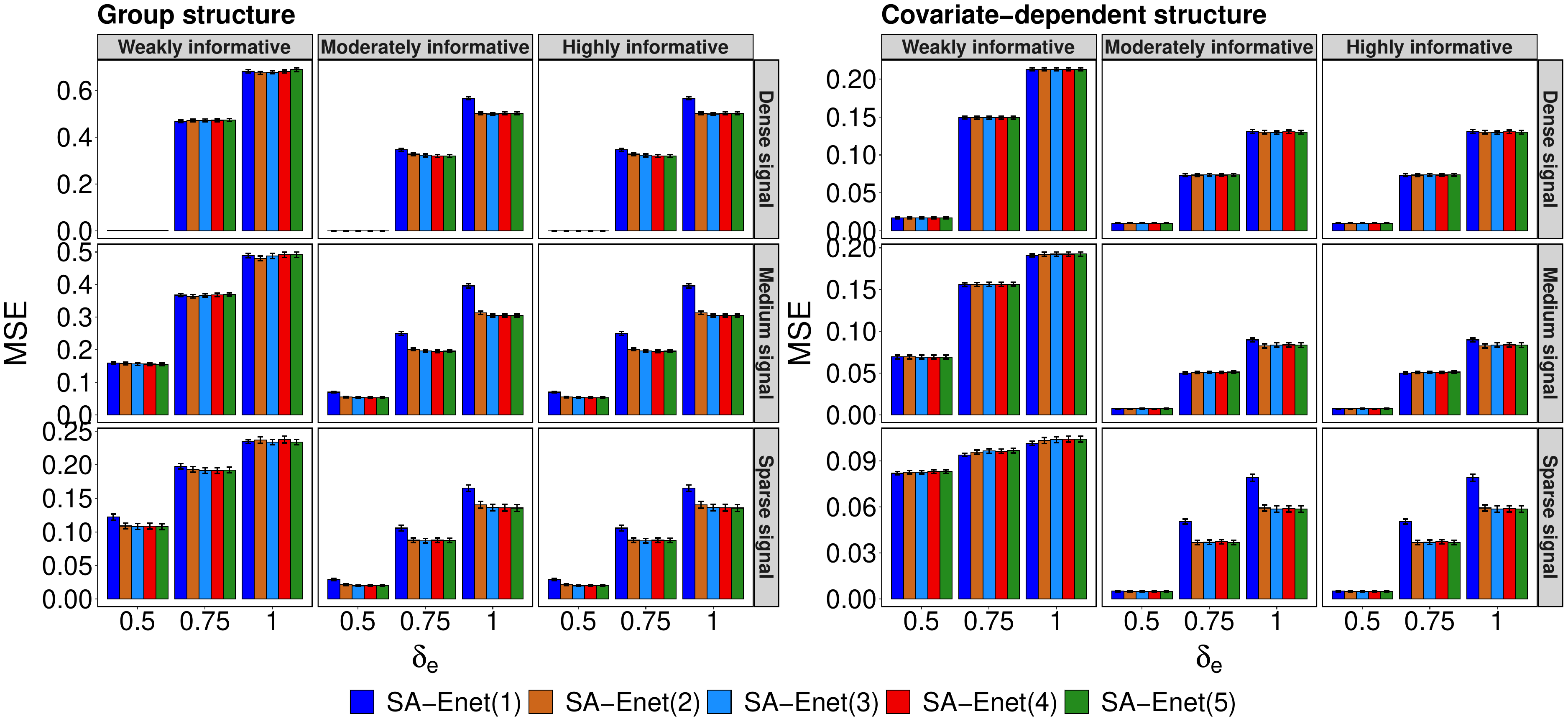}
	\caption{Robustness of the MSE for SA-Enet($T$) in an AR(1) design.}\label{ar-mse-all}
\end{figure}

\begin{figure}[H]
	\centering
    \includegraphics[width=\linewidth]{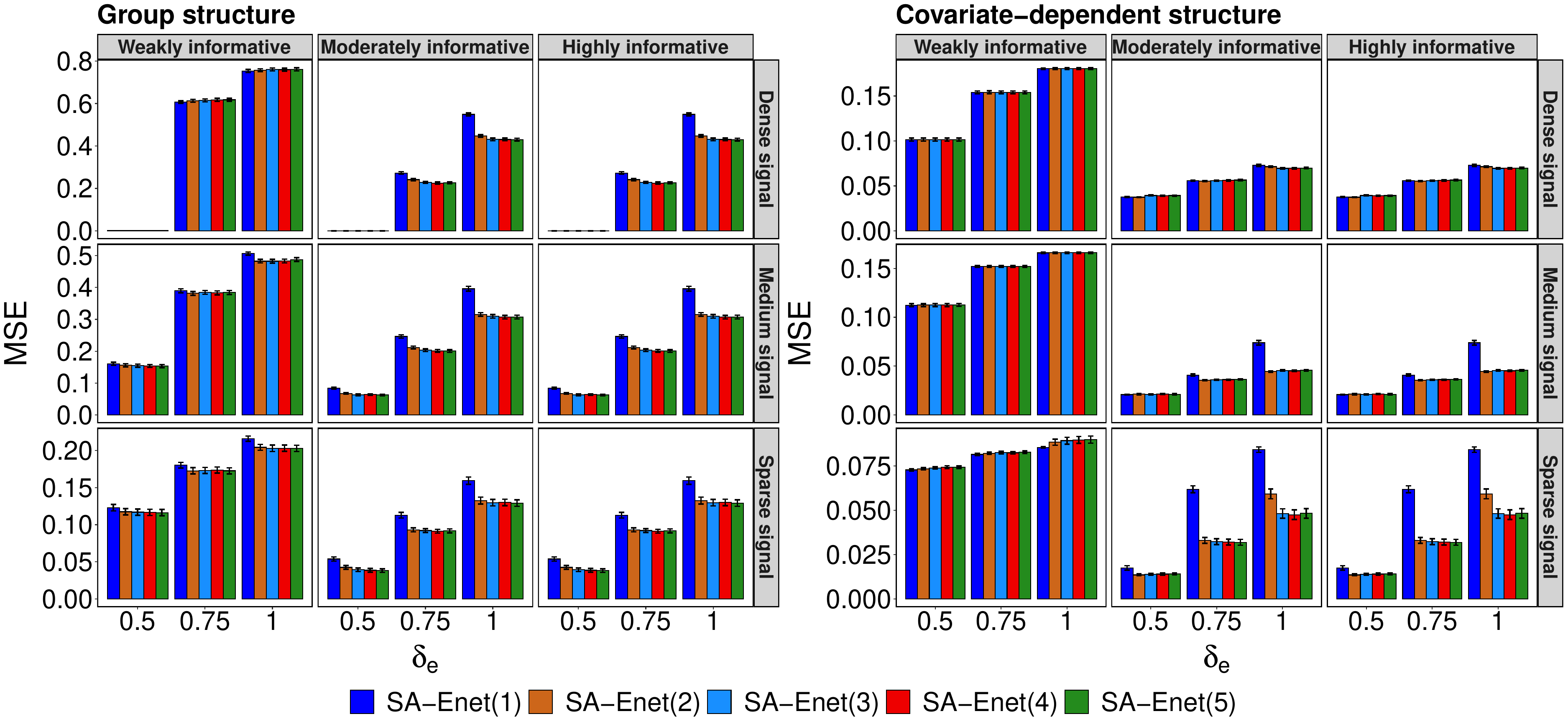}
	\caption{Robustness of the MSE for SA-Enet($T$) in an equicorrelated design.}\label{eq-mse-all}
\end{figure}

\begin{figure}[H]
	\centering
    \includegraphics[width=\linewidth]{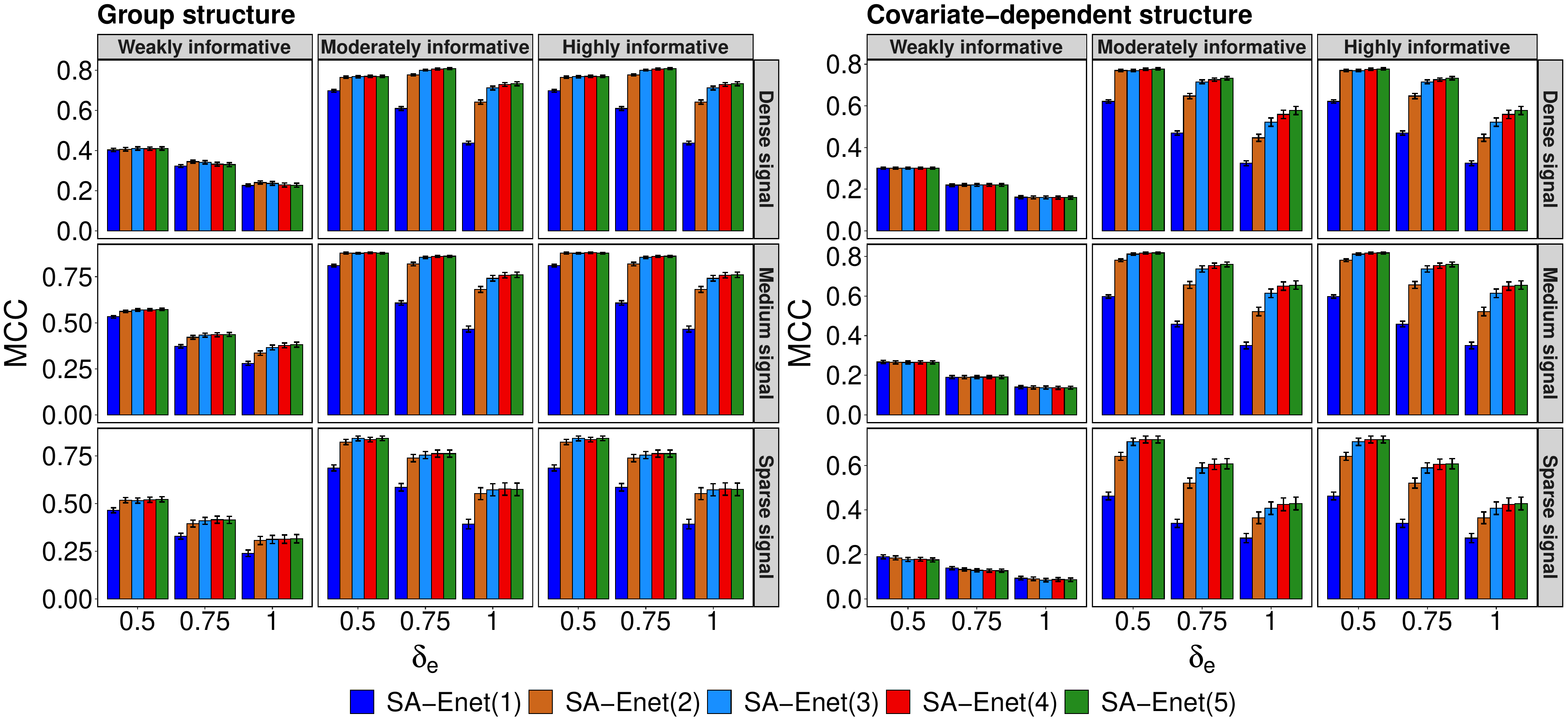}
	\caption{Robustness of the MCC for SA-Enet($T$) in an iid-design.}\label{iid-mcc-all}
\end{figure}

\begin{figure}[H]
	\centering
    \includegraphics[width=\linewidth]{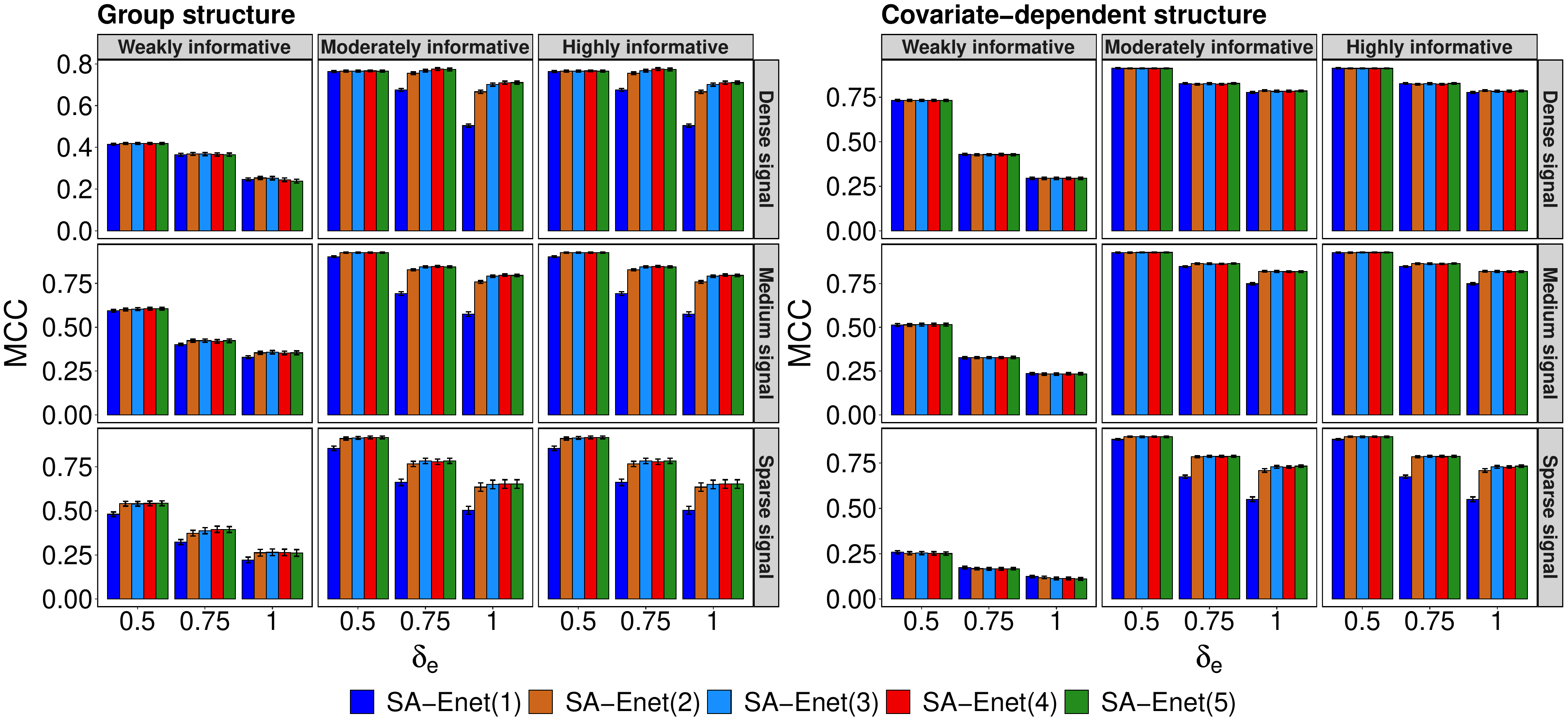}
	\caption{Robustness of the MCC for SA-Enet($T$) in an AR(1) design.}\label{ar-mcc-all}
\end{figure}

\begin{figure}[H]
	\centering
    \includegraphics[width=\linewidth]{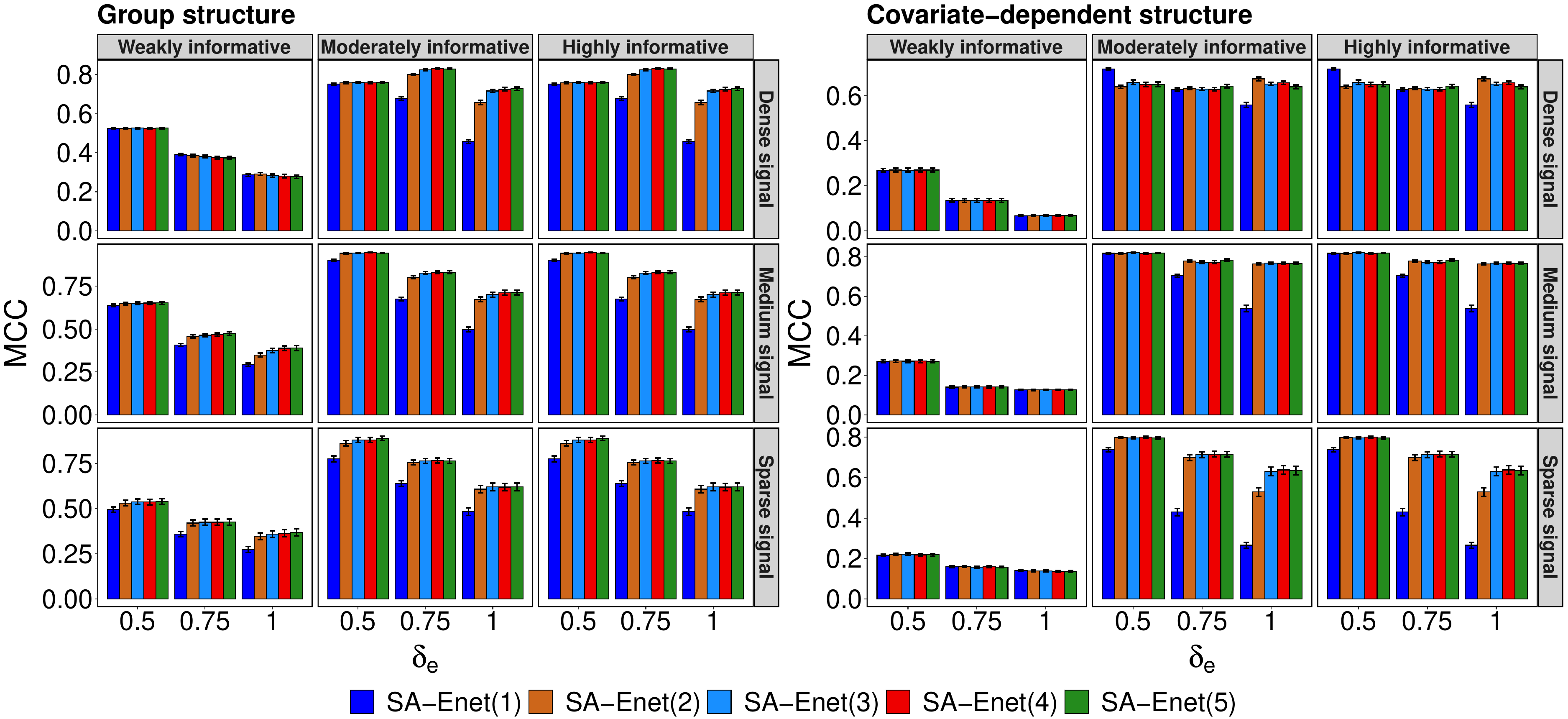}
	\caption{Robustness of the MCC for SA-Enet($T$) in an equicorrelated design.}\label{eq-mcc-all}
\end{figure}

The examples in the main article depicted SA-Enet with 5 iterations as we suggest for default implementations. In this section, we present robustness of the estimator with respect to the the number of iterations $T$. Figures~\ref{iid-mse-all}--\ref{eq-mse-all} and Figures~\ref{iid-mcc-all}--\ref{eq-mcc-all} compare MSE and MCC, respectively, as $T$ increases from 1 to 5. In case of a weakly informative structure and a dense true signal,
the performance of the estimator does not really change with more iterations. As the structure becomes more informative or the true signal become more sparse, the performance of the estimator improves for more than one iteration and the change in performance improvement becomes negligible as $T$ increases to 5. So after combining all the figures we can safely conclude that the method is SA-Enet is fairly robust with respect to the number of iterations under both group and covariate-dependent structural information for the wide varieties of simulation scenarios considered here.


\section{Drug response prediction in leukemia samples}\label{sec app: Drug response prediction in leukemia samples}

In this section, we refer back to Section~\ref{sec: Drug response prediction in leukemia samples}. In this real data analysis, our goal is to predict the response from a drug based on several molecular predictors. The CLL data consists of several omic measurements from 121 patients \citep{dietrich18}. In fact, there are 3 different features corresponding to 3 different omic types, and these different omic types have different scales of measurement. So instead of scaling all the feature measurements to make them comparable, we intend to penalize them accordingly in an adaptive fashion. So we intend to use a group structure with 3 groups representing 3 omic types.

\subsection{Consistency in variable selection}\label{subsec app: Consistency in variable selection}

In this section we present the results for the five methods in case of the four regression regression problems where the goal is to predict response to the Ibrutinib drug with $10 \, \mu M$, $2.5 \, \mu M$, $0.625 \, \mu M$ and $0.156 \, \mu M$ doses.
\begin{figure}[H]
    \centering
    \includegraphics[width=.9\linewidth]{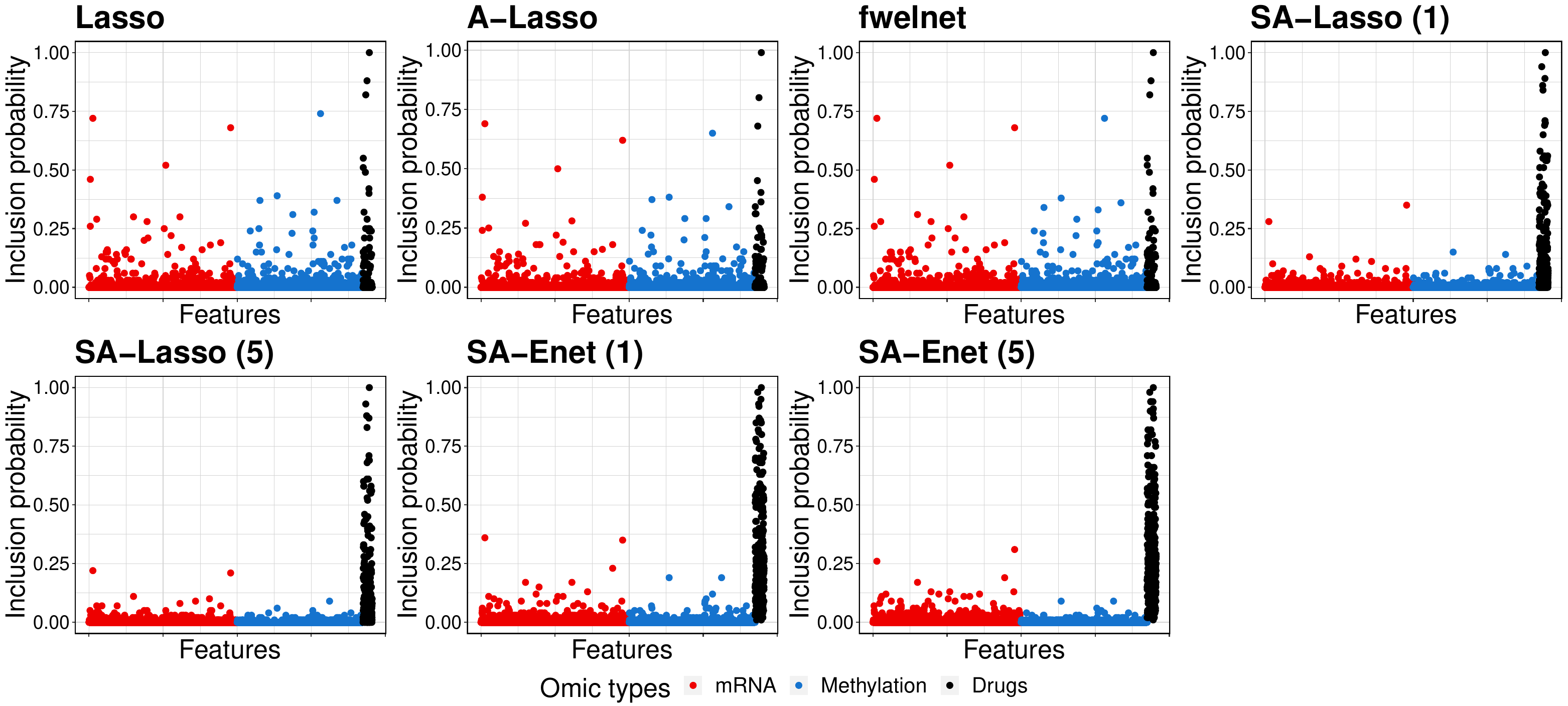}
    \caption{Feature inclusion probability of different methods in Ibrutinib $(10 \, \mu M)$ response prediction. In each panel, the vertical axis shows the proportion of times the method selects each feature in the model. This reflects the consistency of feature selection by the methods. The horizontal axis corresponds to the features color-coded according to the omic types.}
    \label{CLL-feature-inclusion-MOFA2}
\end{figure}
\begin{figure}[H]
    \centering
    \includegraphics[width=.9\linewidth]{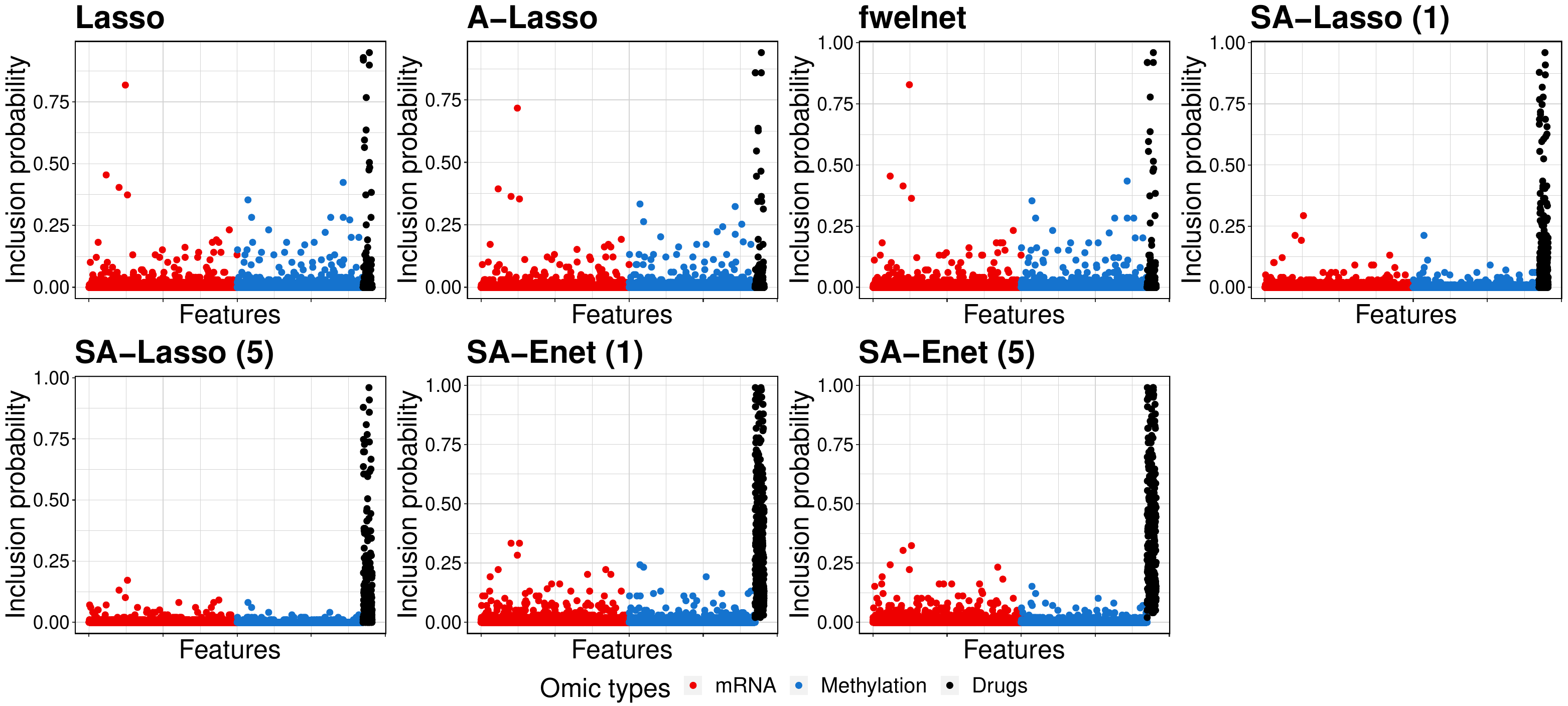}
    \caption{Feature inclusion probability of different methods in Ibrutinib $(2.5 \, \mu M)$ response prediction. In each panel, the vertical axis shows the proportion of times the method selects each feature in the model. This reflects the consistency of feature selection by the methods. The horizontal axis corresponds to the features color-coded according to the omic types.}
    \label{CLL-feature-inclusion-MOFA3}
\end{figure}
\begin{figure}[H]
    \centering
    \includegraphics[width=.9\linewidth]{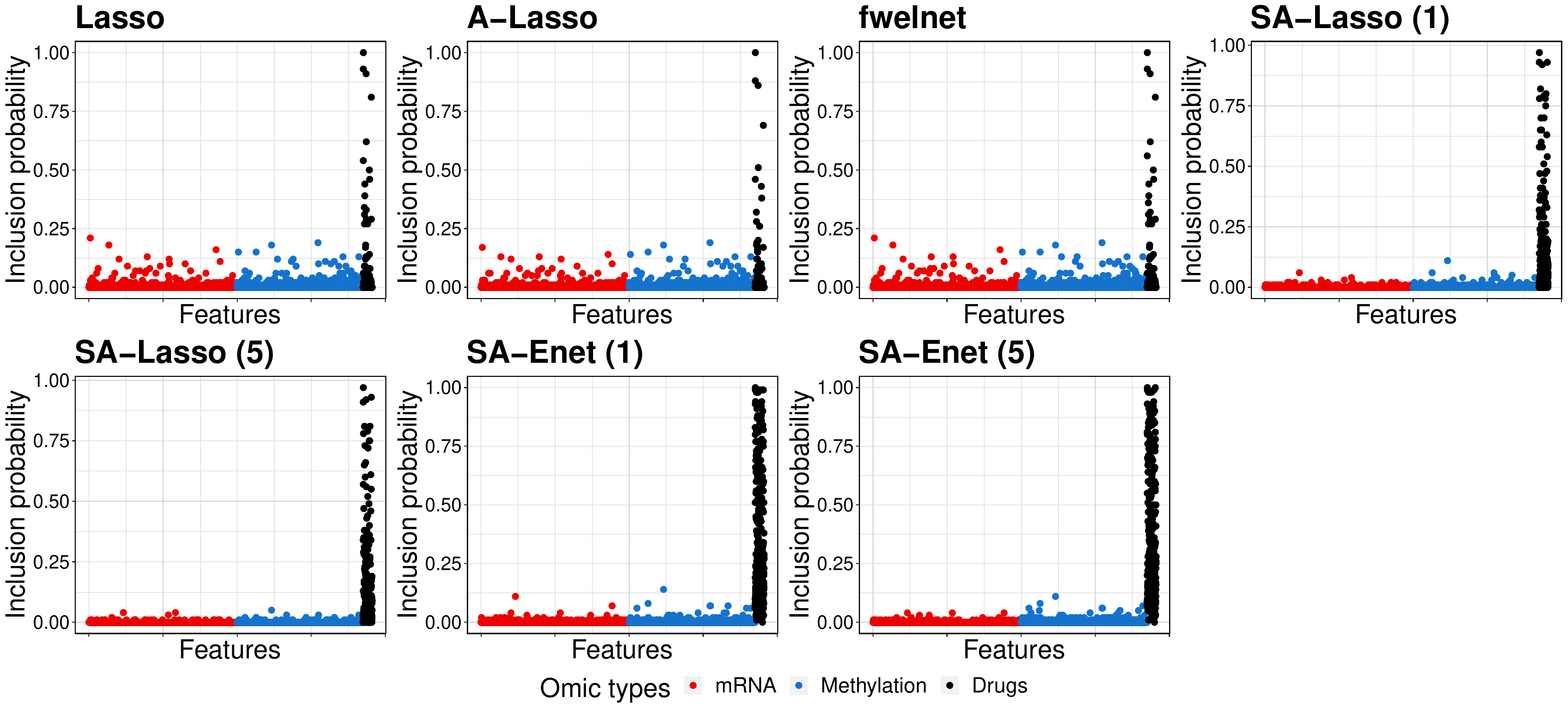}
    \caption{Feature inclusion probability of different methods in Ibrutinib $(0.625 \, \mu M)$ response prediction. In each panel, the vertical axis shows the proportion of times the method selects each feature in the model. This reflects the consistency of feature selection by the methods. The horizontal axis corresponds to the features color-coded according to the omic types.}
    \label{CLL-feature-inclusion-MOFA4}
\end{figure}
\begin{figure}[H]
    \centering
    \includegraphics[width=.9\linewidth]{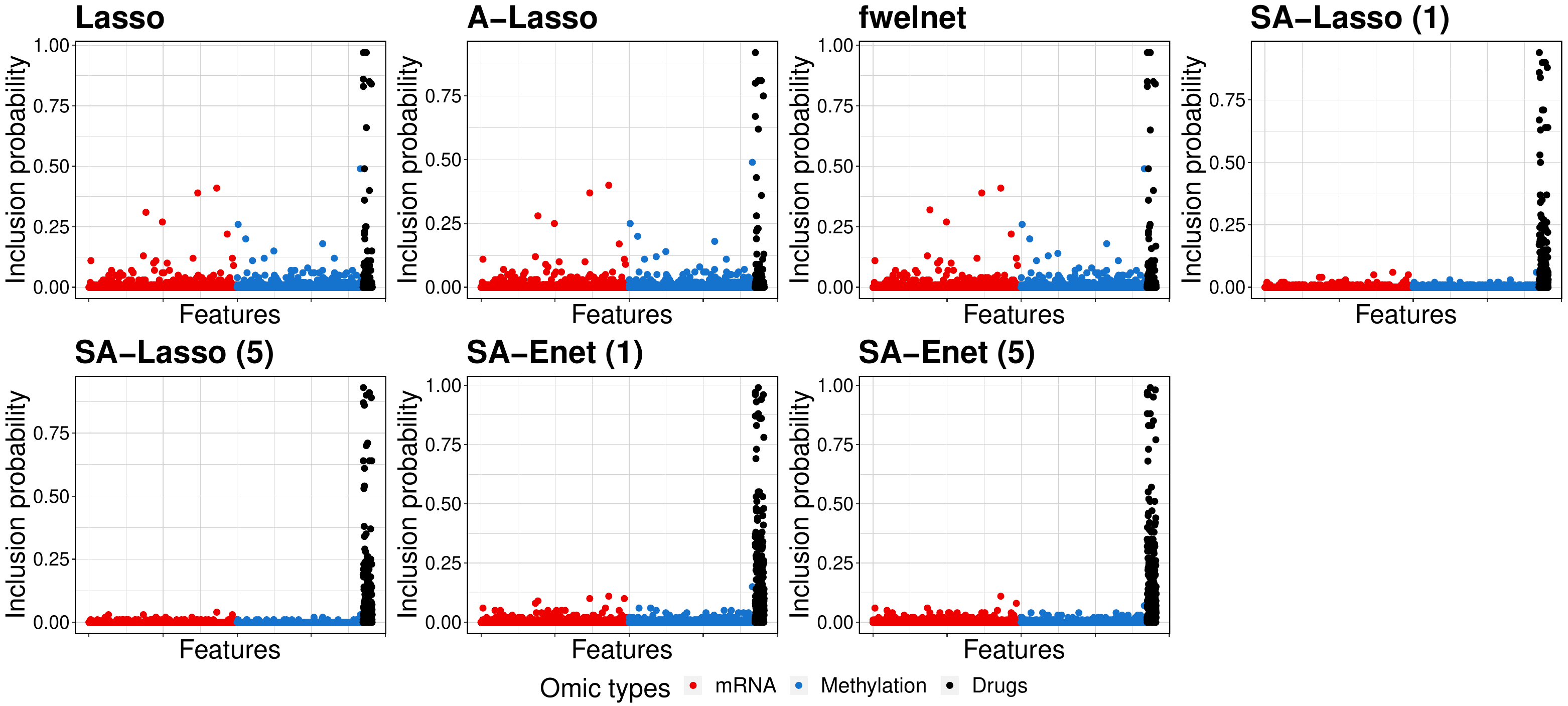}
    \caption{Feature inclusion probability of different methods in Ibrutinib $(0.156 \, \mu M)$ response prediction. In each panel, the vertical axis shows the proportion of times the method selects each feature in the model. This reflects the consistency of feature selection by the methods. The horizontal axis corresponds to the features color-coded according to the omic types.}
    \label{CLL-feature-inclusion-MOFA5}
\end{figure}
The findings from the Figures~\ref{CLL-feature-inclusion-MOFA2}--\ref{CLL-feature-inclusion-MOFA5} are similar to that observed for Ibrutinib $(40 \, \mu M)$ in Section~\ref{sec: Drug response prediction in leukemia samples}.

\subsection{Robustness across iterations}\label{subsec app: Robustness across iterations}

\begin{figure}[h]
	\centering
    \includegraphics[width = .9\linewidth]{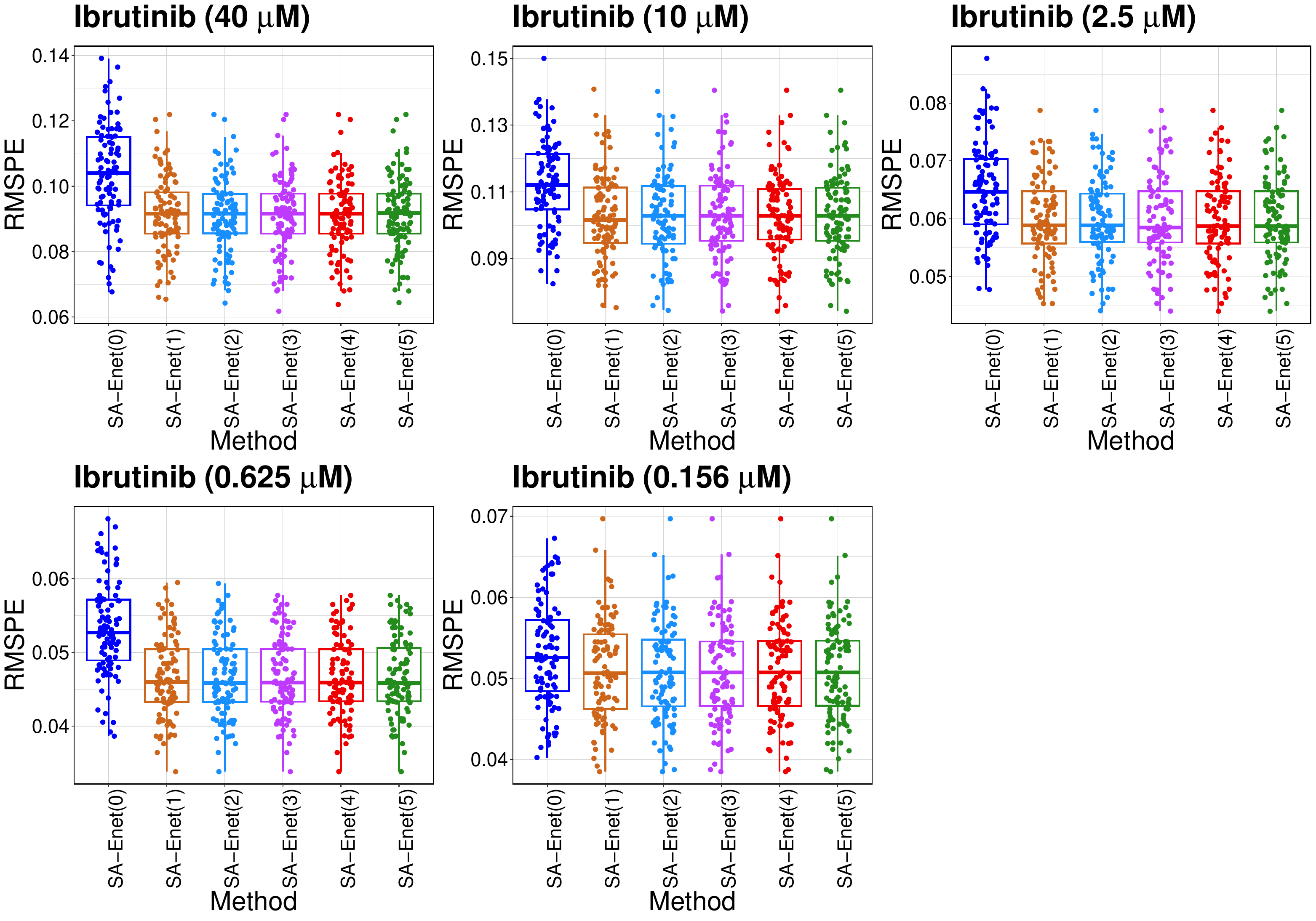}
	\caption{Robustness of the SA-Enet across 5 iterations in each of the 5 regressions.}\label{CLL-robustness wrt iterations}
\end{figure}

In this section, we analyze the robustness of SA-Enet for the real data application presented in Section~\ref{sec: Drug response prediction in leukemia samples}. For each of the five regressions, Figure~\ref{CLL-robustness wrt iterations} compares the RMSPE as we vary the number of iterations $T$ from 1 to 5.
We find that for all regressions the biggest decrease in RMSPE occur at $T=1$. The change in performance afterward is negligible.

\bibliographystyle{apalike}
\bibliography{saenet-ref}


\end{document}